\documentclass[reprint,superscriptaddress,amsmath,amssymb,pra,noeprint]{revtex4-2}
\pdfoutput=1
\usepackage{graphicx}
\usepackage{graphicx}
\usepackage{setspace}
\usepackage[usenames, dvipsnames]{color}
\usepackage{braket}
\usepackage{upgreek}
\usepackage{ulem}
\usepackage[english]{babel}

\newcommand{\fref}[1]{Fig.~\ref{#1}}
\newcommand{\eref}[1]{Eq.~\ref{#1}}

\begin{document}

\title{Controlling the ac Stark effect of RbCs with dc electric and magnetic fields}

\author{Jacob A. Blackmore}
\email[email: ]{j.a.blackmore@durham.ac.uk}
\affiliation{Joint Quantum Centre (JQC) Durham-Newcastle, Department of Physics, Durham University, South Road, Durham, DH1 3LE}

\author{Rahul Sawant}
\altaffiliation[Present Address: ]{Midlands Ultracold Atom Research Centre, School of Physics and Astronomy, Birmingham University, Edgbaston Park Rd, Birmingham, B15 2TT}
\affiliation{Joint Quantum Centre (JQC) Durham-Newcastle, Department of Physics, Durham University, South Road, Durham, DH1 3LE}

\author{Philip D. Gregory}
\affiliation{Joint Quantum Centre (JQC) Durham-Newcastle, Department of Physics, Durham University, South Road, Durham, DH1 3LE}

\author{Sarah L. Bromley}
\affiliation{Joint Quantum Centre (JQC) Durham-Newcastle, Department of Physics, Durham University, South Road, Durham, DH1 3LE}

\author{Jes\'{u}s Aldegunde}
\affiliation{Departamento de Quimica Fisica, Universidad de Salamanca, 37008 Salamanca, Spain}

\author{Jeremy M. Hutson}
\email[email: ]{j.m.hutson@durham.ac.uk}
\affiliation{Joint Quantum Centre (JQC) Durham-Newcastle, Department of Chemistry, Durham University, South Road, Durham, DH1 3LE}

\author{Simon L. Cornish}
\email[email: ]{s.l.cornish@durham.ac.uk}
\affiliation{Joint Quantum Centre (JQC) Durham-Newcastle, Department of Physics, Durham University, South Road, Durham, DH1 3LE}

\begin{abstract}
We investigate the effects of static electric and magnetic fields on the differential ac Stark shifts for microwave transitions in ultracold bosonic $^{87}$Rb$^{133}$Cs molecules, for light of wavelength $\lambda = 1064~\mathrm{nm}$. Near this wavelength we observe unexpected two-photon transitions that may cause trap loss. We measure the ac Stark effect in external magnetic and electric fields, using microwave spectroscopy of the first rotational transition. We quantify the isotropic and anisotropic parts of the molecular polarizability at this wavelength. We demonstrate that a modest electric field can decouple the nuclear spins from the rotational angular momentum, greatly simplifying the ac Stark effect. We use this simplification to control the ac Stark shift using the polarization angle of the trapping laser. 
\end{abstract}

\maketitle

\section{Introduction}

Experimental interest in ultracold molecules is growing rapidly, spurred on by applications spanning precision measurement~\cite{Zelevinsky2008,Hudson2011,Salumbides2011,Salumbides2013,Schiller2014,ACME2014,Hanneke2016,Cairncross2017,Borkowski2018,ACME2018,Borkowski2019}, state-resolved chemistry~\cite{Krems2008,Bell2009,Ospelkaus2010,Dulieu2011,Balakrishnan2016,Hu2019}, dipolar quantum matter~\cite{Santos2000,Ni2009,Carr2009,Baranov2012,Marco2018}, quantum simulation~\cite{Barnett2006,Micheli2006,Buchler2007,Macia2012,Manmana2013,Gorshkov2013} and quantum information processing~\cite{Demille2002,Yelin2006,Zhu2013,Herrera2014,Ni2018,Sawant2019,Hughes2020}. Two prominent methods have emerged for producing molecular gases in the ultracold regime. The first relies on association of pre-cooled atoms using magneto-association on a Feshbach resonance followed by coherent optical transfer to the rovibronic ground state. When the initial atomic gases are at or near quantum degeneracy, a molecular gas with high phase-space density is produced. Numerous bialkali molecules have been created in this way, including KRb~\cite{Ni2008}, Cs$_2$~\cite{Danzl2008}, Rb$_2$~\cite{Lang2008}, RbCs~\cite{Takekoshi2014,Molony2014}, NaK~\cite{Park2015,Seesselberg2018b,Yang2019}, NaRb~\cite{Guo2016} and NaLi~\cite{Rvachov2017}. Significant progress is also being made towards producing molecules from mixtures of open-shell and closed-shell atoms~\cite{Hara2011,Roy2016,Guttridge2018,Barbe2018,Frye2019,Green2019}. The second method employs direct laser cooling of molecules. Although molecules have complex level structures that make them difficult to cool, there are a few that have almost closed electronic transitions suitable for laser cooling. So far, laser cooling has been demonstrated for SrF~\cite{Shuman2010,Barry2014,McCarron2015,Norrgard2016}, YO~\cite{Hummon2013}, CaF~\cite{Zhelyazkova2014,Truppe2017,Anderegg2018}, YbF~\cite{Lim2018} and SrOH~\cite{Kozyryev2017}, with several other species being pursued~\cite{Hunter2012,Chen2017,Iwata2017,Aggarwal2019}.
 
There are many proposals that use polar molecules confined in optical lattices for the simulation of novel problems in many-body physics~\cite{Barnett2006,Micheli2006,Buchler2007,Micheli2007,Pollet2010,Capogrosso_Sansone2010,Macia2012,Manmana2013,Lechner2013,Gorshkov2013}. These proposals exploit long-range dipole-dipole interactions between molecules, which can be controlled using dc electric or microwave fields. If the molecules are permitted to tunnel between lattice sites, novel quantum phases, including super-solids and spin glasses, are predicted to emerge~\cite{Buchler2007,Micheli2007,Pollet2010,Capogrosso_Sansone2010,Macia2012,Lechner2013,Gorshkov2013}. Alternatively, if the molecules are pinned to the lattice sites, pseudo-spin excitations encoded in the rotational states of the molecule can propagate in the lattice due to dipolar spin-exchange interactions. 
This allows exploration of Hamiltonians relevant to quantum magnetism~\cite{Barnett2006,Micheli2006,Gorshkov2011,Gorshkov2011b,Zhou2011,Manmana2013,Hazzard2013}.  
The rich rotational and hyperfine structure of molecules expands the range of Hamiltonians that are accessible. In both cases, the majority of proposals require high occupancy in the lattice. For atoms, high occupancy may be achieved via the Mott-insulator transition~\cite{Greiner2002}. 
Extending this to two atomic species and then forming molecules from the atomic pairs can lead to high occupancy for the molecules, as demonstrated for ground-state KRb~\cite{Moses2015} and RbCs Feshbach molecules~\cite{Reichsoellner2017}.

To realise a useful simulator, we need to confine molecules in a lattice or optical tweezers in a way that fulfils several criteria.  First, the molecules must be adequately confined. This requires a peak laser intensity of at least a few kW\,cm$^{-2}$. Second, the one-body lifetime, limited by evaporation and off-resonant light scattering, must be much greater than the timescales associated with the evolution of the Hamiltonian under investigation. Third, the ac Stark effect must be controlled such that the coherence time is longer than the characteristic intersite interaction time. Bialkali molecules typically have permanent electric dipole moments of $\sim1$~D, which for a lattice spacing of $\sim500$~nm leads to a dipole-dipole interaction energy $\sim h\times1$~kHz. Rotational coherence times greater than 10\,ms are therefore needed. Finally, for molecules like RbCs, which are transferred to the rovibronic ground state from a Feshbach state, it is desirable that the Feshbach state and the ground state have very similar polarizabilities. This prevents excitation into higher bands of the lattice during the optical transfer to the ground state~\cite{Lang2008,Danzl2010}. In the case of RbCs molecules, it has been predicted~\cite{Vexiau2017} that this last condition should be satisfied for a lattice wavelength near 1064\,nm, motivating the present work.

In this paper, we investigate the ac Stark effect for $^{87}$Rb$^{133}$Cs molecules (hereafter referred to simply as RbCs) in light of wavelength $\lambda=1064$~nm. We show that application of a dc electric field parallel to the magnetic field can dramatically simplify the microwave spectrum, as predicted theoretically \cite{Aldegunde2008, Aldegunde:spectra:2009, Ran:2010}. We demonstrate that this simplification extends to the ac Stark affect. With a careful choice of laser polarization, RbCs in an optical lattice  will be suitable for quantum simulation using the $N=0$ and $N=1$ rotational states. En route, we measure the lifetime of RbCs in the optical trap at $\lambda = 1064$~nm, revealing several unanticipated two-photon resonances, and accurately determine the isotropic and anisotropic polarizabilities at this wavelength. 

The paper is structured as follows. In Sec.~\ref{sec:theory}, we present the theory describing the energy-level structure of RbCs in magnetic, electric and optical fields. We demonstrate how the use of an electric field simplifies the ac Stark effect and permits the creation of magic-angle traps. In Sec.~\ref{sec:expt}, we briefly present our experimental setup. In Sec.~\ref{sec:trapping}, we investigate the lifetime of RbCs molecules in the 1064\,nm trap. In Sec.~\ref{sec:polarizability} we report measurements of the isotropic and anisotropic polarizabilities at this wavelength. In Sec.~\ref{sec:magic}, we use these results to control the ac Stark effect, with a view to achieving long rotational coherence times for trapped molecules. Finally, in Sec.\ref{sec:outlook} we conclude and consider the implications of our results for quantum science with RbCs molecules.

\section{Theory}\label{sec:theory}
We choose to focus on the rotational transition, $N=0\rightarrow1$, and aim to minimise differential ac Stark shifts of this transition for suitable trap depths. To achieve this, we need a detailed understanding of the internal structure of the molecule in the presence of magnetic, electric and optical fields. In this section we describe a general Hamiltonian appropriate for diatomic molecules in the lowest vibrational state of a $^1\Sigma$ electronic state and then apply it to the specific case of RbCs under experimentally relevant conditions.

\subsection{The Molecular Hamiltonian}
The effective Hamiltonian for a diatomic molecule in a single vibrational level $v$ of a $^1\Sigma$ electronic state is~\cite{BrownandCarrington,Aldegunde2008}
\begin{equation}\label{eq:Hamiltonian}
H = H_\mathrm{rot} + H_\mathrm{hf} + H_\mathrm{ext},
\end{equation}
where $H_\mathrm{rot}$ is the Hamiltonian associated with the rotational degree of freedom, $H_\mathrm{hf}$ is the hyperfine structure, and $H_\mathrm{ext}$ is associated with interaction between the molecule and external fields. We will discuss each of these terms below.

The rotational structure for low-lying rotational states is well described by~\cite{Herzberg,BrownandCarrington,Aldegunde2008}
\begin{equation}
H_\mathrm{rot} = B_{v}\boldsymbol{N}^2 -D_{v} \boldsymbol{N}^2 \boldsymbol{N}^2.
\end{equation}
Here $\boldsymbol{N}$ is the rotational angular momentum operator, $B_v$ is the rotational constant and $D_v$ is the centrifugal distortion constant. The eigenstates of this Hamiltonian are simply the spherical harmonics; we denote these states by kets $\ket{N,M_N}$, where $N$ is the rotational quantum number and $M_N$ is the projection of the rotational angular momentum onto the space-fixed $z$ axis. 

The hyperfine term in \eref{eq:Hamiltonian} has four contributions \cite{Aldegunde2008, Aldegunde2017}
\begin{equation}
    	H_\mathrm{hf}= H_\mathrm{quad} + H^{(0)}_{II} + H^{(2)}_{II} + H_{NI},
\end{equation}
where
\begin{subequations}
	\begin{gather}
	H_\mathrm{quad} = \sum_{j=\mathrm{Rb}, \mathrm{Cs}} e\boldsymbol{Q}_{j} \cdot \boldsymbol{q}_{j},\label{seq:NuclearQuad}\\
	H^{(0)}_{II} =c_{4} \boldsymbol{I}_{\mathrm{Rb}} \cdot \boldsymbol{I}_{\mathrm{Cs}},\\
	H^{(2)}_{II}=-c_{3}\sqrt{6}\boldsymbol{T}^2(C)\cdot \boldsymbol{T}^2\left(\boldsymbol{I}_\mathrm{Cs},\boldsymbol{I}_\mathrm{Rb}\right)\\
	H_{NI} = \sum_{j=\mathrm{Rb}, \mathrm{Cs}} c_{j} \boldsymbol{N} \cdot \boldsymbol{I}_{j}.
	\end{gather}
\end{subequations}
$H_\mathrm{quad}$ represents the interaction between the nuclear electric quadrupole of nucleus $j$ ($e\boldsymbol{Q}_j$) with the electric field gradient at the nucleus ($\boldsymbol{q}_j$). $H^{(0)}_{II}$ and $H^{(2)}_{II}$ are the scalar and tensor nuclear spin-spin interactions, with strengths governed by the coefficients $c_4$ and $c_3$. The second-rank tensors $\boldsymbol{T}^2$ describe the angular dependence and anisotropy of the interactions~\cite{Aldegunde2017}. $H_{NI}$ is the interaction between the nuclear magnetic moments and the magnetic field generated by the rotating molecule and has a coupling constant $c_j$ for each of the two nuclei. 

The term $H_\mathrm{ext}$ describes interactions between the molecule and external fields. Here this includes the Zeeman effect, the dc Stark effect and the ac Stark effect,
\begin{equation}
H_\mathrm{ext}  = H_\mathrm{Zeeman} +H_\mathrm{dc}+H_\mathrm{ac}.\\
\end{equation}

The Zeeman term describes interaction of the rotational and nuclear magnetic moments with the external magnetic field ($\boldsymbol{B}$) and is
\begin{equation}
    \begin{split}
    H_\mathrm{Zeeman} =
    &-g_{\mathrm{r}} \mu_{\mathrm{N}} \boldsymbol{N} \cdot \boldsymbol{B} \\
    &-\sum_{j=\mathrm{Rb}, \mathrm{Cs}} g_{j}\left(1-\sigma_{j}\right) \mu_{\mathrm{N}} \boldsymbol{I}_{j} \cdot \boldsymbol{B}.\label{seq:Zeeman}
    \end{split}
\end{equation}
The first term accounts for the magnetic moment generated by the rotation of the molecule, characterised by the rotational $g$-factor $g_\mathrm{r}$. The second term accounts for the nuclear spin contributions, characterised by the nuclear $g$-factors $g_j$ shielded isotropically by the factor $\sigma_j$~\cite{Aldegunde2008}. In both terms $\mu_\textrm{N}$ is the nuclear magneton. For our analysis we designate the axis of the magnetic field $\boldsymbol{B}$ as the space-fixed $z$ axis and its magnitude as $B_z$.

The dc Stark effect describes the interaction between the molecular dipole moment and a static electric field $\boldsymbol{E}$, and is
\begin{equation}
H_\mathrm{dc}= -\boldsymbol{\mu}\cdot\boldsymbol{E}\label{seq:DC_Stark}.
\end{equation}
The matrix elements of the dipole moment operator, $\boldsymbol{\mu}$, are ~\cite{BrownandCarrington}
\begin{equation}
\begin{split}
&\bra{N,M_N}\boldsymbol{\mu}\ket{N',M_N'}=\\
&\sum_{i=-1,0,+1}\mu_{v} \hat{i}\sqrt{(2 N+1)\left(2 N^{\prime}+1\right)} (-1)^{M_{N}} \\
&\times\left(\begin{array}{ccc}{N} & {1} & {N^{\prime}} \\ {-M_{N}} & {i} & {M_{N}^{\prime}}\end{array}\right)\left(\begin{array}{ccc}{N} & {1} & {N^{\prime}} \\ {0} & {0} & {0}\end{array}\right).
\end{split}
\end{equation}
where $\hat{i}$ is a unit vector along the axes $x-\textrm{i}y$, $z$ and $x+\textrm{i}y$ for $i = -1$, 0, +1 respectively. $\mu_{v}$ is the permanent dipole moment of the molecule in the molecule-fixed frame and the symbols in parentheses are Wigner 3j symbols. $i=-1$, $0$, $1$ correspond to $\sigma_+,\pi,\sigma_-$ transitions, respectively, which can change $M_N$ by $+1$, $0$, $-1$ when a photon is absorbed. We restrict $\boldsymbol{E}$ to lie along $z$, with magnitude $E_z$, so that only the term with $i=0$ contributes to the dc Stark effect. 

The ac Stark effect arises from the interaction of an off-resonant oscillating electric field $\boldsymbol{E}_\mathrm{ac}$ with the frequency-dependent molecular polarizability tensor $\boldsymbol{\alpha}$~\cite{Gregory2017}. It is
\begin{equation}
    H_\mathrm{ac}= -\frac{1}{2} \boldsymbol{E}_\mathrm{ac}\cdot\boldsymbol{\alpha}\cdot\boldsymbol{E}_\mathrm{ac}.\label{seq:AC_Stark}
\end{equation}
For a linear diatomic molecule, the polarizability along the internuclear axis, $\alpha_\parallel$, is greater than that perpendicular to it, $\alpha_\perp$. The overall polarizability of the molecule is therefore anisotropic. For a molecule oriented at an angle $\theta$ to the laser polarization, the polarizability is
\begin{equation}\label{eq:Polarizability}
\begin{split}
\alpha(\theta) &= \alpha_{\parallel}\cos^{2}\theta + \alpha_{\perp}\sin^{2}\theta \\
							 &= \alpha^{(0)} + \alpha^{(2)}P_{2}(\cos\theta),
\end{split}
\end{equation}
where $\alpha^{(0)}=\frac{1}{3}(\alpha_{\parallel}+2\alpha_{\perp})$,
$\alpha^{(2)}=\frac{2}{3}(\alpha_{\parallel}-\alpha_{\perp})$ and $P_2(x) =(3x^2-1)/2$ is the second-order Legendre polynomial~\cite{Gregory2017}. We consider the case where the laser is polarized in the $xz$ plane at an angle $\beta$ to the $z$ axis. To transform the polarizability from the molecular frame to laboratory coordinates requires a rotation through angle $\beta$ about $y$. The matrix elements of \eref{seq:AC_Stark} in the basis set $\ket{N,M_N}$ are \cite{Gregory2017}
\begin{equation}
\begin{split}
&\bra{N,M_N}H_\mathrm{ac}\ket{N',M_N'} = -\frac{I\alpha^{(0)}}{2\epsilon_0c}\delta_{N,N'}\delta_{M_N,M_N'}\\
-&\frac{I \alpha^{(2)}}{2\epsilon_0c} \sum_{M} d_{M 0}^{2}(\beta)(-1)^{M_{N}^{\prime}} \sqrt{(2 N+1)\left(2 N^{\prime}+1\right)}\\
\times&\left(\begin{array}{ccc}{N^{\prime}} & {2} & {N} \\ {0} & {0} & {0}\end{array}\right)\left(\begin{array}{ccc}{N^{\prime}} & {2} & {N} \\ {-M_{N}^{\prime}} & {M} & {M_{N}}\end{array}\right).
\label{eq:ACStark_elements}
\end{split}
\end{equation}
Here $I$ is the laser intensity, $\delta_{A,B}$ is a Kronecker delta and $d^2_{M0}(\beta)$ is a reduced Wigner rotation matrix.

The term proportional to the isotropic part, $\alpha^{(0)}$, produces an equal energy shift of all $(N,M_N)$. 
The term proportional to the anisotropic part, $\alpha^{(2)}$, has more complicated behavior: for $N>0$ it has elements both diagonal and off-diagonal in $M_N$ that depend on $\beta$. If we consider only matrix elements with $N=N'=1$, the matrix representation of the term proportional to $I\alpha^{(2)}$ is
\begin{equation}
\begin{split}
&\left(\frac{I \alpha^{(2)}}{5\times 2\epsilon_0 c}\right) \\
&\times \left(
\begin{array}{ccc}
{P_{2}(\cos \beta)} & {\frac{3}{\sqrt{2}} \sin \beta \cos \beta} & {-\frac{3}{2}\sin^2\beta} \\
{\frac{3}{\sqrt{2}} \sin \beta \cos \beta} & {-2 P_{2}(\cos \beta)} & {-\frac{3}{\sqrt{2}} \sin\beta \cos\beta} \\ 
{-\frac{3}{2} \sin^2 \beta}& -\frac{3}{\sqrt{2}}\sin\beta\cos\beta & {
	P_{2}(\cos \beta)}\\
\end{array}\right),
\end{split}
\end{equation}
where rows and columns are in the order $M_N$, $M_N' = -1$, 0, 1.

We construct the Hamiltonian in a fully uncoupled basis set. The eigenstates are
\begin{equation}\label{eq:Hyperfine states}
    \ket{\psi_\mathrm{hf}} = \sum_{N,X} A_X \ket{N,M_N}\ket{i_\mathrm{Rb},m_\mathrm{Rb}}\ket{i_\mathrm{Cs},m_\mathrm{Cs}},
\end{equation}
where $\ket{i_j,m_j}$ represents a basis function for nucleus $j$ with nuclear spin $i_j$ and projection $m_j$ and $X$ collectively represents the quantum numbers $M_N$, $m_\mathrm{Rb}$ and $m_\mathrm{Cs}$. At the fields we consider here, it is sufficient to include basis functions with $N\le 5$. The mixing of different values of $N$ is relatively weak; $N$ is not a good quantum number, but it remains a useful label for the eigenstates. 

The hyperfine term in the Hamiltonian mixes states with different values of $M_N$, $m_\mathrm{Rb}$ and $m_\mathrm{Cs}$, though it conserves their sum, $M_F = M_N + m_\mathrm{Rb} + m_\mathrm{Cs}$. It is also weakly off-diagonal in $N$. Accordingly, the eigenstates can be mixtures of many different rigid-rotor and nuclear-spin states, even in the absence of external fields.

The Zeeman, dc Stark and ac Stark terms are diagonal in $m_\mathrm{Rb}$ and $m_\mathrm{Cs}$. The Zeeman term is also diagonal in $N$ and $M_N$. For an electric field along $z$, the dc Stark term is diagonal in $M_N$ but is off-diagonal in $N$ with $\Delta N=\pm 1$. When $\beta=0$, so that the trapping laser is polarized along $z$, $H_\mathrm{ac}$ is diagonal in $M_N$, but weakly off-diagonal in $N$ with $\Delta N=0, \pm 2$. Under these circumstances, the entire Hamiltonian is diagonal in $M_F$, so it is a good quantum number. For $\beta\neq0$, however, $H_\mathrm{ac}$ is off-diagonal in $M_N$ and hence in $M_F$. In this case no good quantum numbers exist.
\subsection{Application to RbCs}
\begin{figure}[t]
	\centering
	\includegraphics[width=0.45\textwidth]{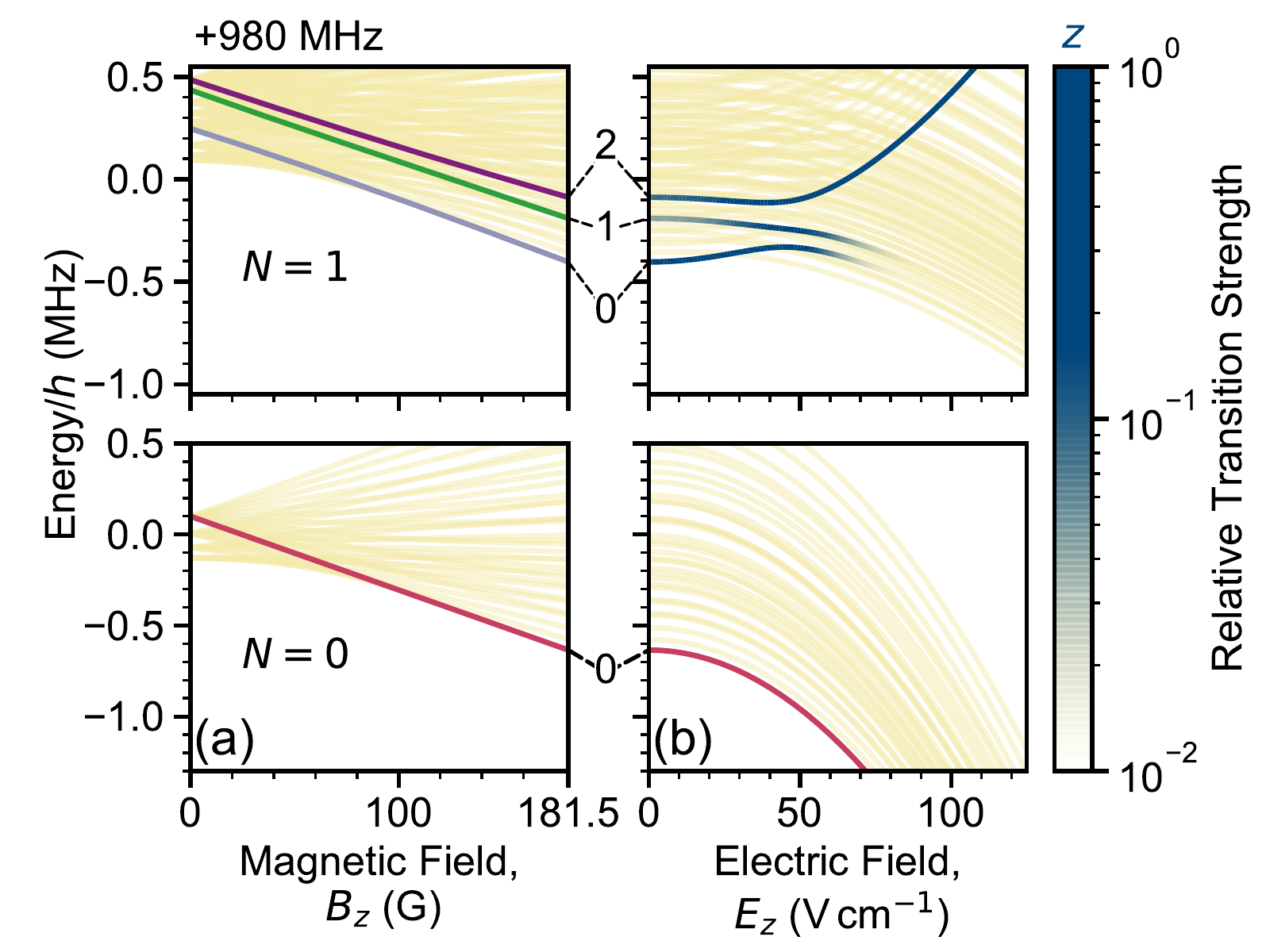}
	\caption{ 
	The hyperfine structure of RbCs.  (a) as function of dc magnetic field. For $N=0$, the state $(N,M_F)_i = (0,5)_0$ is highlighted in red and labeled as state 0. For $N=1$, the states $(1,5)_i$ are highlighted and labeled with $i=0$, 1, 2.	(b) as function of dc electric field with magnetic field fixed at $B_z = 181.5$~G. The relative transition strength for microwaves polarized along $z$ is shown by the blue color map.
	}
	\label{fig:Theory}
\end{figure}
We now apply the full Hamiltonian of \eref{eq:Hamiltonian} to \eref{eq:ACStark_elements} to RbCs molecules in the $v=0$ state of the $^1\Sigma^+$ electronic state, as used in our experiments. 
We use the spectroscopic constants tabulated in ref.\ \cite{Gregory2016}, determined from microwave spectroscopy or from the calculations of ref.\ \cite{Aldegunde2008} where necessary.
The nuclear spins of $^{87}$Rb and $^{133}$Cs are $i_\mathrm{Rb}=3/2$ and $i_\mathrm{Cs}=7/2$, respectively. At small magnetic fields ($B_z < 10~\mathrm{G}$) these couple to one another and to the rotational angular momentum to form a resultant $F$. For $N=0$ there are four zero-field levels with $F=2$, 3, 4, 5; these split into a total of ${(2N+1)(2i_\mathrm{Rb}+1)(2i_\mathrm{Cs}+1)=32}$ Zeeman sub-levels in a magnetic field. For $N=1$, the number of sub-levels is increased to 96 because of the additional rotational angular momentum. We designate individual sub-levels $(N,M_F)_i$, where $i$ distinguishes between those with the same values of $N$ and $M_F$, in order of increasing energy at $181.5$~G; the lowest sub-level for each $(N,M_F)$ is designated $i$=0.

In \fref{fig:Theory}(a) we show the energies of the sub-levels for $N=0$ and 1 as a function of magnetic field in the absence of dc and ac electric fields. In our apparatus, we produce RbCs molecules at a magnetic field of 181.5~G in the lowest hyperfine sublevel, $(N,M_F)_i = (0,5)_0$, which has well-defined nuclear spin projections $m_\mathrm{Rb}=3/2$ and $m_\mathrm{Cs}=7/2$. At this magnetic field, $F$ is no longer a good quantum number but transitions with $\Delta N =1$ and $\Delta M_F = 0,\pm1$  have significant strength. In \fref{fig:Theory}(a), we have highlighted the three  sub-levels of $N=1$ with $M_F=5$. Because of the nuclear quadrupole interaction, \eref{seq:NuclearQuad}, these states have components with different nuclear spin projections; this can be exploited to change the nuclear spin state in the ground rotational state~\cite{Aldegunde:spectra:2009,Ospelkaus2010,Will2016,Gregory2016,Guo2018}.

In \fref{fig:Theory}(b) we show the level energies at 181.5~G as a function of a small electric field $E_z< 150~\mathrm{V\,cm^{-1}}$, parallel to the magnetic field. For $N=0$ the sub-levels are parallel to one another as a function of electric field, because every sub-level has the same rotational projection $M_N=0$. For $N=1$, by contrast, there is a pattern of crossings and avoided crossings as the states split according to $|M_N|$. The branch with $M_N=0$ is higher in energy than that with $M_N=\pm1$. The strength of the blue highlighting in \fref{fig:Theory}(b) indicates the value of the relative transition strength for microwaves polarized along $z$. This is proportional to $|\mu_{0i}^z|^2$, where $\mu_{0i}^z$ is the $z$ component of the transition dipole moment $\boldsymbol{\mu}_{0i}= \bra{0}\boldsymbol{\mu}\ket{i}$; here $\ket{0}$ and $\ket{i}$ are the state vectors for the sub-levels $(0,5)_0$ and $(1,5)_i$. 
Of the three transitions allowed for microwaves polarized along $z$ in parallel fields, only one retains its strength as electric field increases. This is the one that is predominantly $M_N=0$ at high field, and correlates with $(1,5)_2$ at zero field. By applying an electric field, we suppress the effect of the nuclear spins on the internal structure. 

\begin{figure}[t]
	\centering
	\includegraphics[width=0.45\textwidth]{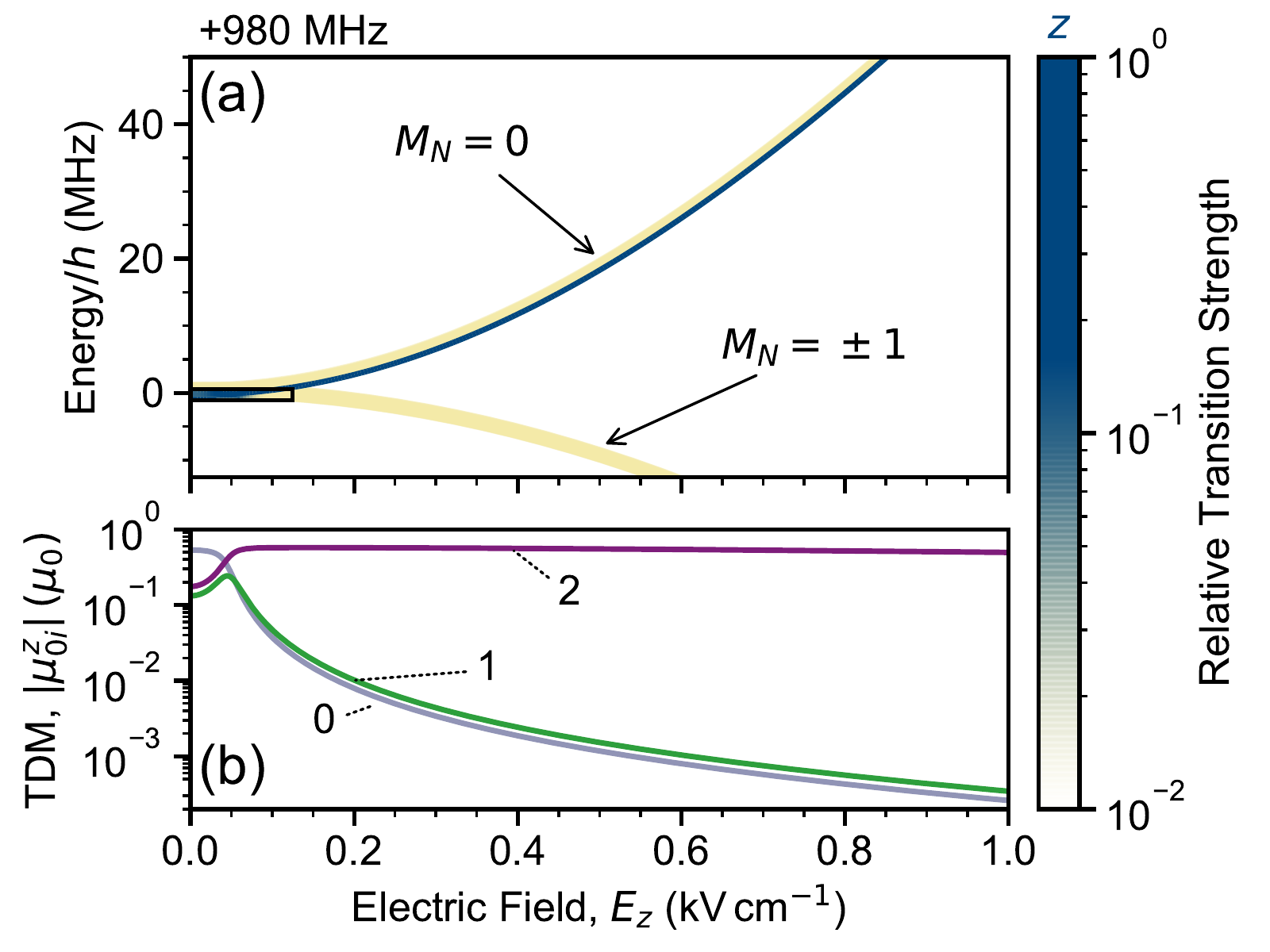}
	\caption{(a) The energy levels of RbCs ($N=1)$ in a dc electric field. The relative transition strengths are coded as in \fref{fig:Theory}(b). The box indicates the region shown in \fref{fig:Theory}(b). (b) The transition dipole moments of the transitions $(0,5)_0\rightarrow (1,5)_i$ for $i$=0,1,2 (as labeled in \fref{fig:Theory}(a)) as a function of the electric field. 
	}
	\label{fig:Theory2}
\end{figure}
In \fref{fig:Theory2}(a) we increase the electric field further to $1000~\mathrm{V\,cm^{-1}}$. The dc Stark shift is approximately quadratic; at 400~V$\,$cm$^{-1}$ only the state $(1,5)_2$ has any appreciable transition dipole moment from $(0,5)_0$. We show the evolution of the relative transition strength in \fref{fig:Theory}(b) by shading the appropriate energy level in the Stark map; we also show the numerical values of $|\mu^z_{0,i}|$ in \fref{fig:Theory2}(b) for the states highlighted in \fref{fig:Theory}(a). Although there are 32 sublevels for each $(N,M_N)$, split by the nuclear Zeeman interaction, the hyperfine coupling no longer has a significant impact on the level structure for states that are predominantly $M_N=0$. This implies that a simpler Hamiltonian can be used to explain the internal structure; for this we drop $H_\mathrm{hf}$ and $H_\mathrm{Zeeman}$ from \eref{eq:Hamiltonian} to leave a simpler hindered-rotor Hamiltonian~\cite{Wei2011,Zhu2013,Li2017}. The eigenstates of this simpler system do not depend on the nuclear spin degrees of freedom and can be constructed from the spherical harmonics alone~\cite{Wei2011,Zhu2013},
\begin{equation}\label{eq:pendular}
\ket{\psi_\textrm{hindered}} = \sum_{N'} A_{N'} \ket{N',M_N}.
\end{equation}
\begin{figure}[t!]
	\centering
	\includegraphics[width=0.45\textwidth]{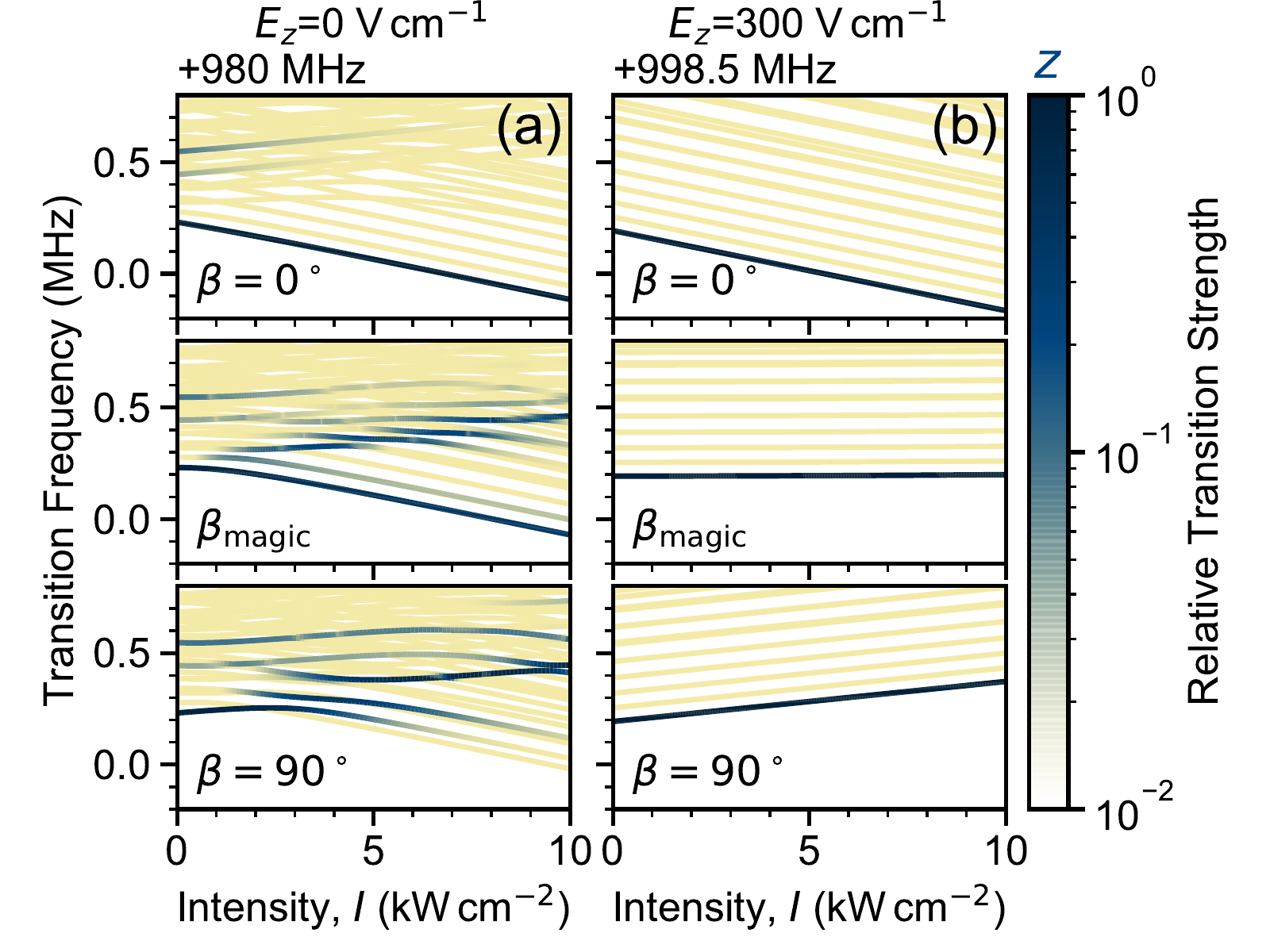}
	\caption{Frequencies of the transitions $(0,5)_0 \rightarrow (1,5)_i$ in a dc magnetic field of 181.5~G, as a function of laser intensity, for three polarization angles $\beta =0^\circ,$ $\beta_\mathrm{magic}$ and $90^\circ$. The relative transition strengths are coded as in \fref{fig:Theory}(b). 
	(a) At zero dc electric field. (b) At an electric field of 300~V$\,$cm$^{-1}$. 
	Note the different offsets on the transition frequency axes due the dc Stark shift. The calculations use the values of $\alpha^{(0)}$ and $\alpha^{(2)}$ determined experimentally in Section~\ref{sec:polarizability}.
	}
	\label{fig:Theory3}
\end{figure}

The application of an electric field greatly simplifies the hyperfine structure, and leads to more linear ac Stark shifts in the presence of $\boldsymbol{E}_\textrm{ac}$. In \fref{fig:Theory3} we show the shifts of the  transitions with $\Delta M_F = 0$ as a function of laser intensity, with a dc magnetic field of 181.5~G, for three laser polarization angles: $\beta = 0^\circ,\, \beta_\mathrm{magic}$ and $90^\circ$. Here $\beta_\mathrm{magic}$ is the ``magic angle'' that occurs at the point where $P_2(\cos \beta)=0$, which is $\beta_\mathrm{magic}\approx54.7^\circ$~\cite{Kotochigova2010,Neyenhuis2012,Seesselberg2018}. At this angle the diagonal elements of \eref{eq:ACStark_elements} are reduced to zero. In \fref{fig:Theory3}(a) we show the ac Stark shift without an electric field. For $\beta\neq0^\circ$, the intensity dependence is non-trivial, with a rich pattern of avoided crossings. We have previously shown that, whilst these avoided crossings can be used to minimise the differential ac Stark shift locally for certain transitions~\cite{Blackmore2018}, their effectiveness in producing an intensity-insensitive trap is limited. In this region the ac Stark shift can be adequately quantified only by the gradient of the transition frequency $f$ with respect to intensity $I$. This gives a measure of the ``local'' differential polarizability, which depends strongly on intensity. We contrast this with the behavior in \fref{fig:Theory3}(b), where an electric field of $300~\mathrm{V\,cm^{-1}}$ has been introduced. The states are now well represented by \eref{eq:pendular}, and this leads to $\mathrm{d}f/\mathrm{d}I$ independent of intensity. Most importantly, $\mathrm{d}f/\mathrm{d}I$ can be reduced to zero at the magic angle~\cite{Kotochigova2010,Neyenhuis2012,Seesselberg2018}.

The simplified internal structure is less sensitive to variation in the external trapping potential. For molecules at a temperature of $1.5~\mathrm{\upmu K}$, a suitable trap has depth $\sim k_\mathrm{B} \times 15~\mathrm{\upmu K} \approx h\times 300~\mathrm{kHz}$. This depth is comparable to the hyperfine splitting and so variations in laser intensity as molecules move around the trap can strongly change the structure. By contrast, when states with  different values of $|M_N|$ are separated by more than 1~MHz, as is the case at $E_z > 100~\mathrm{V\,cm^{-1}}$, the ac Stark effect is much weaker than the energy splitting and so sensitivity to such variations is reduced. 

\section{Details of the Experiment}\label{sec:expt} 
Our experimental apparatus and method for creating ultracold RbCs molecules have been discussed in previous publications~\cite{McCarron2011, Koeppinger2014, Molony2014, Gregory2015, Molony2016,Molony2016a}. In this work we create a sample of up to $\sim 3000$ RbCs molecules in their absolute ground state at a temperature of $1.5(1)~\upmu$K by magnetoassociation on an interspecies Feshbach resonance~\cite{Koeppinger2014}, followed by transfer to the rovibronic and hyperfine ground state by stimulated Raman adiabatic passage (STIRAP)~\cite{Molony2014,Molony2016a}. The magnetoassociation is performed in a crossed optical dipole trap (ODT) operating at $\lambda=1550$~nm, but the STIRAP is performed in free space to avoid spatially varying ac Stark shifts~\cite{Molony2016,Molony2016a}. A uniform magnetic field of 181.5~G is present following magnetoassociation and throughout the STIRAP. Once prepared in the ground state, the molecules may be recaptured in the ODT and/or interrogated with microwave fields to drive transitions to rotationally excited states~\cite{Gregory2016}. Detection is performed by reversing the STIRAP and magnetoassociation steps, breaking the molecules apart into their constituent atoms, which are then imaged using conventional absorption imaging. We are therefore able to detect molecules only when they are in the specific rotational and hyperfine state addressed by STIRAP. 

In this work we consider the ac Stark shift at $\lambda=1064$~nm. The 1064~nm light is derived from a Nd:YAG based master-oscilator-power-amplifier (MOPA) system. We monitor the laser frequency using a HighFinesse \mbox{WS-U} wavemeter. The wavemeter output is used as the input to a software-based servo loop which provides feedback to the temperature of the MOPA's Nd:YAG crystal to stabilise the output frequency. Using this approach we are able to stabilise the frequency of the light to $\sim10$~MHz over a tuning range of $\sim 24$~GHz. For trapping we can split the light into two beams which form a second crossed optical trap. For our spectroscopic work we require only a single beam, which has a waist of $173(1)~\upmu\mathrm{m}$; the large waist ensures the molecules experience a nearly homogeneous intensity.

To control the rotational and hyperfine state of the molecule, we use resonant microwaves~\cite{Gregory2016,Gregory2017,Blackmore2018}. The microwaves are emitted by two omni-directional quarter-wave antennas, tuned to the $N=0\rightarrow1$ transition frequency. Each antenna is connected to an independent signal generator, frequency referenced to a 10~MHz GPS reference. The microwave output can be switched on a nanosecond timescale and is controlled externally by TTL synchronised to the experimental sequence. The antennas are oriented perpendicularly to each other; in free space one would produce microwaves linearly polarized along $z$ and the other would produce microwaves linearly polarized in the $xy$ plane. We find that both antennas actually generate a strong component polarized along $z$; we attribute this to the presence of magnetic field coils that are separated along $z$ by significantly less than a quarter of the microwave wavelength. 

For the generation of dc electric fields, two pairs of electrodes are mounted around the UHV fused silica cell in which the ultracold molecules are prepared~\cite{Molony2014}. The electrodes are arranged such that the electric field maximum is located at the position of the ODT. As the cell is dielectric, some remnant charge can build up on the glass. We find that limiting the maximum electric field we apply to ${500~\mathrm{V\,cm^{-1}}}$ reduces the impact of the electric field variation on our spectroscopic measurements, within a single experimental run, to less than the $\sim 10~\mathrm{kHz}$ Fourier width associated with our square microwave pulses. Additionally, between experimental runs, high-power UV light (up to 3~W  with wavelength  365~$\mathrm{nm}$)   irradiates the cell for 1\,s, removing any accumulated charges. 

\section{Optical trapping in the ground state}\label{sec:trapping} 
\begin{figure}[t]
    \centering
    \includegraphics[width=0.45\textwidth]{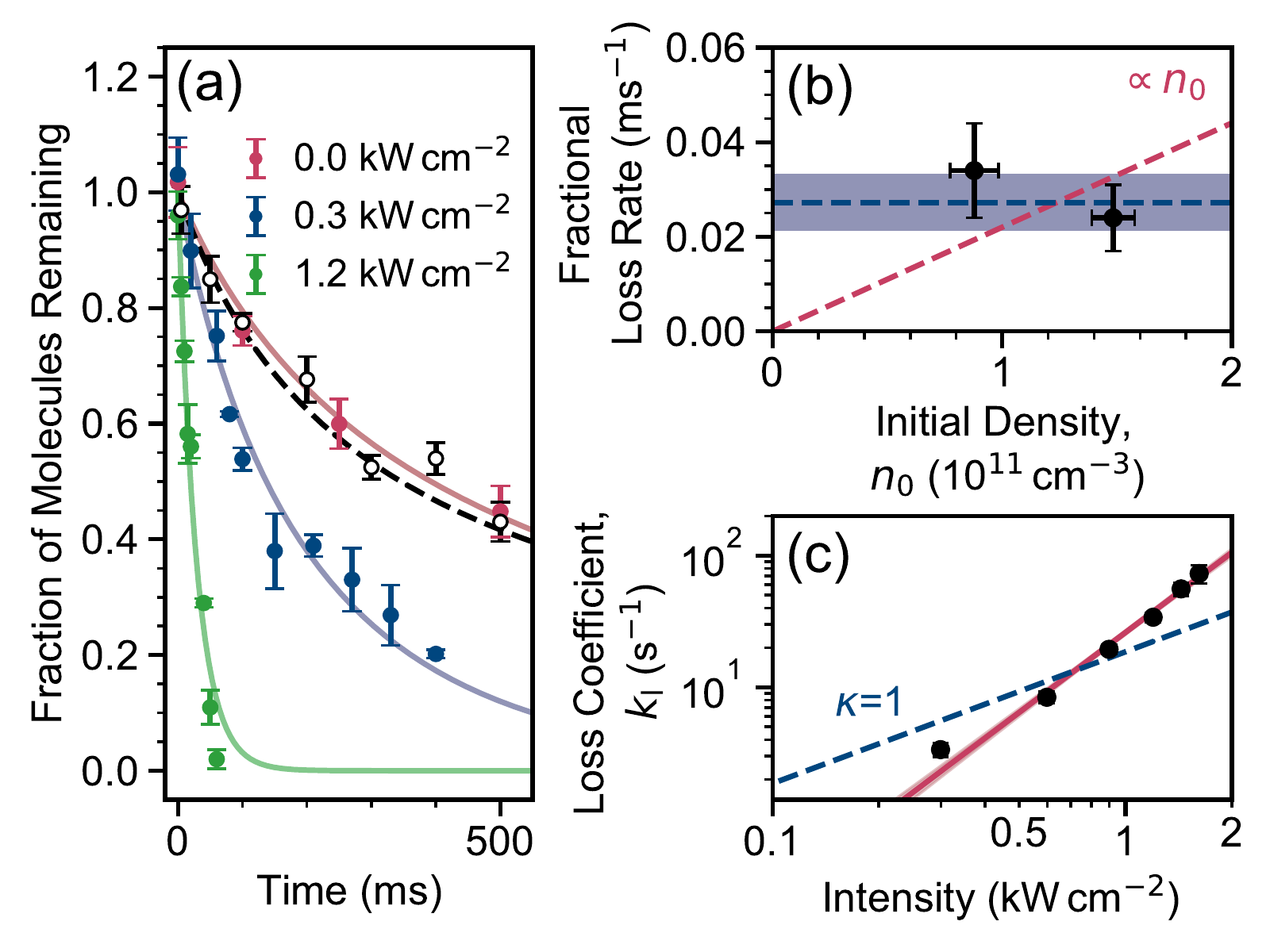}
    \caption{Molecular loss in light of wavelength $\lambda = 1064$~nm. (a)~Molecule loss as a function of time for various laser intensities. The filled points show resonant loss at laser frequency $f_0$; an additional dipole trap with $\lambda=1550$~nm is used for confinement. For the unfilled black points the laser is detuned by $-12.3~\mathrm{GHz}$, and the intensity is increased to 11.2~kW$\,$cm$^{-2}$ without the dipole trap at 1550~nm. (b) Fractional loss rate over the first 15~ms as a function of initial peak density for laser frequency $f_0$. The dashed lines indicate the scaling expected for one-molecule (blue) and two-molecule (red) processes.  (c) Rate coefficient for resonant loss ($k_\textrm{l}$) as a function of laser intensity ($I$). The solid line shows a fit of the form $k_\textrm{l} \propto I^\kappa$, which yields a best-fit value of $\kappa=2.02(8)$; the shaded region indicates the error on the fit. The dashed line shows the scaling expected for a one-photon process. }
    \label{fig:Mechanism}
\end{figure}
\begin{figure*}[t]
    \centering
    \includegraphics[width=0.9\textwidth]{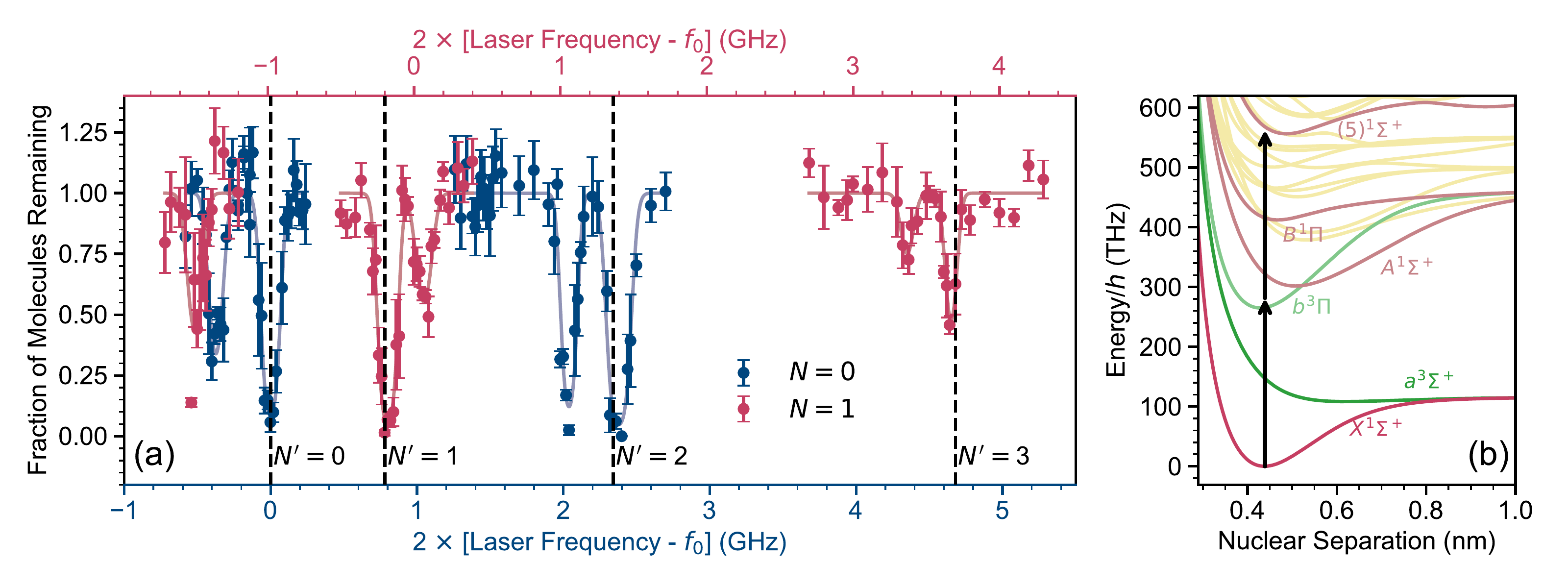}
    \caption{Resonant two-photon loss of ground-state molecules in the presence of light with $\lambda=1064$\,nm. (a) The fraction of molecules trapped with light at $\lambda=1550$\,nm that remain after exposure to light with $\lambda=1064$\,nm, as a function of laser frequency. Measurements are shown for molecules prepared in $N=0$ (blue) or $N=1$ (red). To display the results on a continuous scale corresponding to the excited-state energy, we plot the horizontal axes as twice the laser frequency (offset from $f_0 = 281,634.630(2)$\,GHz) and offset the measurements for $N=1$ by $2\times B_{v=0}/h =980.231$\,MHz on the top axis. Each data point is the average of 3 runs, with error bars indicating the standard deviation; the solid lines are Gaussian fits to the data. The vertical dashed lines indicate the excited-state rotational structure for $B'_{v'}=h\times389.9(4)~\mathrm{MHz}$.  (b) The potential energy curves for low-lying electronic states of RbCs~\cite{Allouche2000}. The black arrows indicate the energies of one- and two-photon transitions at $\lambda =1064$\,nm.   }
    \label{fig:1064_spect}
\end{figure*}
Theoretical analysis shows that the polarizability of RbCs in the ground vibronic state is dominated by transitions to vibrational levels of the $A^1\Sigma$ and $B^1\Pi$ states~\cite{Docenko2010,Vexiau2017}. 
The potential minima of these states lie 300~THz and 410~ THz above the vibronic ground state, respectively. This is higher in energy than the $\sim h\times281~\mathrm{THz}$ provided by one photon with $\lambda = 1064$~nm.
It is worth noting that, due to spin-orbit coupling, the lowest levels of the $b^3\Pi$ electronic state have significant singlet character, which produces strong resonant behavior from $h\times262$~THz to $h\times270$~THz even though singlet-triplet transitions are forbidden. The calculations indicate that the polarizabilities of the ground and Feshbach molecular states vary smoothly from 270 THz to 300 THz (wavelengths from 1020~nm to 1110~nm), and predict that the difference between them crosses zero at 1064.96~nm~\cite{Vexiau2017}. The transitions to vibronic levels of the $b^3\Pi$ state in this wavelength range are suppressed by negligible Franck-Condon factors~\cite{Docenko2010,Vexiau2017}. Nevertheless, during initial attempts to load RbCs molecules into an ODT of $\lambda \approx 1064~\mathrm{nm}$, we observed loss of ground-state molecules that was orders of magnitude faster than the near-universal collisional losses that typically dominate our experiments~\cite{Gregory2019,Gregory2020}.

To investigate the optical losses, we trap ground-state molecules in the crossed optical dipole trap at $\lambda = 1550$~nm and pulse on the light at $\lambda=1064$~nm in a single beam to perform spectroscopy. We find that the loss of molecules is very sensitive to laser frequency, with several resonances in the region accessible to us. To determine the nature of these resonances, we designate one of the resonant frequencies $f_0= 281.634\,630(2)$~THz and stabilise the laser to this frequency; we apply additional modulation of 35~MHz at a rate of 5~kHz to remove error due to frequency tuning. We measure the loss of molecules as a function of time due to the resonant light. Example loss measurements are shown in \fref{fig:Mechanism}(a). 

We examine the density dependence of the loss by reducing the starting number of ground-state molecules, while keeping all other experimental parameters the same. \fref{fig:Mechanism}(b) shows the rate of change of the fraction remaining, over the first 15~ms, for two samples with a factor of 2 difference in the initial density; this is the largest change that we can make whilst still being able to measure an accurate lifetime. The two measurements agree within one standard deviation, so we conclude that the resonant loss is principally a one-body process. 

To investigate the mechanism behind the resonant loss, we model the rate of change of density $n$ as
\begin{equation}
\dot{n}(r,t) = -k_{2} n(r,t)^2 - k_\textrm{l} n(r,t).
\label{Eq:Losseqn}
\end{equation}
We fix the rate coefficient for two-body inelastic loss  ${k_2=4.8\times10^{-11}\mathrm{~cm^{3}\,s^{-1}}}$ at the value measured previously~\cite{Gregory2019} and extract the rate coefficient $k_\textrm{l}$ for resonant loss. We measure $k_\textrm{l}$ as a function of the peak intensity of the beam at $\lambda = 1064~\mathrm{nm}$. The results are shown in ~\fref{fig:Mechanism}(c). We fit the resulting variation with the function $k_\textrm{l}=AI^{\kappa}$, with $A$ as a free parameter. We fix $\kappa=1,2,3$ corresponding to loss due to a one-, two-, or three-photon process, and find $\chi^{2}_{\text{red}}=31$, 4 and 17 respectively. We confirm this fitting by additionally allowing $\kappa$ to vary: we find $\kappa=2.02(8)$ with no significant change in the best value for $\chi^{2}_{\text{red}}$. We conclude that the loss is a two-photon process, with ${A_{\gamma=2}=25(1)~\mathrm{s^{-1}\,(kW\,cm^{-2})^{-2}}}$. 

The resonant loss is a one-body process with quadratic dependence on intensity. We believe it is caused by driving one or more two-photon transitions to an electronically excited state. A pair of photons with $\lambda=$1064~nm has sufficient energy to drive transitions to the low-lying levels of the $(5)^{1}\Sigma^{+}$ state; its potential curve, shown in \fref{fig:1064_spect}(b), has a minimum of energy $h\times557$~THz ($\lambda=538~\mathrm{nm}$) above that of the ground state~\cite{Allouche2000,Fahs2002}. We can suppress the resonant loss, such that it becomes unobservable in our experiments, by tuning the laser frequency several GHz away from the transition. By doing this, we have been able to load RbCs molecules into an ODT made with 1064~nm light, with the lifetime limited by optical excitation of complexes formed in bimolecular collisions~\cite{Christianen2019,Gregory2019,Gregory2020}.

To resolve transitions we reduce the intensity to $\sim 1~\mathrm{kW\,cm^{-2}}$ and dither the frequency over $35~\mathrm{MHz}$, pulsing the light on for 50~ms. The dithering artificially broadens the transitions to the level where we can resolve individual lines with our current apparatus. In \fref{fig:1064_spect}(a) we show the resulting spectra. We fit the observed lines with a Gaussian function to extract center frequencies with uncertainties of a few $\mathrm{MHz}$. From the $(0,5)_0$ sub-level of $N=0$ we observe two doublets with strongest components at laser frequencies $f_0$ and $f_0+1.189(3) \mathrm{GHz}$. This energy splitting is typical for rotational transitions.  

To ensure that we can reliably trap excited-state molecules, we repeat the loss spectroscopy for molecules that have been transferred to the $(1,5)_0$ sub-level of $N=1$ by a microwave $\pi$-pulse \cite{Gregory2016}. The results are shown in \fref{fig:1064_spect}, offset by the microwave transition frequency $980.231~\mathrm{MHz}$. From this state we see markedly different structure, with two doublets and a singlet. The strongest components of the two doublets are at laser frequencies $f_0-0.090\,3(17)~\mathrm{GHz}$ and $f_0+1.834(3)~\mathrm{GHz}$. The singlet is at $f= f_0 -0.749(6)~\mathrm{GHz}$. 

We believe that the lines observed are probably due to two-photon transitions of the form $X^1\Sigma, v=0, N \rightarrow (5)^1\Sigma,v',N'$. The rotational selection rule for a two-photon transition is $\Delta N=0,\ \pm2$. We fit the strongest of each pair of lines in the spectra to the eigenvalues of the rotational Hamiltonian $B'_{v'} N'(N'+1)$, weighted by the error in the center frequencies extracted from the fits. This gives $B'_{v'} = h\times389.9(4)~\mathrm{MHz}$, which compares favorably with the theoretical value of 410~MHz~\cite{Allouche2000}, calculated at the potential minimum of the $(5)^1\Sigma$ state.
\section{Determining the polarizability}\label{sec:polarizability} 
\subsection{Isotropic polarizability in the ground state}\label{sec:isotropic} 

\begin{figure}[t!]
    \centering
    \includegraphics[width=0.45\textwidth]{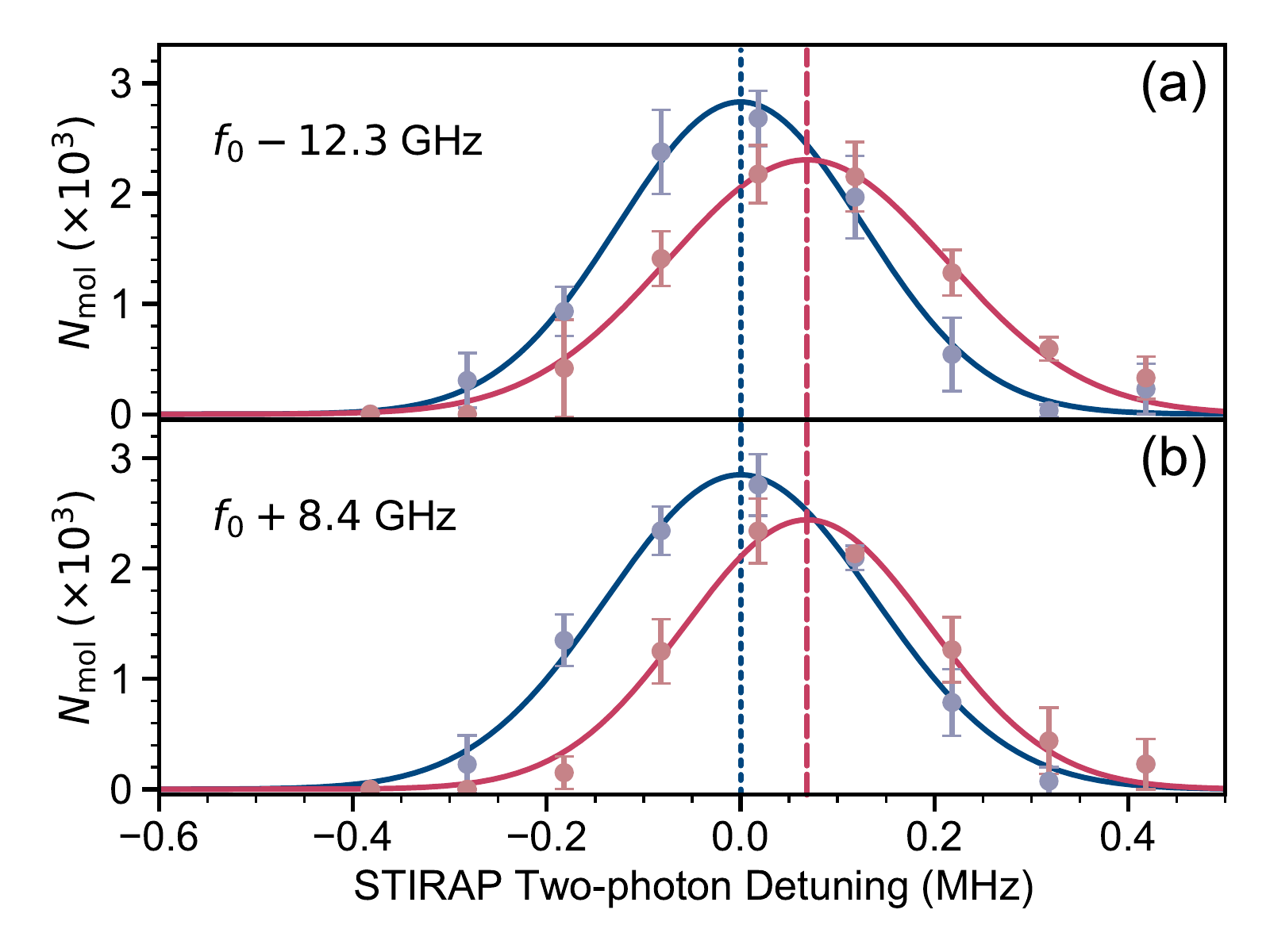}
    \caption{
    Measurement of the polarizability of RbCs in the rovibronic ground state, using the ac Stark shift of the two-photon STIRAP resonance (a) with the laser tuned 12.3~GHz below the resonance at $f_0$ shown in Fig.\ 1(b); (b) with the laser tuned 8.4~GHz above the resonance.  The blue points are measured in free space whilst the red points are measured with the 1064~nm laser at an intensity of 8.60(7)~kW$\,$cm$^{-2}$. The vertical dashed lines indicate the centres of the fits. The shifts for the two detunings agree within experimental error, indicating that there is no impact from the two-photon transitions in \fref{fig:1064_spect}. 
    }
    \label{fig:Stokes shift}
\end{figure}

As described in section~\ref{sec:theory}, the isotropic part of the polarizability, $\alpha^{(0)}$, produces an equal energy shift of all $(N,M_F)_i$ whilst the term proportional to $\alpha^{(2)}$ is non-zero only for $N>0$. Thus only $\alpha^{(0)}$ contributes to the trapping potential for $N=0$. The value of this component can be found either by direct trap frequency measurements, \textit{e.g.} through parametric heating~\cite{Gregory2017}, or by the differential ac Stark shift between two vibronic states. We choose the second method and measure the intensity-dependent energy shift of the ground state with respect to a weakly-bound Feshbach state. In \fref{fig:Stokes shift} we show the two-photon transition used in STIRAP~\cite{Molony2016}, measured with and without the light of wavelength 1064~nm. Hereafter we refer to this light as the ``trapping light'', even when its intensity is too low to form an actual trap. It is delivered in a single beam with a waist of 173(1)~$\upmu$m, and peak intensity of 8.60(7)~kW~cm$^{-2}$.

Two-photon transitions might produce a term in the Hamiltonian proportional to $I^2$, corresponding to a 4th-order hyperpolarizability. To ascertain whether such effects are significant, we perform measurements with the trapping laser tuned approximately $\pm10$~GHz from the loss features observed in Fig.~\ref{fig:1064_spect}. The energy shifts measured above and below the transitions are ${h\times68(8)~\mathrm{kHz}}$ and ${h\times68(9)~\mathrm{kHz}}$ respectively. These identical shifts confirm that the laser frequency is far from resonance with the two-photon transitions. 

The measured energy shift gives the difference in polarizability between the ground and Feshbach states. The molecule's constituent atoms interact very weakly in the Feshbach state, so its polarizability is just the sum of the atomic polarizabilities $\alpha_\text{Rb}$ and $\alpha_{\text{Cs}}$, which are well known~\cite{Safronova2006}. The weighted average of our two measurements yields the polarizability of the rovibronic ground state. Because of rotational averaging, this is equivalent to the isotropic polarizability of the ground vibronic state, 
\begin{equation*}
\begin{aligned}
\alpha^{(0)} &= \alpha_\text{Rb} + \alpha_{\text{Cs}} + 4\pi \epsilon_0 \times 1.7(4)\times10^{2}~a_{0}^{3} \\
&= 4\pi \epsilon_0 \times 2.02(4)\times10^{3}~a_{0}^{3}. 
\end{aligned}
\end{equation*}
Our measurement corresponds to a ratio of the ground-state and Feshbach polarizabilities (${\alpha^{(0)}/(\alpha_\mathrm{Rb}+\alpha_\mathrm{Cs})}$) of 1.09(2). This is significantly different from the value of 1.00006 predicted by Vexiau \textit{et al.}~\cite{Vexiau2017}. Our previous work~\cite{Gregory2017} at a wavelength of 1550~nm found good agreement between the experimental ratio of 0.88(1) and the predicted value of 0.874. The difference observed here may arise because the polarizability at $\lambda=$1550~nm is dominated by far-detuned transitions to the $A^1\Sigma$ state, whereas $\lambda=$1064~nm is closer to resonance and the polarizability is more sensitive to the frequencies and strengths of individual transitions. 
\begin{figure}[t]
    \centering
    \includegraphics[width=0.45\textwidth]{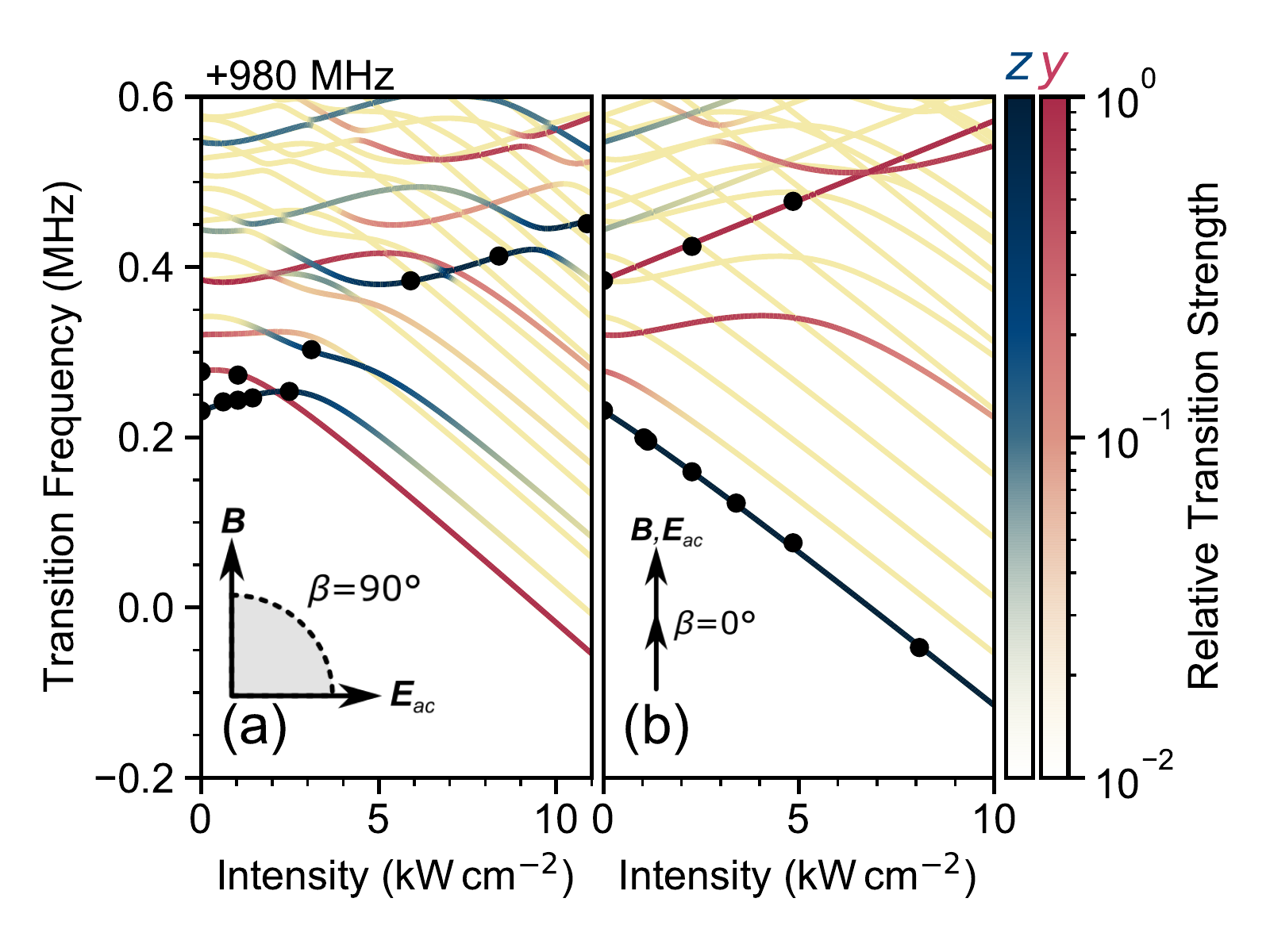}
    \caption{
    Frequencies of the transitions from $(0,5)_0$ to hyperfine sub-levels of $N=1$ in a dc magnetic field of 181.5~G, as a function of laser intensity, for laser polarization angle ($\beta$) (a) perpendicular to and (b) parallel to the uniform 181.5~G magnetic field. Each point is the fitted center frequency from a measured microwave spectrum; the uncertainties are a few kHz, which is too small to be seen at this scale. The relative transition strengths for microwaves polarized along $z$ and $y$ are shown as blue and red color maps respectively.}
    \label{fig:anisotropic_0_90}
\end{figure}
\subsection{Anisotropic polarizability}\label{sec:anisotropic} 
\begin{figure}[t]
    \centering
    \includegraphics[width=0.45\textwidth]{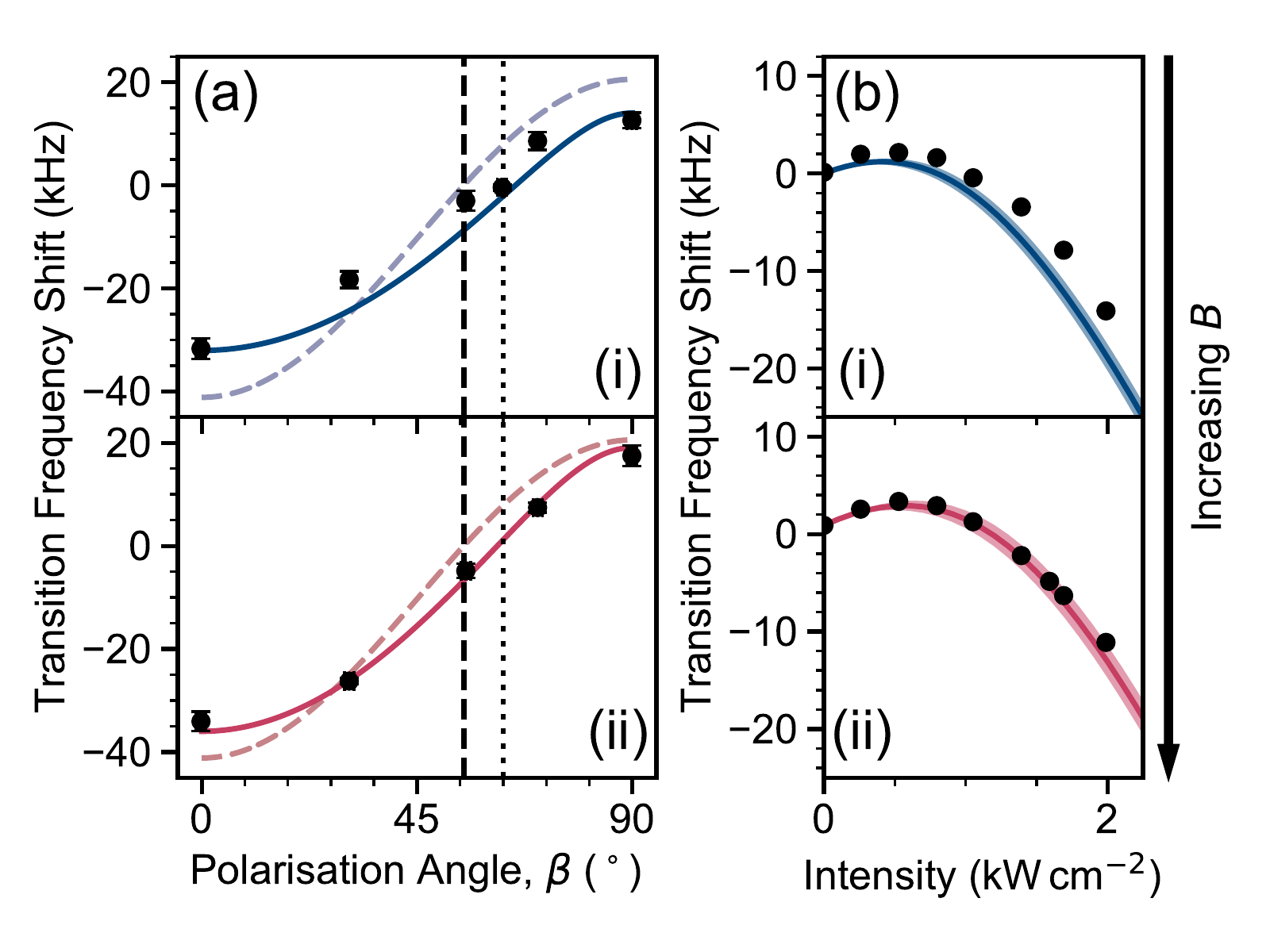}
    \caption{The anisotropic light shift of the transition $(0,5)_0 \rightarrow (1,5)_0$ in a magnetic field of (i) 181.5~G and (ii) 355~G as a function of (a) polarization angle $\beta$, for a fixed laser intensity of 1.05(1)~kW$\,$cm$^{-2}$ and (b) laser intensity, with $\beta$ set to the zero crossing measured in (a), as indicated by the dotted line. Each point is the fitted center frequency from a measured microwave spectrum; the uncertainties are a few kHz.
    The solid lines are calculated from the full Hamiltonian, including the hyperfine and Zeeman structure. The shaded regions in (b) indicate the $\pm1^\circ$ uncertainty in setting the polarization angle to the zero crossing. The colored dashed line shows the result of the hyperfine-free hindered-rotor Hamiltonian, with the vertical dashed line indicating the position of the associated zero crossing. 
    }
    \label{fig:anisotropic_noE}
\end{figure}
\begin{figure}[t]
    \centering
    \includegraphics[width=0.45\textwidth]{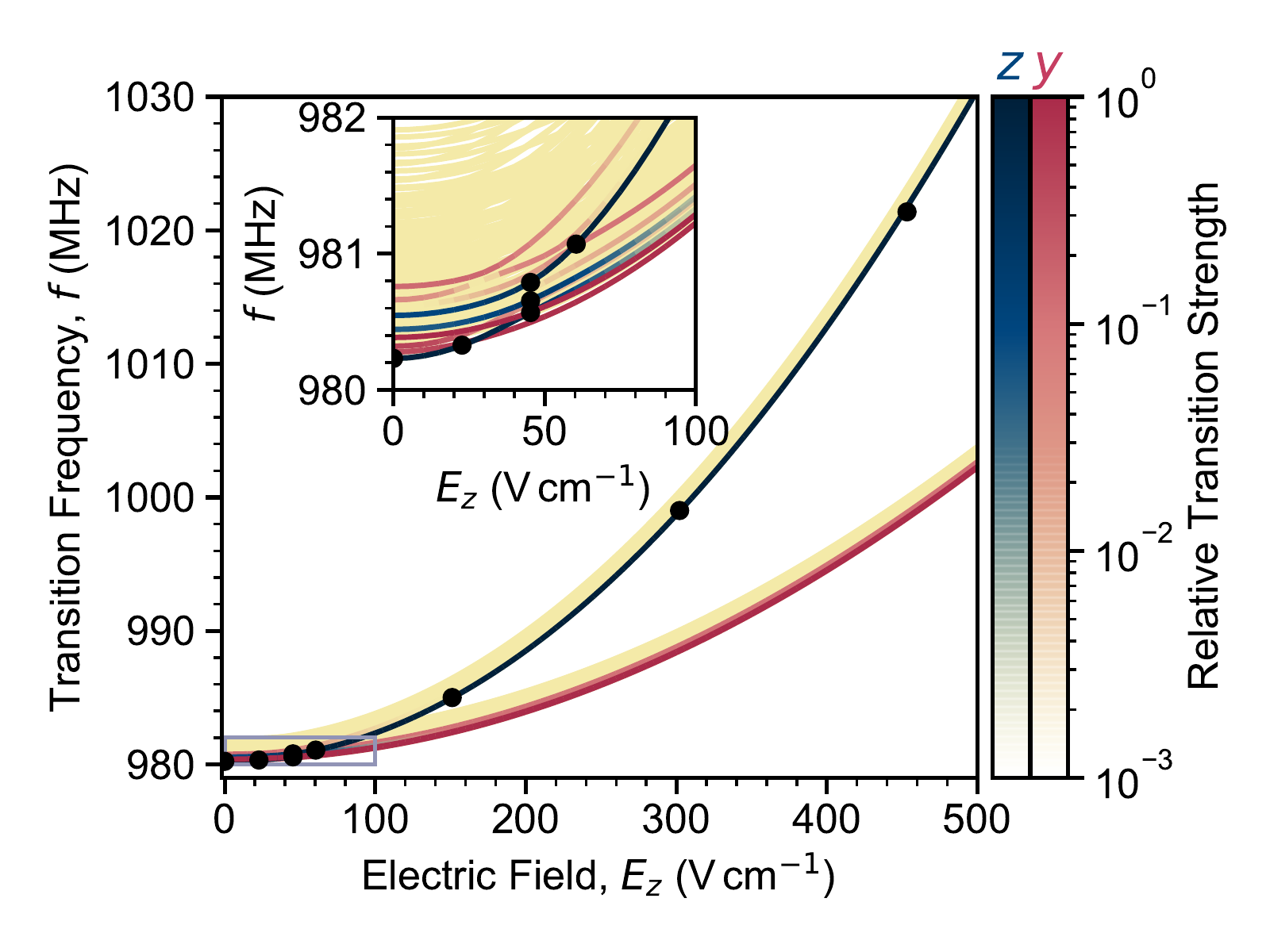}
    \caption{The dc Stark shift of transitions from $(0,5)_0$ to hyperfine sub-levels of $N=1$ as a function of electric field. The inset shows the highlighted region at low electric field. The relative transition strengths for microwaves polarized along $z$ and $y$ are coded as in \fref{fig:anisotropic_0_90}.  Each point is the fitted center frequency from a measured microwave spectrum; the uncertainties are a few kHz, which is too small to be seen at this scale. At higher electric fields, the energy levels for $N=1$ are split into branches with $M_N=0$ and $|M_N|=1$. In the high-field limit,  only one transition can be driven with microwaves polarized along $z$. At lower fields, hyperfine mixing allows multiple transitions; this can be seen clearly at $E_z \lesssim 50~\mathrm{V\,cm^{-1}}$.
    }
    \label{fig:DCStark}
\end{figure}
\begin{figure}[t]
    \centering
    \includegraphics[width=0.45\textwidth]{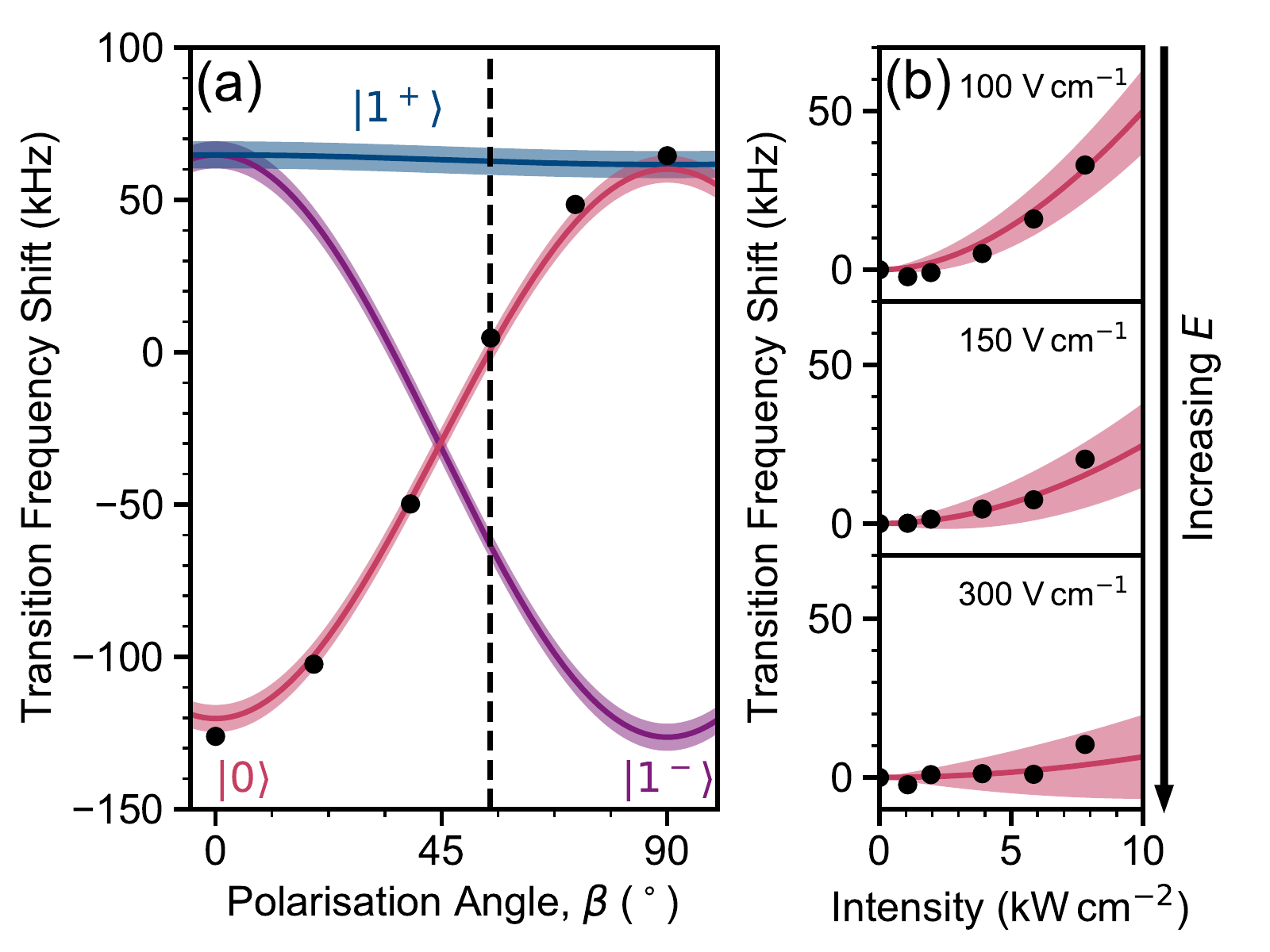}
    \caption{
    The ac Stark shifts of transitions $N=0\rightarrow N=1$ with a dc electric field along the uniform 181.5~G magnetic field. The shifts are shown as a function of (a) the polarization angle ($\beta$) of the trapping laser with an electric field of 300~V$\,$cm$^{-1}$ and a laser intensity of 3.12~kW$\,$cm$^{-2}$; and (b) the laser intensity with $\beta$ fixed to the magic angle [indicated in (a) by the dashed line]. Each point is the fitted center frequency from a measured microwave spectrum for the $(0,5)_0 \rightarrow (1,5)_2$ line; the uncertainties are a few kHz, which is too small to be seen at this scale. The solid lines show the results of the hyperfine-free hindered-rotor Hamiltonian. The lines are labeled with $|M_N|$ for $N=1$; the ac Stark term mixes states with $M_N=\pm1$, and the resulting combinations are labeled $1^\pm$. The shaded regions indicate the error due to the uncertainty in $\alpha^{(2)}$, though the scatter in the experimental points is dominated by electric field noise.
    }
    \label{fig:Magic Angle}
\end{figure}
In this section we determine the value of the anisotropic polarizability, $\alpha^{(2)}$. We use microwave spectroscopy of the $N=0\rightarrow 1$ rotational transition in RbCs, with an applied magnetic field of 181.5~G. The measurements in the previous subsection are insensitive to $\alpha^{(2)}$, because it does not affect trapping in the rotational ground state. However, $\alpha^{(2)}$ is always important for $N>0$. Moreover, states with permanent laboratory-frame dipole moments always involve mixtures of rotational states and so $\alpha^{(2)}$ is needed whenever an electric field is applied.

We measure the value of $\alpha^{(2)}$ for RbCs at $\lambda=1064~\mathrm{nm}$ by considering the 
frequency shift of the rotational transitions as a function of laser intensity. In \fref{fig:anisotropic_0_90} we show the ac Stark maps for $\beta=0^\circ$ and $90^\circ$. The mixing of hyperfine states causes complex patterns of crossings and avoided crossings. We calculate these patterns by diagonalising the full Hamiltonian of \eref{eq:Hamiltonian} to \eref{eq:ACStark_elements}. We fit simultaneously to the experimental microwave spectra for $\beta=0^\circ$ and $90^\circ$ and obtain a good fit with the single value ${\alpha^{(2)}/4\pi\epsilon_0 =1997(6) ~\mathrm{a_0}^3}$. This contrasts with previous work \cite{Gregory2017} at $\lambda=$1550~nm that required separate values of $\alpha^{(2)}$  at $\beta =0^\circ$ and $\beta = 90^\circ$.

With knowledge of both parts of the polarizability, we have determined all the parameters in both the full hyperfine and hindered-rotor Hamiltonians described in Section~\ref{sec:theory}. 
Accordingly, all results in the remainder of the paper are presented without free parameters. 

\section{Controlling the ac Stark effect}\label{sec:magic} 

\subsection{In a magnetic field} 
We first investigate the effect of the polarization of the trapping light on the ac Stark effect in the absence of an electric field. We fix the intensity of the trapping light at
${1.05(1)~\mathrm{kW\,cm^{-2}}}$ 
and vary the polarization angle. \fref{fig:anisotropic_noE}(a) shows the polarization-dependent Stark shift at (i) 181.5~G and (ii) 355~G. In both cases the ac Stark shift deviates from the form proportional to $P_2(\cos \beta)$ expected from models that ignore hyperfine structure; the frequency shift actually crosses zero at $\beta = 63(1)^\circ$. To investigate the intensity dependence at this zero-crossing we fix the polarization at this angle and vary the intensity at both magnetic fields; these measurements are shown in \fref{fig:anisotropic_noE}(b). In both cases the light shift shows a maximum as a function of intensity and crosses zero with substantial gradient.

The results in \fref{fig:anisotropic_noE} indicate that hyperfine effects play a significant role in the ac Stark effect under these conditions. The hyperfine interactions mix states with different values of $M_N$ in \eref{eq:Hyperfine states}, so that the ac Stark effect has a different dependence on angle. It is expected that, at higher magnetic fields, the nuclear spins will decouple from the rotational angular momentum.
However, this does not happen for RbCs at the magnetic fields we use; this is unsurprising in view of the small magnetic moments in the molecule, compared to the hyperfine structure and ac Stark effect. Our calculations including hyperfine structure show that the field would need to be $\sim700~\mathrm{G}$ for the $\ket{N=1,M_N=0,m_\mathrm{Rb}=3/2,m_\mathrm{Cs}=7/2}$ component of the state $(1,5)_0$ to be greater than 99\%.

\subsection{In combined electric and magnetic fields}
In the molecular frame RbCs has a permanent electric dipole moment of 1.23~D. Because of this, there is stronger coupling to dc electric fields than to a dc magnetic field. A modest electric field is sufficient to decouple the rotational and nuclear angular momenta. In \fref{fig:DCStark} we show the calculated Stark shifts and relative strengths of the microwave transitions from the state $(0,5)_0$ over a similar range to \fref{fig:Theory2}(a). A field of $100~\mathrm{V\,cm^{-1}}$ is sufficient to split the $N=1$ rotational level into two branches with $M_N=0$ and $\pm1$ and to reach 99\% state purity for $(1,5)_2$. 

In \fref{fig:Magic Angle}(a) we show the polarization dependence of the ac Stark shift in a dc electric field of 300~V$\,\mathrm{cm}^{-1}$, using a fixed intensity of 3.12~$\mathrm{kW\,cm^{-2}}$. In this case the measured points agree within one standard deviation with the results of the hyperfine-free hindered-rotor Hamiltonian, using the previously measured value of $\alpha^{(2)}$. This shows that even a modest electric field can decouple $\boldsymbol{N}$, $\boldsymbol{i}_\mathrm{Rb}$ and $\boldsymbol{i}_\mathrm{Cs}$.

To test the remaining ac Stark shift, we fix the polarization angle $\beta$ of the trapping laser to the predicted magic angle of $54.7^\circ$ and vary the intensity of the light. The results are shown in \fref{fig:Magic Angle}(b) for three electric fields. We see that the ac Stark shift scales approximately as $I^2$; this phenomenon is often termed ``hyperpolarizability'', but it is important to recognise that it is completely described within \eref{eq:ACStark_elements}, which includes only \emph{electronic} polarizability and not hyperpolarizability. At higher electric fields the effect of $\alpha^{(2)}$ is reduced; at $300~\mathrm{V\,cm^{-1}}$ the maximum frequency shift due to $\alpha^{(2)}$ is 10.0(1.6)~kHz, although that due to the dc Stark effect is $\sim20$~MHz. This simplified ac Stark shift comes at the cost of increased sensitivity to electric field noise, which we believe is at the level of one part in a thousand in the present experiments, and is responsible for the remaining scatter in \fref{fig:Magic Angle}(b).

\begin{figure}[t]
    \centering
    \includegraphics[width=0.45\textwidth]{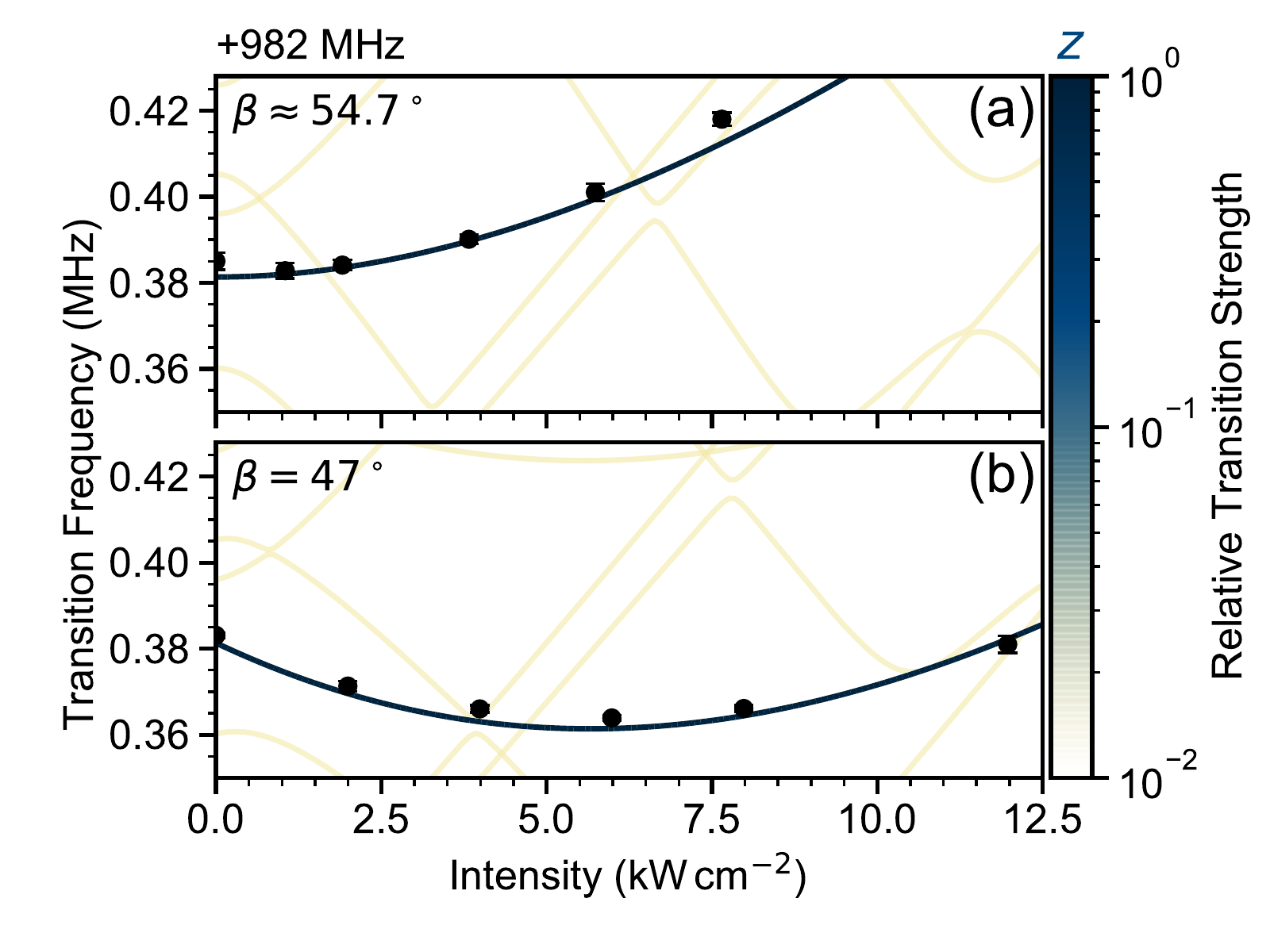}
    \caption{The ac Stark shift of the transition $(0,5)_0 \rightarrow (1,5)_0$  in an electric field of $100.8~\mathrm{V\,cm^{-1}}$ as a function of the intensity of the trapping laser. The shifts are shown for two values of the polarization angle, $\beta$. Each point is the fitted center frequency from a measured microwave spectrum; the uncertainties are a few kHz. The continuous lines are the results from the full Hamiltonian including hyperfine and Zeeman terms. Transition strengths for microwaves polarized along $z$ are coded as in \fref{fig:anisotropic_noE}.
    }
    \label{fig:BetterThanMagic}
\end{figure}

\subsection{Beyond the magic angle}
At the magic angle, $\beta_\mathrm{magic} \approx 54.7^\circ$, the diagonal terms of \eref{eq:ACStark_elements} are zero. In a state that is well described by a single basis function $\ket{N=1,M_N}$, this removes the component of the ac Stark shift due to $\alpha^{(2)}$. Previous work has focused on finding the magic angle \cite{Neyenhuis2012,Seesselberg2018}. However due to the quadratic intensity dependence of the remnant ac Stark shift, as seen in \fref{fig:Magic Angle}(b) and \fref{fig:BetterThanMagic}(a), the differential polarizability is reduced to zero at $I=0$. Here we describe a tunable arrangement in which the total frequency shift is reduced to near zero at an intensity suitable for trapping. This technique does not require full decoupling of the nuclear spins from the rotation, so it can be used at much smaller electric fields, reducing the sensitivity to electric field noise.

In \fref{fig:BetterThanMagic} we compare the ac Stark shifts of the $(N,M_N)=(0,0)\rightarrow(1,0)$ transition at two different polarization angles with the results of the full Hamiltonian including hyperfine structure. At $\beta_\mathrm{magic}$, shown in \fref{fig:BetterThanMagic},(a) the ac Stark shift increases almost quadratically with intensity. However, by tuning the polarization angle away from $\beta_\mathrm{magic}$ we can engineer a broad minimum at a required intensity; for ${I \approx 6~\mathrm{kW\,cm^{-2}}}$, this occurs at $\beta=47^\circ$, as seen in \fref{fig:BetterThanMagic}(b). This intensity corresponds to a trap depth of $27~\mathrm{\upmu K}$ in the rotational ground state, so would be suitable for trapping molecules at our current molecular temperature of $1.5~\mathrm{\upmu K}$. 

We quantify the expected transition frequency spread $\Delta f$ for a 4\% spread in intensity by evaluating the Taylor expansion of the ac Stark map at $I=I_0$ and $I_0-\Delta I$ for the two cases shown in \fref{fig:BetterThanMagic}. For $\beta = 54.7^\circ$ we obtain $\Delta f = 1.5~\mathrm{kHz}$. By contrast, for $\beta = 47^\circ$, where we have optimised the intensity to minimise differential Stark shifts, we obtain $\Delta f = 33~\mathrm{Hz}$. The coherence time in this arrangement should scale with $1/\Delta f \sim 30~\mathrm{ms}$. For comparison, See{\ss}elberg \textit{et al.}~\cite{Seesselberg2018} trapped $^{23}$Na$^{40}$K in an electric field in a spin-decoupled optical lattice and observed a coherence time of 8.7(6)\,ms.
\section{Outlook and Summary}\label{sec:outlook}

In this work we have characterized the behavior of the RbCs molecule in light of wavelength 1064~nm. We have found that the lifetime of the molecule in traps of this wavelength can be severely impacted by two-photon transitions to the $(5)\Sigma^+$ state, which were not predicted by theory.
However, there are regions between these transitions where the losses are not significantly enhanced. We have determined the isotropic polarizability of the molecule in these regions with optical spectroscopy of the rovibronic ground state. We have also found the anisotropic polarizability by observing the ac Stark effect on hyperfine components of the $N=0\rightarrow1$ rotational transition using microwave spectroscopy with kHz precision.

We have studied the impact of applying additional dc electric and magnetic fields on the ac Stark shift. A magnetic field beyond our current experimental capability would be required to reduce the impact of the hyperfine structure. However, we have found that a modest electric field is sufficient to decouple the rotational motion from the nuclear spins and simplify the hyperfine structure. Under these circumstances we can model the structure using a hindered-rotor Hamiltonian, without hyperfine or Zeeman terms.

Our medium-term goal is to achieve long coherence times for molecules in optical lattices and tweezers. To achieve this, we need to reduce the dependence of the ac Stark shift on laser intensity, at intensities high enough for trapping. We have found that this may be achieved for RbCs by applying a moderate dc electric field and adjusting the polarization angle of the laser. The optimum polarization angle is close to the ``magic angle" of $54.7^\circ$ for low laser intensities, but different polarization angles are needed at the higher intensities necessary for trapping. This result opens the way to using arrays of RbCs molecules as a platform for quantum simulation.

\subsection*{Acknowledgements}
This work was supported by U.K. Engineering and Physical Sciences Research Council (EPSRC) Grants EP/P01058X/1 and EP/P008275/1.

The data, code and analysis associated with this work are available at: \textbf{DOI to be added}. The python code for hyperfine structure calculations can be found at \textbf{DOI:10.5281/ZENODO.3755881}.

\bibliography{Polarizability}

\begin{thebibliography}{106}%
\makeatletter
\providecommand \@ifxundefined [1]{%
 \@ifx{#1\undefined}
}%
\providecommand \@ifnum [1]{%
 \ifnum #1\expandafter \@firstoftwo
 \else \expandafter \@secondoftwo
 \fi
}%
\providecommand \@ifx [1]{%
 \ifx #1\expandafter \@firstoftwo
 \else \expandafter \@secondoftwo
 \fi
}%
\providecommand \natexlab [1]{#1}%
\providecommand \enquote  [1]{``#1''}%
\providecommand \bibnamefont  [1]{#1}%
\providecommand \bibfnamefont [1]{#1}%
\providecommand \citenamefont [1]{#1}%
\providecommand \href@noop [0]{\@secondoftwo}%
\providecommand \href [0]{\begingroup \@sanitize@url \@href}%
\providecommand \@href[1]{\@@startlink{#1}\@@href}%
\providecommand \@@href[1]{\endgroup#1\@@endlink}%
\providecommand \@sanitize@url [0]{\catcode `\\12\catcode `\$12\catcode
  `\&12\catcode `\#12\catcode `\^12\catcode `\_12\catcode `\%12\relax}%
\providecommand \@@startlink[1]{}%
\providecommand \@@endlink[0]{}%
\providecommand \url  [0]{\begingroup\@sanitize@url \@url }%
\providecommand \@url [1]{\endgroup\@href {#1}{\urlprefix }}%
\providecommand \urlprefix  [0]{URL }%
\providecommand \Eprint [0]{\href }%
\providecommand \doibase [0]{https://doi.org/}%
\providecommand \selectlanguage [0]{\@gobble}%
\providecommand \bibinfo  [0]{\@secondoftwo}%
\providecommand \bibfield  [0]{\@secondoftwo}%
\providecommand \translation [1]{[#1]}%
\providecommand \BibitemOpen [0]{}%
\providecommand \bibitemStop [0]{}%
\providecommand \bibitemNoStop [0]{.\EOS\space}%
\providecommand \EOS [0]{\spacefactor3000\relax}%
\providecommand \BibitemShut  [1]{\csname bibitem#1\endcsname}%
\let\auto@bib@innerbib\@empty
\bibitem [{\citenamefont {Zelevinsky}\ \emph {et~al.}(2008)\citenamefont
  {Zelevinsky}, \citenamefont {Kotochigova},\ and\ \citenamefont
  {Ye}}]{Zelevinsky2008}%
  \BibitemOpen
  \bibfield  {author} {\bibinfo {author} {\bibfnamefont {T.}~\bibnamefont
  {Zelevinsky}}, \bibinfo {author} {\bibfnamefont {S.}~\bibnamefont
  {Kotochigova}},\ and\ \bibinfo {author} {\bibfnamefont {J.}~\bibnamefont
  {Ye}},\ }\bibfield  {title} {\bibinfo {title} {Precision test of mass-ratio
  variations with lattice-confined ultracold molecules},\ }\href
  {https://doi.org/10.1103/physrevlett.100.043201} {\bibfield  {journal}
  {\bibinfo  {journal} {Phys. Rev. Lett.}\ }\textbf {\bibinfo {volume} {100}},\
  \bibinfo {pages} {043201} (\bibinfo {year} {2008})}\BibitemShut {NoStop}%
\bibitem [{\citenamefont {Hudson}\ \emph {et~al.}(2011)\citenamefont {Hudson},
  \citenamefont {Kara}, \citenamefont {Smallman}, \citenamefont {Sauer},
  \citenamefont {Tarbutt},\ and\ \citenamefont {Hinds}}]{Hudson2011}%
  \BibitemOpen
  \bibfield  {author} {\bibinfo {author} {\bibfnamefont {J.~J.}\ \bibnamefont
  {Hudson}}, \bibinfo {author} {\bibfnamefont {D.~M.}\ \bibnamefont {Kara}},
  \bibinfo {author} {\bibfnamefont {I.~J.}\ \bibnamefont {Smallman}}, \bibinfo
  {author} {\bibfnamefont {B.~E.}\ \bibnamefont {Sauer}}, \bibinfo {author}
  {\bibfnamefont {M.~R.}\ \bibnamefont {Tarbutt}},\ and\ \bibinfo {author}
  {\bibfnamefont {E.~A.}\ \bibnamefont {Hinds}},\ }\bibfield  {title} {\bibinfo
  {title} {Improved measurement of the shape of the electron},\ }\href
  {https://doi.org/10.1038/nature10104} {\bibfield  {journal} {\bibinfo
  {journal} {Nature}\ }\textbf {\bibinfo {volume} {473}},\ \bibinfo {pages}
  {493} (\bibinfo {year} {2011})}\BibitemShut {NoStop}%
\bibitem [{\citenamefont {Salumbides}\ \emph {et~al.}(2011)\citenamefont
  {Salumbides}, \citenamefont {Dickenson}, \citenamefont {Ivanov},\ and\
  \citenamefont {Ubachs}}]{Salumbides2011}%
  \BibitemOpen
  \bibfield  {author} {\bibinfo {author} {\bibfnamefont {E.~J.}\ \bibnamefont
  {Salumbides}}, \bibinfo {author} {\bibfnamefont {G.~D.}\ \bibnamefont
  {Dickenson}}, \bibinfo {author} {\bibfnamefont {T.~I.}\ \bibnamefont
  {Ivanov}},\ and\ \bibinfo {author} {\bibfnamefont {W.}~\bibnamefont
  {Ubachs}},\ }\bibfield  {title} {\bibinfo {title} {{QED} effects in
  molecules: Test on rotational quantum states of {H$_2$}},\ }\href
  {https://doi.org/10.1103/physrevlett.107.043005} {\bibfield  {journal}
  {\bibinfo  {journal} {Phys. Rev. Lett.}\ }\textbf {\bibinfo {volume} {107}},\
  \bibinfo {pages} {043005} (\bibinfo {year} {2011})}\BibitemShut {NoStop}%
\bibitem [{\citenamefont {Salumbides}\ \emph {et~al.}(2013)\citenamefont
  {Salumbides}, \citenamefont {Koelemeij}, \citenamefont {Komasa},
  \citenamefont {Pachucki}, \citenamefont {Eikema},\ and\ \citenamefont
  {Ubachs}}]{Salumbides2013}%
  \BibitemOpen
  \bibfield  {author} {\bibinfo {author} {\bibfnamefont {E.~J.}\ \bibnamefont
  {Salumbides}}, \bibinfo {author} {\bibfnamefont {J.~C.~J.}\ \bibnamefont
  {Koelemeij}}, \bibinfo {author} {\bibfnamefont {J.}~\bibnamefont {Komasa}},
  \bibinfo {author} {\bibfnamefont {K.}~\bibnamefont {Pachucki}}, \bibinfo
  {author} {\bibfnamefont {K.~S.~E.}\ \bibnamefont {Eikema}},\ and\ \bibinfo
  {author} {\bibfnamefont {W.}~\bibnamefont {Ubachs}},\ }\bibfield  {title}
  {\bibinfo {title} {Bounds on fifth forces from precision measurements on
  molecules},\ }\href {https://doi.org/10.1103/physrevd.87.112008} {\bibfield
  {journal} {\bibinfo  {journal} {Phys. Rev. D}\ }\textbf {\bibinfo {volume}
  {87}},\ \bibinfo {pages} {112008} (\bibinfo {year} {2013})}\BibitemShut
  {NoStop}%
\bibitem [{\citenamefont {Schiller}\ \emph {et~al.}(2014)\citenamefont
  {Schiller}, \citenamefont {Bakalov},\ and\ \citenamefont
  {Korobov}}]{Schiller2014}%
  \BibitemOpen
  \bibfield  {author} {\bibinfo {author} {\bibfnamefont {S.}~\bibnamefont
  {Schiller}}, \bibinfo {author} {\bibfnamefont {D.}~\bibnamefont {Bakalov}},\
  and\ \bibinfo {author} {\bibfnamefont {V.}~\bibnamefont {Korobov}},\
  }\bibfield  {title} {\bibinfo {title} {Simplest molecules as candidates for
  precise optical clocks},\ }\href
  {https://doi.org/10.1103/physrevlett.113.023004} {\bibfield  {journal}
  {\bibinfo  {journal} {Phys. Rev. Lett.}\ }\textbf {\bibinfo {volume} {113}},\
  \bibinfo {pages} {023004} (\bibinfo {year} {2014})}\BibitemShut {NoStop}%
\bibitem [{\citenamefont {{The ACME Collaboration}}\ \emph
  {et~al.}(2014)\citenamefont {{The ACME Collaboration}}, \citenamefont
  {Baron}, \citenamefont {Campbell}, \citenamefont {DeMille}, \citenamefont
  {Doyle}, \citenamefont {Gabrielse}, \citenamefont {Gurevich}, \citenamefont
  {Hess}, \citenamefont {Hutzler}, \citenamefont {Kirilov}, \citenamefont
  {Kozyryev}, \citenamefont {O'Leary}, \citenamefont {Panda}, \citenamefont
  {Parsons}, \citenamefont {Petrik}, \citenamefont {Spaun}, \citenamefont
  {Vutha},\ and\ \citenamefont {West}}]{ACME2014}%
  \BibitemOpen
  \bibfield  {author} {\bibinfo {author} {\bibnamefont {{The ACME
  Collaboration}}}, \bibinfo {author} {\bibfnamefont {J.}~\bibnamefont
  {Baron}}, \bibinfo {author} {\bibfnamefont {W.~C.}\ \bibnamefont {Campbell}},
  \bibinfo {author} {\bibfnamefont {D.}~\bibnamefont {DeMille}}, \bibinfo
  {author} {\bibfnamefont {J.~M.}\ \bibnamefont {Doyle}}, \bibinfo {author}
  {\bibfnamefont {G.}~\bibnamefont {Gabrielse}}, \bibinfo {author}
  {\bibfnamefont {Y.~V.}\ \bibnamefont {Gurevich}}, \bibinfo {author}
  {\bibfnamefont {P.~W.}\ \bibnamefont {Hess}}, \bibinfo {author}
  {\bibfnamefont {N.~R.}\ \bibnamefont {Hutzler}}, \bibinfo {author}
  {\bibfnamefont {E.}~\bibnamefont {Kirilov}}, \bibinfo {author} {\bibfnamefont
  {I.}~\bibnamefont {Kozyryev}}, \bibinfo {author} {\bibfnamefont {B.~R.}\
  \bibnamefont {O'Leary}}, \bibinfo {author} {\bibfnamefont {C.~D.}\
  \bibnamefont {Panda}}, \bibinfo {author} {\bibfnamefont {M.~F.}\ \bibnamefont
  {Parsons}}, \bibinfo {author} {\bibfnamefont {E.~S.}\ \bibnamefont {Petrik}},
  \bibinfo {author} {\bibfnamefont {B.}~\bibnamefont {Spaun}}, \bibinfo
  {author} {\bibfnamefont {A.~C.}\ \bibnamefont {Vutha}},\ and\ \bibinfo
  {author} {\bibfnamefont {A.~D.}\ \bibnamefont {West}},\ }\bibfield  {title}
  {\bibinfo {title} {Order of magnitude smaller limit on the electric dipole
  moment of the electron},\ }\href {https://doi.org/10.1126/science.1248213}
  {\bibfield  {journal} {\bibinfo  {journal} {Science}\ }\textbf {\bibinfo
  {volume} {343}},\ \bibinfo {pages} {269} (\bibinfo {year}
  {2014})}\BibitemShut {NoStop}%
\bibitem [{\citenamefont {Hanneke}\ \emph {et~al.}(2016)\citenamefont
  {Hanneke}, \citenamefont {Carollo},\ and\ \citenamefont
  {Lane}}]{Hanneke2016}%
  \BibitemOpen
  \bibfield  {author} {\bibinfo {author} {\bibfnamefont {D.}~\bibnamefont
  {Hanneke}}, \bibinfo {author} {\bibfnamefont {R.~A.}\ \bibnamefont
  {Carollo}},\ and\ \bibinfo {author} {\bibfnamefont {D.~A.}\ \bibnamefont
  {Lane}},\ }\bibfield  {title} {\bibinfo {title} {High sensitivity to
  variation in the proton-to-electron mass ratio in {O$_2^+$}},\ }\href
  {https://doi.org/10.1103/physreva.94.050101} {\bibfield  {journal} {\bibinfo
  {journal} {Phys. Rev. A}\ }\textbf {\bibinfo {volume} {94}},\ \bibinfo
  {pages} {050101} (\bibinfo {year} {2016})}\BibitemShut {NoStop}%
\bibitem [{\citenamefont {Cairncross}\ \emph {et~al.}(2017)\citenamefont
  {Cairncross}, \citenamefont {Gresh}, \citenamefont {Grau}, \citenamefont
  {Cossel}, \citenamefont {Roussy}, \citenamefont {Ni}, \citenamefont {Zhou},
  \citenamefont {Ye},\ and\ \citenamefont {Cornell}}]{Cairncross2017}%
  \BibitemOpen
  \bibfield  {author} {\bibinfo {author} {\bibfnamefont {W.~B.}\ \bibnamefont
  {Cairncross}}, \bibinfo {author} {\bibfnamefont {D.~N.}\ \bibnamefont
  {Gresh}}, \bibinfo {author} {\bibfnamefont {M.}~\bibnamefont {Grau}},
  \bibinfo {author} {\bibfnamefont {K.~C.}\ \bibnamefont {Cossel}}, \bibinfo
  {author} {\bibfnamefont {T.~S.}\ \bibnamefont {Roussy}}, \bibinfo {author}
  {\bibfnamefont {Y.}~\bibnamefont {Ni}}, \bibinfo {author} {\bibfnamefont
  {Y.}~\bibnamefont {Zhou}}, \bibinfo {author} {\bibfnamefont {J.}~\bibnamefont
  {Ye}},\ and\ \bibinfo {author} {\bibfnamefont {E.~A.}\ \bibnamefont
  {Cornell}},\ }\bibfield  {title} {\bibinfo {title} {Precision measurement of
  the electron's electric dipole moment using trapped molecular ions},\ }\href
  {https://doi.org/10.1103/physrevlett.119.153001} {\bibfield  {journal}
  {\bibinfo  {journal} {Phys. Rev. Lett.}\ }\textbf {\bibinfo {volume} {119}},\
  \bibinfo {pages} {153001} (\bibinfo {year} {2017})}\BibitemShut {NoStop}%
\bibitem [{\citenamefont {Borkowski}(2018)}]{Borkowski2018}%
  \BibitemOpen
  \bibfield  {author} {\bibinfo {author} {\bibfnamefont {M.}~\bibnamefont
  {Borkowski}},\ }\bibfield  {title} {\bibinfo {title} {Optical lattice clocks
  with weakly bound molecules},\ }\href
  {https://doi.org/10.1103/physrevlett.120.083202} {\bibfield  {journal}
  {\bibinfo  {journal} {Phys. Rev. Lett.}\ }\textbf {\bibinfo {volume} {120}},\
  \bibinfo {pages} {083202} (\bibinfo {year} {2018})}\BibitemShut {NoStop}%
\bibitem [{\citenamefont {{The ACME Collaboration}}(2018)}]{ACME2018}%
  \BibitemOpen
  \bibfield  {author} {\bibinfo {author} {\bibnamefont {{The ACME
  Collaboration}}},\ }\bibfield  {title} {\bibinfo {title} {Improved limit on
  the electric dipole moment of the electron},\ }\href
  {https://doi.org/10.1038/s41586-018-0599-8} {\bibfield  {journal} {\bibinfo
  {journal} {Nature}\ }\textbf {\bibinfo {volume} {562}},\ \bibinfo {pages}
  {355} (\bibinfo {year} {2018})}\BibitemShut {NoStop}%
\bibitem [{\citenamefont {Borkowski}\ \emph {et~al.}(2019)\citenamefont
  {Borkowski}, \citenamefont {Buchachenko}, \citenamefont {Ciury{\l}o},
  \citenamefont {Julienne}, \citenamefont {Yamada}, \citenamefont {Kikuchi},
  \citenamefont {Takasu},\ and\ \citenamefont {Takahashi}}]{Borkowski2019}%
  \BibitemOpen
  \bibfield  {author} {\bibinfo {author} {\bibfnamefont {M.}~\bibnamefont
  {Borkowski}}, \bibinfo {author} {\bibfnamefont {A.~A.}\ \bibnamefont
  {Buchachenko}}, \bibinfo {author} {\bibfnamefont {R.}~\bibnamefont
  {Ciury{\l}o}}, \bibinfo {author} {\bibfnamefont {P.~S.}\ \bibnamefont
  {Julienne}}, \bibinfo {author} {\bibfnamefont {H.}~\bibnamefont {Yamada}},
  \bibinfo {author} {\bibfnamefont {Y.}~\bibnamefont {Kikuchi}}, \bibinfo
  {author} {\bibfnamefont {Y.}~\bibnamefont {Takasu}},\ and\ \bibinfo {author}
  {\bibfnamefont {Y.}~\bibnamefont {Takahashi}},\ }\bibfield  {title} {\bibinfo
  {title} {Weakly bound molecules as sensors of new gravitylike forces},\
  }\bibfield  {journal} {\bibinfo  {journal} {Sci. Rep.}\ }\textbf {\bibinfo
  {volume} {9}},\ \href {https://doi.org/10.1038/s41598-019-51346-y}
  {10.1038/s41598-019-51346-y} (\bibinfo {year} {2019})\BibitemShut {NoStop}%
\bibitem [{\citenamefont {Krems}(2008)}]{Krems2008}%
  \BibitemOpen
  \bibfield  {author} {\bibinfo {author} {\bibfnamefont {R.~V.}\ \bibnamefont
  {Krems}},\ }\bibfield  {title} {\bibinfo {title} {Cold controlled
  chemistry},\ }\href {https://doi.org/10.1039/b802322k} {\bibfield  {journal}
  {\bibinfo  {journal} {Phys. Chem. Chem. Phys.}\ }\textbf {\bibinfo {volume}
  {10}},\ \bibinfo {pages} {4079} (\bibinfo {year} {2008})}\BibitemShut
  {NoStop}%
\bibitem [{\citenamefont {Bell}\ and\ \citenamefont
  {Softley}(2009)}]{Bell2009}%
  \BibitemOpen
  \bibfield  {author} {\bibinfo {author} {\bibfnamefont {M.~T.}\ \bibnamefont
  {Bell}}\ and\ \bibinfo {author} {\bibfnamefont {T.~P.}\ \bibnamefont
  {Softley}},\ }\bibfield  {title} {\bibinfo {title} {Ultracold molecules and
  ultracold chemistry},\ }\href@noop {} {\bibfield  {journal} {\bibinfo
  {journal} {Mol. Phys.}\ }\textbf {\bibinfo {volume} {107}},\ \bibinfo {pages}
  {99} (\bibinfo {year} {2009})}\BibitemShut {NoStop}%
\bibitem [{\citenamefont {Ospelkaus}\ \emph {et~al.}(2010)\citenamefont
  {Ospelkaus}, \citenamefont {Ni}, \citenamefont {Wang}, \citenamefont
  {de~Miranda}, \citenamefont {Neyenhuis}, \citenamefont {Qu{\'e}m{\'e}ner},
  \citenamefont {Julienne}, \citenamefont {Bohn}, \citenamefont {Jin},\ and\
  \citenamefont {Ye}}]{Ospelkaus2010}%
  \BibitemOpen
  \bibfield  {author} {\bibinfo {author} {\bibfnamefont {S.}~\bibnamefont
  {Ospelkaus}}, \bibinfo {author} {\bibfnamefont {K.-K.}\ \bibnamefont {Ni}},
  \bibinfo {author} {\bibfnamefont {D.}~\bibnamefont {Wang}}, \bibinfo {author}
  {\bibfnamefont {M.~H.~G.}\ \bibnamefont {de~Miranda}}, \bibinfo {author}
  {\bibfnamefont {B.}~\bibnamefont {Neyenhuis}}, \bibinfo {author}
  {\bibfnamefont {G.}~\bibnamefont {Qu{\'e}m{\'e}ner}}, \bibinfo {author}
  {\bibfnamefont {P.~S.}\ \bibnamefont {Julienne}}, \bibinfo {author}
  {\bibfnamefont {J.~L.}\ \bibnamefont {Bohn}}, \bibinfo {author}
  {\bibfnamefont {D.~S.}\ \bibnamefont {Jin}},\ and\ \bibinfo {author}
  {\bibfnamefont {J.}~\bibnamefont {Ye}},\ }\bibfield  {title} {\bibinfo
  {title} {Quantum-state controlled chemical reactions of ultracold
  potassium-rubidium molecules},\ }\href
  {https://doi.org/10.1126/science.1184121} {\bibfield  {journal} {\bibinfo
  {journal} {Science}\ }\textbf {\bibinfo {volume} {327}},\ \bibinfo {pages}
  {853} (\bibinfo {year} {2010})}\BibitemShut {NoStop}%
\bibitem [{\citenamefont {Dulieu}\ \emph {et~al.}(2011)\citenamefont {Dulieu},
  \citenamefont {Krems}, \citenamefont {Weidem\"{u}ller},\ and\ \citenamefont
  {Willitsch}}]{Dulieu2011}%
  \BibitemOpen
  \bibfield  {author} {\bibinfo {author} {\bibfnamefont {O.}~\bibnamefont
  {Dulieu}}, \bibinfo {author} {\bibfnamefont {R.}~\bibnamefont {Krems}},
  \bibinfo {author} {\bibfnamefont {M.}~\bibnamefont {Weidem\"{u}ller}},\ and\
  \bibinfo {author} {\bibfnamefont {S.}~\bibnamefont {Willitsch}},\ }\bibfield
  {title} {\bibinfo {title} {Physics and chemistry of cold molecules},\
  }\href@noop {} {\bibfield  {journal} {\bibinfo  {journal} {Phys. Chem. Chem.
  Phys.}\ }\textbf {\bibinfo {volume} {13}},\ \bibinfo {pages} {18703}
  (\bibinfo {year} {2011})}\BibitemShut {NoStop}%
\bibitem [{\citenamefont {Balakrishnan}(2016)}]{Balakrishnan2016}%
  \BibitemOpen
  \bibfield  {author} {\bibinfo {author} {\bibfnamefont {N.}~\bibnamefont
  {Balakrishnan}},\ }\bibfield  {title} {\bibinfo {title} {Perpective:
  Ultracold molecules and the dawn of cold controlled chemistry},\ }\href
  {https://doi.org/10.1063/1.4964096} {\bibfield  {journal} {\bibinfo
  {journal} {J. Chem. Phys.}\ }\textbf {\bibinfo {volume} {145}},\ \bibinfo
  {pages} {150901} (\bibinfo {year} {2016})}\BibitemShut {NoStop}%
\bibitem [{\citenamefont {Hu}\ \emph {et~al.}(2019)\citenamefont {Hu},
  \citenamefont {Liu}, \citenamefont {Grimes}, \citenamefont {Lin},
  \citenamefont {Gheorghe}, \citenamefont {Vexiau}, \citenamefont
  {Bouloufa-Maafa}, \citenamefont {Dulieu}, \citenamefont {Rosenband},\ and\
  \citenamefont {Ni}}]{Hu2019}%
  \BibitemOpen
  \bibfield  {author} {\bibinfo {author} {\bibfnamefont {M.-G.}\ \bibnamefont
  {Hu}}, \bibinfo {author} {\bibfnamefont {Y.}~\bibnamefont {Liu}}, \bibinfo
  {author} {\bibfnamefont {D.~D.}\ \bibnamefont {Grimes}}, \bibinfo {author}
  {\bibfnamefont {Y.-W.}\ \bibnamefont {Lin}}, \bibinfo {author} {\bibfnamefont
  {A.~H.}\ \bibnamefont {Gheorghe}}, \bibinfo {author} {\bibfnamefont
  {R.}~\bibnamefont {Vexiau}}, \bibinfo {author} {\bibfnamefont
  {N.}~\bibnamefont {Bouloufa-Maafa}}, \bibinfo {author} {\bibfnamefont
  {O.}~\bibnamefont {Dulieu}}, \bibinfo {author} {\bibfnamefont
  {T.}~\bibnamefont {Rosenband}},\ and\ \bibinfo {author} {\bibfnamefont
  {K.-K.}\ \bibnamefont {Ni}},\ }\bibfield  {title} {\bibinfo {title} {Direct
  observation of bimolecular reactions of ultracold {KRb} molecules},\ }\href
  {https://doi.org/10.1126/science.aay9531} {\bibfield  {journal} {\bibinfo
  {journal} {Science}\ }\textbf {\bibinfo {volume} {366}},\ \bibinfo {pages}
  {1111} (\bibinfo {year} {2019})}\BibitemShut {NoStop}%
\bibitem [{\citenamefont {Santos}\ \emph {et~al.}(2000)\citenamefont {Santos},
  \citenamefont {Shlyapnikov}, \citenamefont {Zoller},\ and\ \citenamefont
  {Lewenstein}}]{Santos2000}%
  \BibitemOpen
  \bibfield  {author} {\bibinfo {author} {\bibfnamefont {L.}~\bibnamefont
  {Santos}}, \bibinfo {author} {\bibfnamefont {G.~V.}\ \bibnamefont
  {Shlyapnikov}}, \bibinfo {author} {\bibfnamefont {P.}~\bibnamefont
  {Zoller}},\ and\ \bibinfo {author} {\bibfnamefont {M.}~\bibnamefont
  {Lewenstein}},\ }\bibfield  {title} {\bibinfo {title} {{Bose-Einstein}
  condensation in trapped dipolar gases},\ }\href
  {https://doi.org/10.1103/PhysRevLett.85.1791} {\bibfield  {journal} {\bibinfo
   {journal} {Phys. Rev. Lett.}\ }\textbf {\bibinfo {volume} {85}},\ \bibinfo
  {pages} {1791} (\bibinfo {year} {2000})}\BibitemShut {NoStop}%
\bibitem [{\citenamefont {Ni}\ \emph {et~al.}(2009)\citenamefont {Ni},
  \citenamefont {Ospelkaus}, \citenamefont {Nesbitt}, \citenamefont {Ye},\ and\
  \citenamefont {Jin}}]{Ni2009}%
  \BibitemOpen
  \bibfield  {author} {\bibinfo {author} {\bibfnamefont {K.-K.}\ \bibnamefont
  {Ni}}, \bibinfo {author} {\bibfnamefont {S.}~\bibnamefont {Ospelkaus}},
  \bibinfo {author} {\bibfnamefont {D.~J.}\ \bibnamefont {Nesbitt}}, \bibinfo
  {author} {\bibfnamefont {J.}~\bibnamefont {Ye}},\ and\ \bibinfo {author}
  {\bibfnamefont {D.~S.}\ \bibnamefont {Jin}},\ }\bibfield  {title} {\bibinfo
  {title} {A dipolar gas of ultracold molecules},\ }\href
  {https://doi.org/10.1039/b911779b} {\bibfield  {journal} {\bibinfo  {journal}
  {Phys. Chem. Chem. Phys.}\ }\textbf {\bibinfo {volume} {322}},\ \bibinfo
  {pages} {231} (\bibinfo {year} {2009})}\BibitemShut {NoStop}%
\bibitem [{\citenamefont {Carr}\ \emph {et~al.}(2009)\citenamefont {Carr},
  \citenamefont {{DeMille}}, \citenamefont {Krems},\ and\ \citenamefont
  {Ye}}]{Carr2009}%
  \BibitemOpen
  \bibfield  {author} {\bibinfo {author} {\bibfnamefont {L.~D.}\ \bibnamefont
  {Carr}}, \bibinfo {author} {\bibfnamefont {D.}~\bibnamefont {{DeMille}}},
  \bibinfo {author} {\bibfnamefont {R.~V.}\ \bibnamefont {Krems}},\ and\
  \bibinfo {author} {\bibfnamefont {J.}~\bibnamefont {Ye}},\ }\bibfield
  {title} {\bibinfo {title} {Cold and ultracold molecules: science, technology
  and applications},\ }\href {https://doi.org/10.1088/1367-2630/11/5/055049}
  {\bibfield  {journal} {\bibinfo  {journal} {New J. Phys.}\ }\textbf {\bibinfo
  {volume} {11}},\ \bibinfo {pages} {055049} (\bibinfo {year}
  {2009})}\BibitemShut {NoStop}%
\bibitem [{\citenamefont {Baranov}\ \emph {et~al.}(2012)\citenamefont
  {Baranov}, \citenamefont {Dalmonte}, \citenamefont {Pupillo},\ and\
  \citenamefont {Zoller}}]{Baranov2012}%
  \BibitemOpen
  \bibfield  {author} {\bibinfo {author} {\bibfnamefont {M.~A.}\ \bibnamefont
  {Baranov}}, \bibinfo {author} {\bibfnamefont {M.}~\bibnamefont {Dalmonte}},
  \bibinfo {author} {\bibfnamefont {G.}~\bibnamefont {Pupillo}},\ and\ \bibinfo
  {author} {\bibfnamefont {P.}~\bibnamefont {Zoller}},\ }\bibfield  {title}
  {\bibinfo {title} {Condensed matter theory of dipolar quantum gases},\ }\href
  {https://doi.org/10.1021/cr2003568} {\bibfield  {journal} {\bibinfo
  {journal} {Chem. Rev.}\ }\textbf {\bibinfo {volume} {112}},\ \bibinfo {pages}
  {5012} (\bibinfo {year} {2012})}\BibitemShut {NoStop}%
\bibitem [{\citenamefont {Marco}\ \emph {et~al.}(2018)\citenamefont {Marco},
  \citenamefont {Valtolina}, \citenamefont {Matsuda}, \citenamefont {Tobias},
  \citenamefont {Covey},\ and\ \citenamefont {Ye}}]{Marco2018}%
  \BibitemOpen
  \bibfield  {author} {\bibinfo {author} {\bibfnamefont {L.~D.}\ \bibnamefont
  {Marco}}, \bibinfo {author} {\bibfnamefont {G.}~\bibnamefont {Valtolina}},
  \bibinfo {author} {\bibfnamefont {K.}~\bibnamefont {Matsuda}}, \bibinfo
  {author} {\bibfnamefont {W.~G.}\ \bibnamefont {Tobias}}, \bibinfo {author}
  {\bibfnamefont {J.~P.}\ \bibnamefont {Covey}},\ and\ \bibinfo {author}
  {\bibfnamefont {J.}~\bibnamefont {Ye}},\ }\bibfield  {title} {\bibinfo
  {title} {A fermi degenerate gas of polar molecules},\ }\href
  {https://doi.org/10.1126/science.aau723} {\bibfield  {journal} {\bibinfo
  {journal} {Science}\ }\textbf {\bibinfo {volume} {363}},\ \bibinfo {pages}
  {853} (\bibinfo {year} {2018})}\BibitemShut {NoStop}%
\bibitem [{\citenamefont {Barnett}\ \emph {et~al.}(2006)\citenamefont
  {Barnett}, \citenamefont {Petrov}, \citenamefont {Lukin},\ and\ \citenamefont
  {Demler}}]{Barnett2006}%
  \BibitemOpen
  \bibfield  {author} {\bibinfo {author} {\bibfnamefont {R.}~\bibnamefont
  {Barnett}}, \bibinfo {author} {\bibfnamefont {D.}~\bibnamefont {Petrov}},
  \bibinfo {author} {\bibfnamefont {M.}~\bibnamefont {Lukin}},\ and\ \bibinfo
  {author} {\bibfnamefont {E.}~\bibnamefont {Demler}},\ }\bibfield  {title}
  {\bibinfo {title} {Quantum magnetism with multicomponent dipolar molecules in
  an optical lattice},\ }\href {https://doi.org/10.1103/PhysRevLett.96.190401}
  {\bibfield  {journal} {\bibinfo  {journal} {Phys. Rev. Lett.}\ }\textbf
  {\bibinfo {volume} {96}},\ \bibinfo {pages} {190401} (\bibinfo {year}
  {2006})}\BibitemShut {NoStop}%
\bibitem [{\citenamefont {Micheli}\ \emph {et~al.}(2006)\citenamefont
  {Micheli}, \citenamefont {Brennen},\ and\ \citenamefont
  {Zoller}}]{Micheli2006}%
  \BibitemOpen
  \bibfield  {author} {\bibinfo {author} {\bibfnamefont {A.}~\bibnamefont
  {Micheli}}, \bibinfo {author} {\bibfnamefont {G.~K.}\ \bibnamefont
  {Brennen}},\ and\ \bibinfo {author} {\bibfnamefont {P.}~\bibnamefont
  {Zoller}},\ }\bibfield  {title} {\bibinfo {title} {A toolbox for lattice-spin
  mmodel with polar molecules},\ }\href {https://doi.org/10.1038/nphys287}
  {\bibfield  {journal} {\bibinfo  {journal} {Nat. Phys.}\ }\textbf {\bibinfo
  {volume} {2}},\ \bibinfo {pages} {341} (\bibinfo {year} {2006})}\BibitemShut
  {NoStop}%
\bibitem [{\citenamefont {B\"{u}chler}\ \emph {et~al.}(2007)\citenamefont
  {B\"{u}chler}, \citenamefont {Demler}, \citenamefont {Lukin}, \citenamefont
  {Micheli}, \citenamefont {Prokof'ev}, \citenamefont {Pupillo},\ and\
  \citenamefont {Zoller}}]{Buchler2007}%
  \BibitemOpen
  \bibfield  {author} {\bibinfo {author} {\bibfnamefont {H.~P.}\ \bibnamefont
  {B\"{u}chler}}, \bibinfo {author} {\bibfnamefont {E.}~\bibnamefont {Demler}},
  \bibinfo {author} {\bibfnamefont {M.}~\bibnamefont {Lukin}}, \bibinfo
  {author} {\bibfnamefont {A.}~\bibnamefont {Micheli}}, \bibinfo {author}
  {\bibfnamefont {N.}~\bibnamefont {Prokof'ev}}, \bibinfo {author}
  {\bibfnamefont {G.}~\bibnamefont {Pupillo}},\ and\ \bibinfo {author}
  {\bibfnamefont {P.}~\bibnamefont {Zoller}},\ }\bibfield  {title} {\bibinfo
  {title} {Strongly correlated 2d quantum phases with cold polar molecules:
  Controlling the shape of the interaction potential},\ }\href
  {https://doi.org/10.1103/PhysRevLett.98.060404} {\bibfield  {journal}
  {\bibinfo  {journal} {Phys. Rev. Lett.}\ }\textbf {\bibinfo {volume} {98}},\
  \bibinfo {pages} {060404} (\bibinfo {year} {2007})}\BibitemShut {NoStop}%
\bibitem [{\citenamefont {Maci{\`{a}}}\ \emph {et~al.}(2012)\citenamefont
  {Maci{\`{a}}}, \citenamefont {Hufnagl}, \citenamefont {Mazzanti},
  \citenamefont {Boronat},\ and\ \citenamefont {Zillich}}]{Macia2012}%
  \BibitemOpen
  \bibfield  {author} {\bibinfo {author} {\bibfnamefont {A.}~\bibnamefont
  {Maci{\`{a}}}}, \bibinfo {author} {\bibfnamefont {D.}~\bibnamefont
  {Hufnagl}}, \bibinfo {author} {\bibfnamefont {F.}~\bibnamefont {Mazzanti}},
  \bibinfo {author} {\bibfnamefont {J.}~\bibnamefont {Boronat}},\ and\ \bibinfo
  {author} {\bibfnamefont {R.~E.}\ \bibnamefont {Zillich}},\ }\bibfield
  {title} {\bibinfo {title} {{Excitations and stripe phase formation in a
  two-dimensional dipolar bose gas with tilted polarization}},\ }\href
  {https://doi.org/10.1103/PhysRevLett.109.235307} {\bibfield  {journal}
  {\bibinfo  {journal} {Phys. Rev. Lett.}\ }\textbf {\bibinfo {volume} {109}},\
  \bibinfo {pages} {235307} (\bibinfo {year} {2012})}\BibitemShut {NoStop}%
\bibitem [{\citenamefont {Manmana}\ \emph {et~al.}(2013)\citenamefont
  {Manmana}, \citenamefont {Stoudenmire}, \citenamefont {Hazzard},
  \citenamefont {Rey},\ and\ \citenamefont {Gorshkov}}]{Manmana2013}%
  \BibitemOpen
  \bibfield  {author} {\bibinfo {author} {\bibfnamefont {S.~R.}\ \bibnamefont
  {Manmana}}, \bibinfo {author} {\bibfnamefont {E.~M.}\ \bibnamefont
  {Stoudenmire}}, \bibinfo {author} {\bibfnamefont {K.~R.~A.}\ \bibnamefont
  {Hazzard}}, \bibinfo {author} {\bibfnamefont {A.~M.}\ \bibnamefont {Rey}},\
  and\ \bibinfo {author} {\bibfnamefont {A.~V.}\ \bibnamefont {Gorshkov}},\
  }\bibfield  {title} {\bibinfo {title} {Topological phases in ultracold
  polar-molecule quantum magnets},\ }\href
  {https://doi.org/10.1103/physrevb.87.081106} {\bibfield  {journal} {\bibinfo
  {journal} {Phys. Rev. B}\ }\textbf {\bibinfo {volume} {87}},\ \bibinfo
  {pages} {081106} (\bibinfo {year} {2013})}\BibitemShut {NoStop}%
\bibitem [{\citenamefont {Gorshkov}\ \emph {et~al.}(2013)\citenamefont
  {Gorshkov}, \citenamefont {Hazzard},\ and\ \citenamefont
  {Rey}}]{Gorshkov2013}%
  \BibitemOpen
  \bibfield  {author} {\bibinfo {author} {\bibfnamefont {A.~V.}\ \bibnamefont
  {Gorshkov}}, \bibinfo {author} {\bibfnamefont {K.~R.~A.}\ \bibnamefont
  {Hazzard}},\ and\ \bibinfo {author} {\bibfnamefont {A.~M.}\ \bibnamefont
  {Rey}},\ }\bibfield  {title} {\bibinfo {title} {Kitaev honeycomb and other
  exotic spin models with polar molecules},\ }\href
  {https://doi.org/10.1080/00268976.2013.800604} {\bibfield  {journal}
  {\bibinfo  {journal} {Mol. Phys.}\ }\textbf {\bibinfo {volume} {111}},\
  \bibinfo {pages} {1908} (\bibinfo {year} {2013})}\BibitemShut {NoStop}%
\bibitem [{\citenamefont {DeMille}(2002)}]{Demille2002}%
  \BibitemOpen
  \bibfield  {author} {\bibinfo {author} {\bibfnamefont {D.}~\bibnamefont
  {DeMille}},\ }\bibfield  {title} {\bibinfo {title} {Quantum computation with
  trapped polar molecules},\ }\href
  {https://doi.org/10.1103/PhysRevLett.88.067901} {\bibfield  {journal}
  {\bibinfo  {journal} {Phys. Rev. Lett.}\ }\textbf {\bibinfo {volume} {88}},\
  \bibinfo {pages} {067901} (\bibinfo {year} {2002})}\BibitemShut {NoStop}%
\bibitem [{\citenamefont {Yelin}\ \emph {et~al.}(2006)\citenamefont {Yelin},
  \citenamefont {Kirby},\ and\ \citenamefont {C{\^{o}}t{\'{e}}}}]{Yelin2006}%
  \BibitemOpen
  \bibfield  {author} {\bibinfo {author} {\bibfnamefont {S.~F.}\ \bibnamefont
  {Yelin}}, \bibinfo {author} {\bibfnamefont {K.}~\bibnamefont {Kirby}},\ and\
  \bibinfo {author} {\bibfnamefont {R.}~\bibnamefont {C{\^{o}}t{\'{e}}}},\
  }\bibfield  {title} {\bibinfo {title} {Schemes for robust quantum computation
  with polar molecules},\ }\href {https://doi.org/10.1103/physreva.74.050301}
  {\bibfield  {journal} {\bibinfo  {journal} {Phys. Rev. A}\ }\textbf {\bibinfo
  {volume} {74}},\ \bibinfo {pages} {050301} (\bibinfo {year}
  {2006})}\BibitemShut {NoStop}%
\bibitem [{\citenamefont {Zhu}\ \emph {et~al.}(2013)\citenamefont {Zhu},
  \citenamefont {Kais}, \citenamefont {Wei}, \citenamefont {Herschbach},\ and\
  \citenamefont {Friedrich}}]{Zhu2013}%
  \BibitemOpen
  \bibfield  {author} {\bibinfo {author} {\bibfnamefont {J.}~\bibnamefont
  {Zhu}}, \bibinfo {author} {\bibfnamefont {S.}~\bibnamefont {Kais}}, \bibinfo
  {author} {\bibfnamefont {Q.}~\bibnamefont {Wei}}, \bibinfo {author}
  {\bibfnamefont {D.}~\bibnamefont {Herschbach}},\ and\ \bibinfo {author}
  {\bibfnamefont {B.}~\bibnamefont {Friedrich}},\ }\bibfield  {title} {\bibinfo
  {title} {Implementation of quantum logic gates using polar molecules in
  pendular states},\ }\href {https://doi.org/10.1063/1.4774058} {\bibfield
  {journal} {\bibinfo  {journal} {J. Chem. Phys.}\ }\textbf {\bibinfo {volume}
  {138}},\ \bibinfo {pages} {024104} (\bibinfo {year} {2013})}\BibitemShut
  {NoStop}%
\bibitem [{\citenamefont {Herrera}\ \emph {et~al.}(2014)\citenamefont
  {Herrera}, \citenamefont {Cao}, \citenamefont {Kais},\ and\ \citenamefont
  {Whaley}}]{Herrera2014}%
  \BibitemOpen
  \bibfield  {author} {\bibinfo {author} {\bibfnamefont {F.}~\bibnamefont
  {Herrera}}, \bibinfo {author} {\bibfnamefont {Y.}~\bibnamefont {Cao}},
  \bibinfo {author} {\bibfnamefont {S.}~\bibnamefont {Kais}},\ and\ \bibinfo
  {author} {\bibfnamefont {K.~B.}\ \bibnamefont {Whaley}},\ }\bibfield  {title}
  {\bibinfo {title} {Infrared-dressed entanglement of cold open-shell polar
  molecules for universal matchgate quantum computing},\ }\href
  {https://doi.org/10.1088/1367-2630/16/7/075001} {\bibfield  {journal}
  {\bibinfo  {journal} {New J. Phys.}\ }\textbf {\bibinfo {volume} {16}},\
  \bibinfo {pages} {075001} (\bibinfo {year} {2014})}\BibitemShut {NoStop}%
\bibitem [{\citenamefont {Ni}\ \emph {et~al.}(2018)\citenamefont {Ni},
  \citenamefont {Rosenband},\ and\ \citenamefont {Grimes}}]{Ni2018}%
  \BibitemOpen
  \bibfield  {author} {\bibinfo {author} {\bibfnamefont {K.-K.}\ \bibnamefont
  {Ni}}, \bibinfo {author} {\bibfnamefont {T.}~\bibnamefont {Rosenband}},\ and\
  \bibinfo {author} {\bibfnamefont {D.~D.}\ \bibnamefont {Grimes}},\ }\bibfield
   {title} {\bibinfo {title} {Dipolar exchange quantum logic gate with polar
  molecules},\ }\href {https://doi.org/10.1039/C8SC02355G} {\bibfield
  {journal} {\bibinfo  {journal} {Chem. Sci.}\ }\textbf {\bibinfo {volume}
  {9}},\ \bibinfo {pages} {6830} (\bibinfo {year} {2018})}\BibitemShut
  {NoStop}%
\bibitem [{\citenamefont {Sawant}\ \emph {et~al.}(2019)\citenamefont {Sawant},
  \citenamefont {Blackmore}, \citenamefont {Gregory}, \citenamefont
  {Mur-Petit}, \citenamefont {Jaksch}, \citenamefont {Aldegunde}, \citenamefont
  {Hutson}, \citenamefont {Tarbutt},\ and\ \citenamefont
  {Cornish}}]{Sawant2019}%
  \BibitemOpen
  \bibfield  {author} {\bibinfo {author} {\bibfnamefont {R.}~\bibnamefont
  {Sawant}}, \bibinfo {author} {\bibfnamefont {J.~A.}\ \bibnamefont
  {Blackmore}}, \bibinfo {author} {\bibfnamefont {P.~D.}\ \bibnamefont
  {Gregory}}, \bibinfo {author} {\bibfnamefont {J.}~\bibnamefont {Mur-Petit}},
  \bibinfo {author} {\bibfnamefont {D.}~\bibnamefont {Jaksch}}, \bibinfo
  {author} {\bibfnamefont {J.}~\bibnamefont {Aldegunde}}, \bibinfo {author}
  {\bibfnamefont {J.~M.}\ \bibnamefont {Hutson}}, \bibinfo {author}
  {\bibfnamefont {M.~R.}\ \bibnamefont {Tarbutt}},\ and\ \bibinfo {author}
  {\bibfnamefont {S.~L.}\ \bibnamefont {Cornish}},\ }\bibfield  {title}
  {\bibinfo {title} {Ultracold molecules as qudits},\ }\href@noop {} {\bibfield
   {journal} {\bibinfo  {journal} {New J. Phys.}\ } (\bibinfo {year}
  {2019})}\BibitemShut {NoStop}%
\bibitem [{\citenamefont {Hughes}\ \emph {et~al.}(2020)\citenamefont {Hughes},
  \citenamefont {Frye}, \citenamefont {Sawant}, \citenamefont {Bhole},
  \citenamefont {Jones}, \citenamefont {Cornish}, \citenamefont {Tarbutt},
  \citenamefont {Hutson}, \citenamefont {Jaksch},\ and\ \citenamefont
  {Mur-Petit}}]{Hughes2020}%
  \BibitemOpen
  \bibfield  {author} {\bibinfo {author} {\bibfnamefont {M.}~\bibnamefont
  {Hughes}}, \bibinfo {author} {\bibfnamefont {M.~D.}\ \bibnamefont {Frye}},
  \bibinfo {author} {\bibfnamefont {R.}~\bibnamefont {Sawant}}, \bibinfo
  {author} {\bibfnamefont {G.}~\bibnamefont {Bhole}}, \bibinfo {author}
  {\bibfnamefont {J.~A.}\ \bibnamefont {Jones}}, \bibinfo {author}
  {\bibfnamefont {S.~L.}\ \bibnamefont {Cornish}}, \bibinfo {author}
  {\bibfnamefont {M.~R.}\ \bibnamefont {Tarbutt}}, \bibinfo {author}
  {\bibfnamefont {J.~M.}\ \bibnamefont {Hutson}}, \bibinfo {author}
  {\bibfnamefont {D.}~\bibnamefont {Jaksch}},\ and\ \bibinfo {author}
  {\bibfnamefont {J.}~\bibnamefont {Mur-Petit}},\ }\bibfield  {title} {\bibinfo
  {title} {Robust entangling gate for polar molecules using magnetic and
  microwave fields},\ }\href {https://doi.org/10.1103/PhysRevA.101.062308}
  {\bibfield  {journal} {\bibinfo  {journal} {Phys. Rev. A}\ }\textbf {\bibinfo
  {volume} {101}},\ \bibinfo {pages} {062308} (\bibinfo {year}
  {2020})}\BibitemShut {NoStop}%
\bibitem [{\citenamefont {Ni}\ \emph {et~al.}(2008)\citenamefont {Ni},
  \citenamefont {Ospelkaus}, \citenamefont {de~Miranda}, \citenamefont {Pe'er},
  \citenamefont {Neyenhuis}, \citenamefont {Zirbel}, \citenamefont
  {Kotochigova}, \citenamefont {Julienne}, \citenamefont {Jin},\ and\
  \citenamefont {Ye}}]{Ni2008}%
  \BibitemOpen
  \bibfield  {author} {\bibinfo {author} {\bibfnamefont {K.-K.}\ \bibnamefont
  {Ni}}, \bibinfo {author} {\bibfnamefont {S.}~\bibnamefont {Ospelkaus}},
  \bibinfo {author} {\bibfnamefont {M.~H.~G.}\ \bibnamefont {de~Miranda}},
  \bibinfo {author} {\bibfnamefont {A.}~\bibnamefont {Pe'er}}, \bibinfo
  {author} {\bibfnamefont {B.}~\bibnamefont {Neyenhuis}}, \bibinfo {author}
  {\bibfnamefont {J.~J.}\ \bibnamefont {Zirbel}}, \bibinfo {author}
  {\bibfnamefont {S.}~\bibnamefont {Kotochigova}}, \bibinfo {author}
  {\bibfnamefont {P.~S.}\ \bibnamefont {Julienne}}, \bibinfo {author}
  {\bibfnamefont {D.~S.}\ \bibnamefont {Jin}},\ and\ \bibinfo {author}
  {\bibfnamefont {J.}~\bibnamefont {Ye}},\ }\bibfield  {title} {\bibinfo
  {title} {A high phase-space-density gas of polar molecules},\ }\href
  {https://doi.org/10.1126/science.1163861} {\bibfield  {journal} {\bibinfo
  {journal} {Science}\ }\textbf {\bibinfo {volume} {322}},\ \bibinfo {pages}
  {231} (\bibinfo {year} {2008})}\BibitemShut {NoStop}%
\bibitem [{\citenamefont {Danzl}\ \emph {et~al.}(2008)\citenamefont {Danzl},
  \citenamefont {Haller}, \citenamefont {Gustavsson}, \citenamefont {Mark},
  \citenamefont {Hart}, \citenamefont {Bouloufa}, \citenamefont {Dulieu},
  \citenamefont {Ritsch},\ and\ \citenamefont {N\"{a}gerl}}]{Danzl2008}%
  \BibitemOpen
  \bibfield  {author} {\bibinfo {author} {\bibfnamefont {J.~G.}\ \bibnamefont
  {Danzl}}, \bibinfo {author} {\bibfnamefont {E.}~\bibnamefont {Haller}},
  \bibinfo {author} {\bibfnamefont {M.}~\bibnamefont {Gustavsson}}, \bibinfo
  {author} {\bibfnamefont {M.~J.}\ \bibnamefont {Mark}}, \bibinfo {author}
  {\bibfnamefont {R.}~\bibnamefont {Hart}}, \bibinfo {author} {\bibfnamefont
  {N.}~\bibnamefont {Bouloufa}}, \bibinfo {author} {\bibfnamefont
  {O.}~\bibnamefont {Dulieu}}, \bibinfo {author} {\bibfnamefont
  {H.}~\bibnamefont {Ritsch}},\ and\ \bibinfo {author} {\bibfnamefont {H.-C.}\
  \bibnamefont {N\"{a}gerl}},\ }\bibfield  {title} {\bibinfo {title} {Quantum
  gas of deeply bound ground state molecules},\ }\href
  {https://doi.org/10.1126/science.1159909} {\bibfield  {journal} {\bibinfo
  {journal} {Science}\ }\textbf {\bibinfo {volume} {321}},\ \bibinfo {pages}
  {1062} (\bibinfo {year} {2008})}\BibitemShut {NoStop}%
\bibitem [{\citenamefont {Lang}\ \emph {et~al.}(2008)\citenamefont {Lang},
  \citenamefont {Winkler}, \citenamefont {Strauss}, \citenamefont {Grimm},\
  and\ \citenamefont {Hecker~Denschlag}}]{Lang2008}%
  \BibitemOpen
  \bibfield  {author} {\bibinfo {author} {\bibfnamefont {F.}~\bibnamefont
  {Lang}}, \bibinfo {author} {\bibfnamefont {K.}~\bibnamefont {Winkler}},
  \bibinfo {author} {\bibfnamefont {C.}~\bibnamefont {Strauss}}, \bibinfo
  {author} {\bibfnamefont {R.}~\bibnamefont {Grimm}},\ and\ \bibinfo {author}
  {\bibfnamefont {J.}~\bibnamefont {Hecker~Denschlag}},\ }\bibfield  {title}
  {\bibinfo {title} {Ultracold triplet molecules in the rovibrational ground
  state},\ }\href {https://doi.org/10.1103/PhysRevLett.101.133005} {\bibfield
  {journal} {\bibinfo  {journal} {Phys. Rev. Lett.}\ }\textbf {\bibinfo
  {volume} {101}},\ \bibinfo {pages} {133005} (\bibinfo {year}
  {2008})}\BibitemShut {NoStop}%
\bibitem [{\citenamefont {Takekoshi}\ \emph {et~al.}(2014)\citenamefont
  {Takekoshi}, \citenamefont {Reichs\"{o}llner}, \citenamefont {Schindewolf},
  \citenamefont {Hutson}, \citenamefont {Le~Sueur}, \citenamefont {Dulieu},
  \citenamefont {Ferlaino}, \citenamefont {Grimm},\ and\ \citenamefont
  {N\"{a}gerl}}]{Takekoshi2014}%
  \BibitemOpen
  \bibfield  {author} {\bibinfo {author} {\bibfnamefont {T.}~\bibnamefont
  {Takekoshi}}, \bibinfo {author} {\bibfnamefont {L.}~\bibnamefont
  {Reichs\"{o}llner}}, \bibinfo {author} {\bibfnamefont {A.}~\bibnamefont
  {Schindewolf}}, \bibinfo {author} {\bibfnamefont {J.~M.}\ \bibnamefont
  {Hutson}}, \bibinfo {author} {\bibfnamefont {C.~R.}\ \bibnamefont
  {Le~Sueur}}, \bibinfo {author} {\bibfnamefont {O.}~\bibnamefont {Dulieu}},
  \bibinfo {author} {\bibfnamefont {F.}~\bibnamefont {Ferlaino}}, \bibinfo
  {author} {\bibfnamefont {R.}~\bibnamefont {Grimm}},\ and\ \bibinfo {author}
  {\bibfnamefont {H.-C.}\ \bibnamefont {N\"{a}gerl}},\ }\bibfield  {title}
  {\bibinfo {title} {Ultracold dense samples of dipolar {RbCs} molecules in the
  rovibrational and hyperfine ground state},\ }\href
  {https://doi.org/10.1103/PhysRevLett.113.205301} {\bibfield  {journal}
  {\bibinfo  {journal} {Phys. Rev. Lett.}\ }\textbf {\bibinfo {volume} {113}},\
  \bibinfo {pages} {205301} (\bibinfo {year} {2014})}\BibitemShut {NoStop}%
\bibitem [{\citenamefont {Molony}\ \emph {et~al.}(2014)\citenamefont {Molony},
  \citenamefont {Gregory}, \citenamefont {Ji}, \citenamefont {Lu},
  \citenamefont {K\"{o}ppinger}, \citenamefont {Le~Sueur}, \citenamefont
  {Blackley}, \citenamefont {Hutson},\ and\ \citenamefont
  {Cornish}}]{Molony2014}%
  \BibitemOpen
  \bibfield  {author} {\bibinfo {author} {\bibfnamefont {P.~K.}\ \bibnamefont
  {Molony}}, \bibinfo {author} {\bibfnamefont {P.~D.}\ \bibnamefont {Gregory}},
  \bibinfo {author} {\bibfnamefont {Z.}~\bibnamefont {Ji}}, \bibinfo {author}
  {\bibfnamefont {B.}~\bibnamefont {Lu}}, \bibinfo {author} {\bibfnamefont
  {M.~P.}\ \bibnamefont {K\"{o}ppinger}}, \bibinfo {author} {\bibfnamefont
  {C.~R.}\ \bibnamefont {Le~Sueur}}, \bibinfo {author} {\bibfnamefont {C.~L.}\
  \bibnamefont {Blackley}}, \bibinfo {author} {\bibfnamefont {J.~M.}\
  \bibnamefont {Hutson}},\ and\ \bibinfo {author} {\bibfnamefont {S.~L.}\
  \bibnamefont {Cornish}},\ }\bibfield  {title} {\bibinfo {title} {Creation of
  ultracold $^{87}${Rb}$^{133}${Cs} molecules in the rovibrational ground
  state},\ }\href {https://doi.org/10.1103/PhysRevLett.113.255301} {\bibfield
  {journal} {\bibinfo  {journal} {Phys. Rev. Lett.}\ }\textbf {\bibinfo
  {volume} {113}},\ \bibinfo {pages} {255301} (\bibinfo {year}
  {2014})}\BibitemShut {NoStop}%
\bibitem [{\citenamefont {Park}\ \emph {et~al.}(2015)\citenamefont {Park},
  \citenamefont {Will},\ and\ \citenamefont {Zwierlein}}]{Park2015}%
  \BibitemOpen
  \bibfield  {author} {\bibinfo {author} {\bibfnamefont {J.~W.}\ \bibnamefont
  {Park}}, \bibinfo {author} {\bibfnamefont {S.~A.}\ \bibnamefont {Will}},\
  and\ \bibinfo {author} {\bibfnamefont {M.~W.}\ \bibnamefont {Zwierlein}},\
  }\bibfield  {title} {\bibinfo {title} {Ultracold dipolar gas of fermionic
  $^{23}${Na}$^{40}${K} molecules in their absolute ground state},\ }\href
  {https://doi.org/10.1103/PhysRevLett.114.205302} {\bibfield  {journal}
  {\bibinfo  {journal} {Phys. Rev. Lett.}\ }\textbf {\bibinfo {volume} {114}},\
  \bibinfo {pages} {205302} (\bibinfo {year} {2015})}\BibitemShut {NoStop}%
\bibitem [{\citenamefont {See{\ss}elberg}\ \emph {et~al.}(2018)\citenamefont
  {See{\ss}elberg}, \citenamefont {Buchheim}, \citenamefont {Lu}, \citenamefont
  {Schneider}, \citenamefont {Luo}, \citenamefont {Tiemann}, \citenamefont
  {Bloch},\ and\ \citenamefont {Gohle}}]{Seesselberg2018b}%
  \BibitemOpen
  \bibfield  {author} {\bibinfo {author} {\bibfnamefont {F.}~\bibnamefont
  {See{\ss}elberg}}, \bibinfo {author} {\bibfnamefont {N.}~\bibnamefont
  {Buchheim}}, \bibinfo {author} {\bibfnamefont {Z.-K.}\ \bibnamefont {Lu}},
  \bibinfo {author} {\bibfnamefont {T.}~\bibnamefont {Schneider}}, \bibinfo
  {author} {\bibfnamefont {X.-Y.}\ \bibnamefont {Luo}}, \bibinfo {author}
  {\bibfnamefont {E.}~\bibnamefont {Tiemann}}, \bibinfo {author} {\bibfnamefont
  {I.}~\bibnamefont {Bloch}},\ and\ \bibinfo {author} {\bibfnamefont
  {C.}~\bibnamefont {Gohle}},\ }\bibfield  {title} {\bibinfo {title} {Modeling
  the adiabatic creation of ultracold polar {$^{23}$}{Na}{$^{40}$}{K}
  molecules},\ }\href {https://doi.org/10.1103/physreva.97.013405} {\bibfield
  {journal} {\bibinfo  {journal} {Phys. Rev. A}\ }\textbf {\bibinfo {volume}
  {97}},\ \bibinfo {pages} {013405} (\bibinfo {year} {2018})}\BibitemShut
  {NoStop}%
\bibitem [{\citenamefont {Yang}\ \emph {et~al.}(2019)\citenamefont {Yang},
  \citenamefont {Zhang}, \citenamefont {Liu}, \citenamefont {Liu},
  \citenamefont {Nan}, \citenamefont {Zhao},\ and\ \citenamefont
  {Pan}}]{Yang2019}%
  \BibitemOpen
  \bibfield  {author} {\bibinfo {author} {\bibfnamefont {H.}~\bibnamefont
  {Yang}}, \bibinfo {author} {\bibfnamefont {D.-C.}\ \bibnamefont {Zhang}},
  \bibinfo {author} {\bibfnamefont {L.}~\bibnamefont {Liu}}, \bibinfo {author}
  {\bibfnamefont {Y.-X.}\ \bibnamefont {Liu}}, \bibinfo {author} {\bibfnamefont
  {J.}~\bibnamefont {Nan}}, \bibinfo {author} {\bibfnamefont {B.}~\bibnamefont
  {Zhao}},\ and\ \bibinfo {author} {\bibfnamefont {J.-W.}\ \bibnamefont
  {Pan}},\ }\bibfield  {title} {\bibinfo {title} {Observation of magnetically
  tunable feshbach resonances in ultracold $^{23}${Na}$^{40}${K} + $^{40}${K}
  collisions},\ }\href {https://doi.org/10.1126/science.aau5322} {\bibfield
  {journal} {\bibinfo  {journal} {Science}\ }\textbf {\bibinfo {volume}
  {363}},\ \bibinfo {pages} {261} (\bibinfo {year} {2019})}\BibitemShut
  {NoStop}%
\bibitem [{\citenamefont {Guo}\ \emph {et~al.}(2016)\citenamefont {Guo},
  \citenamefont {Zhu}, \citenamefont {Lu}, \citenamefont {Ye}, \citenamefont
  {Wang}, \citenamefont {Vexiau}, \citenamefont {{Bouloufa-Maafa}},
  \citenamefont {Qu\'{e}m\'{e}ner}, \citenamefont {Dulieu},\ and\ \citenamefont
  {Wang}}]{Guo2016}%
  \BibitemOpen
  \bibfield  {author} {\bibinfo {author} {\bibfnamefont {M.}~\bibnamefont
  {Guo}}, \bibinfo {author} {\bibfnamefont {B.}~\bibnamefont {Zhu}}, \bibinfo
  {author} {\bibfnamefont {B.}~\bibnamefont {Lu}}, \bibinfo {author}
  {\bibfnamefont {X.}~\bibnamefont {Ye}}, \bibinfo {author} {\bibfnamefont
  {F.}~\bibnamefont {Wang}}, \bibinfo {author} {\bibfnamefont {R.}~\bibnamefont
  {Vexiau}}, \bibinfo {author} {\bibfnamefont {N.}~\bibnamefont
  {{Bouloufa-Maafa}}}, \bibinfo {author} {\bibfnamefont {G.}~\bibnamefont
  {Qu\'{e}m\'{e}ner}}, \bibinfo {author} {\bibfnamefont {O.}~\bibnamefont
  {Dulieu}},\ and\ \bibinfo {author} {\bibfnamefont {D.}~\bibnamefont {Wang}},\
  }\bibfield  {title} {\bibinfo {title} {Creation of an ultracold gas of
  ground-state dipolar $^{23}${Na}$^{87}${Rb} molecules},\ }\href
  {https://doi.org/10.1103/PhysRevLett.116.205303} {\bibfield  {journal}
  {\bibinfo  {journal} {Phys. Rev. Lett.}\ }\textbf {\bibinfo {volume} {116}},\
  \bibinfo {pages} {205303} (\bibinfo {year} {2016})}\BibitemShut {NoStop}%
\bibitem [{\citenamefont {Rvachov}\ \emph {et~al.}(2017)\citenamefont
  {Rvachov}, \citenamefont {Son}, \citenamefont {Sommer}, \citenamefont
  {Ebadi}, \citenamefont {Park}, \citenamefont {Zwierlein}, \citenamefont
  {Ketterle},\ and\ \citenamefont {Jamison}}]{Rvachov2017}%
  \BibitemOpen
  \bibfield  {author} {\bibinfo {author} {\bibfnamefont {T.~M.}\ \bibnamefont
  {Rvachov}}, \bibinfo {author} {\bibfnamefont {H.}~\bibnamefont {Son}},
  \bibinfo {author} {\bibfnamefont {A.~T.}\ \bibnamefont {Sommer}}, \bibinfo
  {author} {\bibfnamefont {S.}~\bibnamefont {Ebadi}}, \bibinfo {author}
  {\bibfnamefont {J.~J.}\ \bibnamefont {Park}}, \bibinfo {author}
  {\bibfnamefont {M.~W.}\ \bibnamefont {Zwierlein}}, \bibinfo {author}
  {\bibfnamefont {W.}~\bibnamefont {Ketterle}},\ and\ \bibinfo {author}
  {\bibfnamefont {A.~O.}\ \bibnamefont {Jamison}},\ }\bibfield  {title}
  {\bibinfo {title} {Long-lived ultracold molecules with electric and magnetic
  dipole moments},\ }\href {https://doi.org/10.1103/PhysRevLett.119.143001}
  {\bibfield  {journal} {\bibinfo  {journal} {Phys. Rev. Lett.}\ }\textbf
  {\bibinfo {volume} {119}},\ \bibinfo {pages} {143001} (\bibinfo {year}
  {2017})}\BibitemShut {NoStop}%
\bibitem [{\citenamefont {Hara}\ \emph {et~al.}(2011)\citenamefont {Hara},
  \citenamefont {Takasu}, \citenamefont {Yamaoka}, \citenamefont {Doyle},\ and\
  \citenamefont {Takahashi}}]{Hara2011}%
  \BibitemOpen
  \bibfield  {author} {\bibinfo {author} {\bibfnamefont {H.}~\bibnamefont
  {Hara}}, \bibinfo {author} {\bibfnamefont {Y.}~\bibnamefont {Takasu}},
  \bibinfo {author} {\bibfnamefont {Y.}~\bibnamefont {Yamaoka}}, \bibinfo
  {author} {\bibfnamefont {J.~M.}\ \bibnamefont {Doyle}},\ and\ \bibinfo
  {author} {\bibfnamefont {Y.}~\bibnamefont {Takahashi}},\ }\bibfield  {title}
  {\bibinfo {title} {Quantum degenerate mixtures of alkali and
  alkaline-earth-like atoms},\ }\href
  {https://doi.org/10.1103/physrevlett.106.205304} {\bibfield  {journal}
  {\bibinfo  {journal} {Phys. Rev. Lett.}\ }\textbf {\bibinfo {volume} {106}},\
  \bibinfo {pages} {205304} (\bibinfo {year} {2011})}\BibitemShut {NoStop}%
\bibitem [{\citenamefont {Roy}\ \emph {et~al.}(2016)\citenamefont {Roy},
  \citenamefont {Shrestha}, \citenamefont {Green}, \citenamefont {Gupta},
  \citenamefont {Li}, \citenamefont {Kotochigova}, \citenamefont {Petrov},\
  and\ \citenamefont {Yuen}}]{Roy2016}%
  \BibitemOpen
  \bibfield  {author} {\bibinfo {author} {\bibfnamefont {R.}~\bibnamefont
  {Roy}}, \bibinfo {author} {\bibfnamefont {R.}~\bibnamefont {Shrestha}},
  \bibinfo {author} {\bibfnamefont {A.}~\bibnamefont {Green}}, \bibinfo
  {author} {\bibfnamefont {S.}~\bibnamefont {Gupta}}, \bibinfo {author}
  {\bibfnamefont {M.}~\bibnamefont {Li}}, \bibinfo {author} {\bibfnamefont
  {S.}~\bibnamefont {Kotochigova}}, \bibinfo {author} {\bibfnamefont
  {A.}~\bibnamefont {Petrov}},\ and\ \bibinfo {author} {\bibfnamefont {C.~H.}\
  \bibnamefont {Yuen}},\ }\bibfield  {title} {\bibinfo {title}
  {Photoassociative production of ultracold heteronuclear {Yb}{Li}$^*$
  molecules},\ }\href {https://doi.org/10.1103/physreva.94.033413} {\bibfield
  {journal} {\bibinfo  {journal} {Phys. Rev. A}\ }\textbf {\bibinfo {volume}
  {94}},\ \bibinfo {pages} {033413} (\bibinfo {year} {2016})}\BibitemShut
  {NoStop}%
\bibitem [{\citenamefont {Guttridge}\ \emph {et~al.}(2018)\citenamefont
  {Guttridge}, \citenamefont {Frye}, \citenamefont {Yang}, \citenamefont
  {Hutson},\ and\ \citenamefont {Cornish}}]{Guttridge2018}%
  \BibitemOpen
  \bibfield  {author} {\bibinfo {author} {\bibfnamefont {A.}~\bibnamefont
  {Guttridge}}, \bibinfo {author} {\bibfnamefont {M.~D.}\ \bibnamefont {Frye}},
  \bibinfo {author} {\bibfnamefont {B.~C.}\ \bibnamefont {Yang}}, \bibinfo
  {author} {\bibfnamefont {J.~M.}\ \bibnamefont {Hutson}},\ and\ \bibinfo
  {author} {\bibfnamefont {S.~L.}\ \bibnamefont {Cornish}},\ }\bibfield
  {title} {\bibinfo {title} {Two-photon photoassociation spectroscopy of
  {CsYb}: Ground-state interaction potential and interspecies scattering
  lengths},\ }\href {https://doi.org/10.1103/physreva.98.022707} {\bibfield
  {journal} {\bibinfo  {journal} {Phys. Rev. A}\ }\textbf {\bibinfo {volume}
  {98}},\ \bibinfo {pages} {022707} (\bibinfo {year} {2018})}\BibitemShut
  {NoStop}%
\bibitem [{\citenamefont {Barb{\'{e}}}\ \emph {et~al.}(2018)\citenamefont
  {Barb{\'{e}}}, \citenamefont {Ciamei}, \citenamefont {Pasquiou},
  \citenamefont {Reichs{\"{o}}llner}, \citenamefont {Schreck}, \citenamefont
  {{\.{Z}}uchowski},\ and\ \citenamefont {Hutson}}]{Barbe2018}%
  \BibitemOpen
  \bibfield  {author} {\bibinfo {author} {\bibfnamefont {V.}~\bibnamefont
  {Barb{\'{e}}}}, \bibinfo {author} {\bibfnamefont {A.}~\bibnamefont {Ciamei}},
  \bibinfo {author} {\bibfnamefont {B.}~\bibnamefont {Pasquiou}}, \bibinfo
  {author} {\bibfnamefont {L.}~\bibnamefont {Reichs{\"{o}}llner}}, \bibinfo
  {author} {\bibfnamefont {F.}~\bibnamefont {Schreck}}, \bibinfo {author}
  {\bibfnamefont {P.~S.}\ \bibnamefont {{\.{Z}}uchowski}},\ and\ \bibinfo
  {author} {\bibfnamefont {J.~M.}\ \bibnamefont {Hutson}},\ }\bibfield  {title}
  {\bibinfo {title} {{Observation of Feshbach resonances between alkali and
  closed-shell atoms}},\ }\href {https://doi.org/10.1038/s41567-018-0169-x}
  {\bibfield  {journal} {\bibinfo  {journal} {Nat. Phys.}\ }\textbf {\bibinfo
  {volume} {14}},\ \bibinfo {pages} {881} (\bibinfo {year} {2018})}\BibitemShut
  {NoStop}%
\bibitem [{\citenamefont {Frye}\ \emph {et~al.}(2019)\citenamefont {Frye},
  \citenamefont {Cornish},\ and\ \citenamefont {Hutson}}]{Frye2019}%
  \BibitemOpen
  \bibfield  {author} {\bibinfo {author} {\bibfnamefont {M.~D.}\ \bibnamefont
  {Frye}}, \bibinfo {author} {\bibfnamefont {S.~L.}\ \bibnamefont {Cornish}},\
  and\ \bibinfo {author} {\bibfnamefont {J.~M.}\ \bibnamefont {Hutson}},\
  }\bibfield  {title} {\bibinfo {title} {Prospects of forming high-spin polar
  molecules from ultracold atoms},\ }\Eprint
  {https://arxiv.org/abs/1910.09641v2} {arXiv:1910.09641v2 [physics.atom-ph]}
  (\bibinfo {year} {2019})\BibitemShut {NoStop}%
\bibitem [{\citenamefont {Green}\ \emph {et~al.}(2019)\citenamefont {Green},
  \citenamefont {Toh}, \citenamefont {Roy}, \citenamefont {Li}, \citenamefont
  {Kotochigova},\ and\ \citenamefont {Gupta}}]{Green2019}%
  \BibitemOpen
  \bibfield  {author} {\bibinfo {author} {\bibfnamefont {A.}~\bibnamefont
  {Green}}, \bibinfo {author} {\bibfnamefont {J.~H.~S.}\ \bibnamefont {Toh}},
  \bibinfo {author} {\bibfnamefont {R.}~\bibnamefont {Roy}}, \bibinfo {author}
  {\bibfnamefont {M.}~\bibnamefont {Li}}, \bibinfo {author} {\bibfnamefont
  {S.}~\bibnamefont {Kotochigova}},\ and\ \bibinfo {author} {\bibfnamefont
  {S.}~\bibnamefont {Gupta}},\ }\bibfield  {title} {\bibinfo {title}
  {Two-photon photoassociation spectroscopy of the {$^2\Sigma^+$} {YbLi}
  molecular ground state},\ }\href {https://doi.org/10.1103/physreva.99.063416}
  {\bibfield  {journal} {\bibinfo  {journal} {Phys. Rev. A}\ }\textbf {\bibinfo
  {volume} {99}},\ \bibinfo {pages} {063416} (\bibinfo {year}
  {2019})}\BibitemShut {NoStop}%
\bibitem [{\citenamefont {Shuman}\ \emph {et~al.}(2010)\citenamefont {Shuman},
  \citenamefont {Barry},\ and\ \citenamefont {DeMille}}]{Shuman2010}%
  \BibitemOpen
  \bibfield  {author} {\bibinfo {author} {\bibfnamefont {E.~S.}\ \bibnamefont
  {Shuman}}, \bibinfo {author} {\bibfnamefont {J.~F.}\ \bibnamefont {Barry}},\
  and\ \bibinfo {author} {\bibfnamefont {D.}~\bibnamefont {DeMille}},\
  }\bibfield  {title} {\bibinfo {title} {{Laser cooling of a diatomic
  molecule}},\ }\href {https://doi.org/10.1038/nature09443} {\bibfield
  {journal} {\bibinfo  {journal} {Nature}\ }\textbf {\bibinfo {volume} {467}},\
  \bibinfo {pages} {820} (\bibinfo {year} {2010})}\BibitemShut {NoStop}%
\bibitem [{\citenamefont {Barry}\ \emph {et~al.}(2014)\citenamefont {Barry},
  \citenamefont {{McCarron}}, \citenamefont {Norrgard}, \citenamefont
  {Steinecker},\ and\ \citenamefont {{DeMille}}}]{Barry2014}%
  \BibitemOpen
  \bibfield  {author} {\bibinfo {author} {\bibfnamefont {J.~F.}\ \bibnamefont
  {Barry}}, \bibinfo {author} {\bibfnamefont {D.~J.}\ \bibnamefont
  {{McCarron}}}, \bibinfo {author} {\bibfnamefont {E.~B.}\ \bibnamefont
  {Norrgard}}, \bibinfo {author} {\bibfnamefont {M.~H.}\ \bibnamefont
  {Steinecker}},\ and\ \bibinfo {author} {\bibfnamefont {D.}~\bibnamefont
  {{DeMille}}},\ }\bibfield  {title} {\bibinfo {title} {Magneto-optical
  trapping of a diatomic molecule},\ }\href
  {https://doi.org/10.1038/nature13634} {\bibfield  {journal} {\bibinfo
  {journal} {Nature}\ }\textbf {\bibinfo {volume} {512}},\ \bibinfo {pages}
  {286} (\bibinfo {year} {2014})}\BibitemShut {NoStop}%
\bibitem [{\citenamefont {{McCarron}}\ \emph {et~al.}(2015)\citenamefont
  {{McCarron}}, \citenamefont {Norrgard}, \citenamefont {Steinecker},\ and\
  \citenamefont {{DeMille}}}]{McCarron2015}%
  \BibitemOpen
  \bibfield  {author} {\bibinfo {author} {\bibfnamefont {D.~J.}\ \bibnamefont
  {{McCarron}}}, \bibinfo {author} {\bibfnamefont {E.~B.}\ \bibnamefont
  {Norrgard}}, \bibinfo {author} {\bibfnamefont {M.~H.}\ \bibnamefont
  {Steinecker}},\ and\ \bibinfo {author} {\bibfnamefont {D.}~\bibnamefont
  {{DeMille}}},\ }\bibfield  {title} {\bibinfo {title} {Improved
  magneto-optical trapping of a diatomic molecule},\ }\href
  {https://doi.org/10.1088/1367-2630/17/3/035014} {\bibfield  {journal}
  {\bibinfo  {journal} {New J. Phys.}\ }\textbf {\bibinfo {volume} {17}},\
  \bibinfo {pages} {035014} (\bibinfo {year} {2015})}\BibitemShut {NoStop}%
\bibitem [{\citenamefont {Norrgard}\ \emph {et~al.}(2016)\citenamefont
  {Norrgard}, \citenamefont {McCarron}, \citenamefont {Steinecker},
  \citenamefont {Tarbutt},\ and\ \citenamefont {DeMille}}]{Norrgard2016}%
  \BibitemOpen
  \bibfield  {author} {\bibinfo {author} {\bibfnamefont {E.~B.}\ \bibnamefont
  {Norrgard}}, \bibinfo {author} {\bibfnamefont {D.~J.}\ \bibnamefont
  {McCarron}}, \bibinfo {author} {\bibfnamefont {M.~H.}\ \bibnamefont
  {Steinecker}}, \bibinfo {author} {\bibfnamefont {M.~R.}\ \bibnamefont
  {Tarbutt}},\ and\ \bibinfo {author} {\bibfnamefont {D.}~\bibnamefont
  {DeMille}},\ }\bibfield  {title} {\bibinfo {title} {{Submillikelvin Dipolar
  Molecules in a Radio-Frequency Magneto-Optical Trap}},\ }\href
  {https://doi.org/10.1103/PhysRevLett.116.063004} {\bibfield  {journal}
  {\bibinfo  {journal} {Phys. Rev. Lett.}\ }\textbf {\bibinfo {volume} {116}},\
  \bibinfo {pages} {063004} (\bibinfo {year} {2016})}\BibitemShut {NoStop}%
\bibitem [{\citenamefont {Hummon}\ \emph {et~al.}(2013)\citenamefont {Hummon},
  \citenamefont {Yeo}, \citenamefont {Stuhl}, \citenamefont {Collopy},
  \citenamefont {Xia},\ and\ \citenamefont {Ye}}]{Hummon2013}%
  \BibitemOpen
  \bibfield  {author} {\bibinfo {author} {\bibfnamefont {M.~T.}\ \bibnamefont
  {Hummon}}, \bibinfo {author} {\bibfnamefont {M.}~\bibnamefont {Yeo}},
  \bibinfo {author} {\bibfnamefont {B.~K.}\ \bibnamefont {Stuhl}}, \bibinfo
  {author} {\bibfnamefont {A.~L.}\ \bibnamefont {Collopy}}, \bibinfo {author}
  {\bibfnamefont {Y.}~\bibnamefont {Xia}},\ and\ \bibinfo {author}
  {\bibfnamefont {J.}~\bibnamefont {Ye}},\ }\bibfield  {title} {\bibinfo
  {title} {{2D magneto-optical trapping of diatomic molecules}},\ }\href
  {https://doi.org/10.1103/PhysRevLett.110.143001} {\bibfield  {journal}
  {\bibinfo  {journal} {Phys. Rev. Lett.}\ }\textbf {\bibinfo {volume} {110}},\
  \bibinfo {pages} {143001} (\bibinfo {year} {2013})}\BibitemShut {NoStop}%
\bibitem [{\citenamefont {Zhelyazkova}\ \emph {et~al.}(2014)\citenamefont
  {Zhelyazkova}, \citenamefont {Cournol}, \citenamefont {Wall}, \citenamefont
  {Matsushima}, \citenamefont {Hudson}, \citenamefont {Hinds}, \citenamefont
  {Tarbutt},\ and\ \citenamefont {Sauer}}]{Zhelyazkova2014}%
  \BibitemOpen
  \bibfield  {author} {\bibinfo {author} {\bibfnamefont {V.}~\bibnamefont
  {Zhelyazkova}}, \bibinfo {author} {\bibfnamefont {A.}~\bibnamefont
  {Cournol}}, \bibinfo {author} {\bibfnamefont {T.~E.}\ \bibnamefont {Wall}},
  \bibinfo {author} {\bibfnamefont {A.}~\bibnamefont {Matsushima}}, \bibinfo
  {author} {\bibfnamefont {J.~J.}\ \bibnamefont {Hudson}}, \bibinfo {author}
  {\bibfnamefont {E.~A.}\ \bibnamefont {Hinds}}, \bibinfo {author}
  {\bibfnamefont {M.~R.}\ \bibnamefont {Tarbutt}},\ and\ \bibinfo {author}
  {\bibfnamefont {B.~E.}\ \bibnamefont {Sauer}},\ }\bibfield  {title} {\bibinfo
  {title} {Laser cooling and slowing of {CaF} molecules},\ }\href
  {https://doi.org/10.1103/PhysRevA.89.053416} {\bibfield  {journal} {\bibinfo
  {journal} {Phys. Rev. A}\ }\textbf {\bibinfo {volume} {89}},\ \bibinfo
  {pages} {053416} (\bibinfo {year} {2014})}\BibitemShut {NoStop}%
\bibitem [{\citenamefont {Truppe}\ \emph {et~al.}(2017)\citenamefont {Truppe},
  \citenamefont {Williams}, \citenamefont {Hambach}, \citenamefont {Caldwell},
  \citenamefont {Fitch}, \citenamefont {Hinds}, \citenamefont {Sauer},\ and\
  \citenamefont {Tarbutt}}]{Truppe2017}%
  \BibitemOpen
  \bibfield  {author} {\bibinfo {author} {\bibfnamefont {S.}~\bibnamefont
  {Truppe}}, \bibinfo {author} {\bibfnamefont {H.~J.}\ \bibnamefont
  {Williams}}, \bibinfo {author} {\bibfnamefont {M.}~\bibnamefont {Hambach}},
  \bibinfo {author} {\bibfnamefont {L.}~\bibnamefont {Caldwell}}, \bibinfo
  {author} {\bibfnamefont {N.~J.}\ \bibnamefont {Fitch}}, \bibinfo {author}
  {\bibfnamefont {E.~A.}\ \bibnamefont {Hinds}}, \bibinfo {author}
  {\bibfnamefont {B.~E.}\ \bibnamefont {Sauer}},\ and\ \bibinfo {author}
  {\bibfnamefont {M.~R.}\ \bibnamefont {Tarbutt}},\ }\bibfield  {title}
  {\bibinfo {title} {{Molecules cooled below the Doppler limit}},\ }\href
  {https://doi.org/10.1038/nphys4241} {\bibfield  {journal} {\bibinfo
  {journal} {Nat. Phys.}\ }\textbf {\bibinfo {volume} {13}},\ \bibinfo {pages}
  {1173} (\bibinfo {year} {2017})}\BibitemShut {NoStop}%
\bibitem [{\citenamefont {Anderegg}\ \emph {et~al.}(2018)\citenamefont
  {Anderegg}, \citenamefont {Augenbraun}, \citenamefont {Bao}, \citenamefont
  {Burchesky}, \citenamefont {Cheuk}, \citenamefont {Ketterle},\ and\
  \citenamefont {Doyle}}]{Anderegg2018}%
  \BibitemOpen
  \bibfield  {author} {\bibinfo {author} {\bibfnamefont {L.}~\bibnamefont
  {Anderegg}}, \bibinfo {author} {\bibfnamefont {B.~L.}\ \bibnamefont
  {Augenbraun}}, \bibinfo {author} {\bibfnamefont {Y.}~\bibnamefont {Bao}},
  \bibinfo {author} {\bibfnamefont {S.}~\bibnamefont {Burchesky}}, \bibinfo
  {author} {\bibfnamefont {L.~W.}\ \bibnamefont {Cheuk}}, \bibinfo {author}
  {\bibfnamefont {W.}~\bibnamefont {Ketterle}},\ and\ \bibinfo {author}
  {\bibfnamefont {J.~M.}\ \bibnamefont {Doyle}},\ }\bibfield  {title} {\bibinfo
  {title} {Laser cooling of optically trapped molecules},\ }\href
  {https://doi.org/10.1038/s41567-018-0191-z} {\bibfield  {journal} {\bibinfo
  {journal} {Nat. Phys.}\ }\textbf {\bibinfo {volume} {14}},\ \bibinfo {pages}
  {890} (\bibinfo {year} {2018})}\BibitemShut {NoStop}%
\bibitem [{\citenamefont {Lim}\ \emph {et~al.}(2018)\citenamefont {Lim},
  \citenamefont {Almond}, \citenamefont {Trigatzis}, \citenamefont {Devlin},
  \citenamefont {Fitch}, \citenamefont {Sauer}, \citenamefont {Tarbutt},\ and\
  \citenamefont {Hinds}}]{Lim2018}%
  \BibitemOpen
  \bibfield  {author} {\bibinfo {author} {\bibfnamefont {J.}~\bibnamefont
  {Lim}}, \bibinfo {author} {\bibfnamefont {J.~R.}\ \bibnamefont {Almond}},
  \bibinfo {author} {\bibfnamefont {M.~A.}\ \bibnamefont {Trigatzis}}, \bibinfo
  {author} {\bibfnamefont {J.~A.}\ \bibnamefont {Devlin}}, \bibinfo {author}
  {\bibfnamefont {N.~J.}\ \bibnamefont {Fitch}}, \bibinfo {author}
  {\bibfnamefont {B.~E.}\ \bibnamefont {Sauer}}, \bibinfo {author}
  {\bibfnamefont {M.~R.}\ \bibnamefont {Tarbutt}},\ and\ \bibinfo {author}
  {\bibfnamefont {E.~A.}\ \bibnamefont {Hinds}},\ }\bibfield  {title} {\bibinfo
  {title} {{Laser Cooled YbF Molecules for Measuring the Electron's Electric
  Dipole Moment}},\ }\href {https://doi.org/10.1103/PhysRevLett.120.123201}
  {\bibfield  {journal} {\bibinfo  {journal} {Phys. Rev. Lett.}\ }\textbf
  {\bibinfo {volume} {120}},\ \bibinfo {pages} {123201} (\bibinfo {year}
  {2018})}\BibitemShut {NoStop}%
\bibitem [{\citenamefont {Kozyryev}\ \emph {et~al.}(2017)\citenamefont
  {Kozyryev}, \citenamefont {Baum}, \citenamefont {Matsuda}, \citenamefont
  {Augenbraun}, \citenamefont {Anderegg}, \citenamefont {Sedlack},\ and\
  \citenamefont {Doyle}}]{Kozyryev2017}%
  \BibitemOpen
  \bibfield  {author} {\bibinfo {author} {\bibfnamefont {I.}~\bibnamefont
  {Kozyryev}}, \bibinfo {author} {\bibfnamefont {L.}~\bibnamefont {Baum}},
  \bibinfo {author} {\bibfnamefont {K.}~\bibnamefont {Matsuda}}, \bibinfo
  {author} {\bibfnamefont {B.~L.}\ \bibnamefont {Augenbraun}}, \bibinfo
  {author} {\bibfnamefont {L.}~\bibnamefont {Anderegg}}, \bibinfo {author}
  {\bibfnamefont {A.~P.}\ \bibnamefont {Sedlack}},\ and\ \bibinfo {author}
  {\bibfnamefont {J.~M.}\ \bibnamefont {Doyle}},\ }\bibfield  {title} {\bibinfo
  {title} {{Sisyphus laser cooling of a polyatomic molecule}},\ }\href
  {https://doi.org/10.1103/physrevlett.118.173201} {\bibfield  {journal}
  {\bibinfo  {journal} {Phys. Rev. Lett.}\ }\textbf {\bibinfo {volume} {118}},\
  \bibinfo {pages} {173201} (\bibinfo {year} {2017})}\BibitemShut {NoStop}%
\bibitem [{\citenamefont {Hunter}\ \emph {et~al.}(2012)\citenamefont {Hunter},
  \citenamefont {Peck}, \citenamefont {Greenspon}, \citenamefont {Alam},\ and\
  \citenamefont {DeMille}}]{Hunter2012}%
  \BibitemOpen
  \bibfield  {author} {\bibinfo {author} {\bibfnamefont {L.~R.}\ \bibnamefont
  {Hunter}}, \bibinfo {author} {\bibfnamefont {S.~K.}\ \bibnamefont {Peck}},
  \bibinfo {author} {\bibfnamefont {A.~S.}\ \bibnamefont {Greenspon}}, \bibinfo
  {author} {\bibfnamefont {S.~S.}\ \bibnamefont {Alam}},\ and\ \bibinfo
  {author} {\bibfnamefont {D.}~\bibnamefont {DeMille}},\ }\bibfield  {title}
  {\bibinfo {title} {Prospects for laser cooling {TlF}},\ }\href
  {https://doi.org/10.1103/physreva.85.012511} {\bibfield  {journal} {\bibinfo
  {journal} {Phys. Rev. A}\ }\textbf {\bibinfo {volume} {85}},\ \bibinfo
  {pages} {012511} (\bibinfo {year} {2012})}\BibitemShut {NoStop}%
\bibitem [{\citenamefont {Chen}\ \emph {et~al.}(2017)\citenamefont {Chen},
  \citenamefont {Bu},\ and\ \citenamefont {Yan}}]{Chen2017}%
  \BibitemOpen
  \bibfield  {author} {\bibinfo {author} {\bibfnamefont {T.}~\bibnamefont
  {Chen}}, \bibinfo {author} {\bibfnamefont {W.}~\bibnamefont {Bu}},\ and\
  \bibinfo {author} {\bibfnamefont {B.}~\bibnamefont {Yan}},\ }\bibfield
  {title} {\bibinfo {title} {Radiative deflection of a {BaF} molecular beam via
  optical cycling},\ }\href {https://doi.org/10.1103/physreva.96.053401}
  {\bibfield  {journal} {\bibinfo  {journal} {Phys. Rev. A}\ }\textbf {\bibinfo
  {volume} {96}},\ \bibinfo {pages} {053401} (\bibinfo {year}
  {2017})}\BibitemShut {NoStop}%
\bibitem [{\citenamefont {Iwata}\ \emph {et~al.}(2017)\citenamefont {Iwata},
  \citenamefont {McNally},\ and\ \citenamefont {Zelevinsky}}]{Iwata2017}%
  \BibitemOpen
  \bibfield  {author} {\bibinfo {author} {\bibfnamefont {G.~Z.}\ \bibnamefont
  {Iwata}}, \bibinfo {author} {\bibfnamefont {R.~L.}\ \bibnamefont {McNally}},\
  and\ \bibinfo {author} {\bibfnamefont {T.}~\bibnamefont {Zelevinsky}},\
  }\bibfield  {title} {\bibinfo {title} {{High-resolution optical spectroscopy
  with a buffer-gas-cooled beam of BaH molecules}},\ }\href
  {https://doi.org/10.1103/physreva.96.022509} {\bibfield  {journal} {\bibinfo
  {journal} {Phys. Rev. A}\ }\textbf {\bibinfo {volume} {96}},\ \bibinfo
  {pages} {022509} (\bibinfo {year} {2017})}\BibitemShut {NoStop}%
\bibitem [{\citenamefont {Aggarwal}\ \emph {et~al.}(2019)\citenamefont
  {Aggarwal}, \citenamefont {Marshall}, \citenamefont {Bethlem}, \citenamefont
  {Boeschoten}, \citenamefont {Borschevsky}, \citenamefont {Denis},
  \citenamefont {Esajas}, \citenamefont {Hao}, \citenamefont {Hoekstra},
  \citenamefont {Jungmann}, \citenamefont {Meijknecht}, \citenamefont {Mooij},
  \citenamefont {Timmermans}, \citenamefont {Touwen}, \citenamefont {Ubachs},
  \citenamefont {Vermeulen}, \citenamefont {Willmann}, \citenamefont {Yin},\
  and\ \citenamefont {Zapara}}]{Aggarwal2019}%
  \BibitemOpen
  \bibfield  {author} {\bibinfo {author} {\bibfnamefont {P.}~\bibnamefont
  {Aggarwal}}, \bibinfo {author} {\bibfnamefont {V.~R.}\ \bibnamefont
  {Marshall}}, \bibinfo {author} {\bibfnamefont {H.~L.}\ \bibnamefont
  {Bethlem}}, \bibinfo {author} {\bibfnamefont {A.}~\bibnamefont {Boeschoten}},
  \bibinfo {author} {\bibfnamefont {A.}~\bibnamefont {Borschevsky}}, \bibinfo
  {author} {\bibfnamefont {M.}~\bibnamefont {Denis}}, \bibinfo {author}
  {\bibfnamefont {K.}~\bibnamefont {Esajas}}, \bibinfo {author} {\bibfnamefont
  {Y.}~\bibnamefont {Hao}}, \bibinfo {author} {\bibfnamefont {S.}~\bibnamefont
  {Hoekstra}}, \bibinfo {author} {\bibfnamefont {K.}~\bibnamefont {Jungmann}},
  \bibinfo {author} {\bibfnamefont {T.~B.}\ \bibnamefont {Meijknecht}},
  \bibinfo {author} {\bibfnamefont {M.~C.}\ \bibnamefont {Mooij}}, \bibinfo
  {author} {\bibfnamefont {R.~G.~E.}\ \bibnamefont {Timmermans}}, \bibinfo
  {author} {\bibfnamefont {A.}~\bibnamefont {Touwen}}, \bibinfo {author}
  {\bibfnamefont {W.}~\bibnamefont {Ubachs}}, \bibinfo {author} {\bibfnamefont
  {S.~M.}\ \bibnamefont {Vermeulen}}, \bibinfo {author} {\bibfnamefont
  {L.}~\bibnamefont {Willmann}}, \bibinfo {author} {\bibfnamefont
  {Y.}~\bibnamefont {Yin}},\ and\ \bibinfo {author} {\bibfnamefont
  {A.}~\bibnamefont {Zapara}} (\bibinfo {collaboration} {NL-eEDM
  Collaboration}),\ }\bibfield  {title} {\bibinfo {title} {Lifetime
  measurements of the {${A}^{2}\Pi_{1/2}$ and ${A}^{2}\Pi_{3/2}$} states in
  {BaF}},\ }\href {https://doi.org/10.1103/PhysRevA.100.052503} {\bibfield
  {journal} {\bibinfo  {journal} {Phys. Rev. A}\ }\textbf {\bibinfo {volume}
  {100}},\ \bibinfo {pages} {052503} (\bibinfo {year} {2019})}\BibitemShut
  {NoStop}%
\bibitem [{\citenamefont {Micheli}\ \emph {et~al.}(2007)\citenamefont
  {Micheli}, \citenamefont {Pupillo}, \citenamefont {B\"{u}chler},\ and\
  \citenamefont {Zoller}}]{Micheli2007}%
  \BibitemOpen
  \bibfield  {author} {\bibinfo {author} {\bibfnamefont {A.}~\bibnamefont
  {Micheli}}, \bibinfo {author} {\bibfnamefont {G.}~\bibnamefont {Pupillo}},
  \bibinfo {author} {\bibfnamefont {H.~P.}\ \bibnamefont {B\"{u}chler}},\ and\
  \bibinfo {author} {\bibfnamefont {P.}~\bibnamefont {Zoller}},\ }\bibfield
  {title} {\bibinfo {title} {Cold polar molecules in two-dimensional traps:
  Tailoring interactions with external fields for novel quantum phases},\
  }\href {https://doi.org/10.1103/physreva.76.043604} {\bibfield  {journal}
  {\bibinfo  {journal} {Phys. Rev. A}\ }\textbf {\bibinfo {volume} {76}},\
  \bibinfo {pages} {043604} (\bibinfo {year} {2007})}\BibitemShut {NoStop}%
\bibitem [{\citenamefont {Pollet}\ \emph {et~al.}(2010)\citenamefont {Pollet},
  \citenamefont {Picon}, \citenamefont {B\"{u}chler},\ and\ \citenamefont
  {Troyer}}]{Pollet2010}%
  \BibitemOpen
  \bibfield  {author} {\bibinfo {author} {\bibfnamefont {L.}~\bibnamefont
  {Pollet}}, \bibinfo {author} {\bibfnamefont {J.~D.}\ \bibnamefont {Picon}},
  \bibinfo {author} {\bibfnamefont {H.~P.}\ \bibnamefont {B\"{u}chler}},\ and\
  \bibinfo {author} {\bibfnamefont {M.}~\bibnamefont {Troyer}},\ }\bibfield
  {title} {\bibinfo {title} {Supersolid phase with cold polar molecules on a
  triangular lattice},\ }\href {https://doi.org/10.1103/physrevlett.104.125302}
  {\bibfield  {journal} {\bibinfo  {journal} {Phys. Rev. Lett.}\ }\textbf
  {\bibinfo {volume} {104}},\ \bibinfo {pages} {125302} (\bibinfo {year}
  {2010})}\BibitemShut {NoStop}%
\bibitem [{\citenamefont {Capogrosso-Sansone}\ \emph
  {et~al.}(2010)\citenamefont {Capogrosso-Sansone}, \citenamefont {Trefzger},
  \citenamefont {Lewenstein}, \citenamefont {Zoller},\ and\ \citenamefont
  {Pupillo}}]{Capogrosso_Sansone2010}%
  \BibitemOpen
  \bibfield  {author} {\bibinfo {author} {\bibfnamefont {B.}~\bibnamefont
  {Capogrosso-Sansone}}, \bibinfo {author} {\bibfnamefont {C.}~\bibnamefont
  {Trefzger}}, \bibinfo {author} {\bibfnamefont {M.}~\bibnamefont
  {Lewenstein}}, \bibinfo {author} {\bibfnamefont {P.}~\bibnamefont {Zoller}},\
  and\ \bibinfo {author} {\bibfnamefont {G.}~\bibnamefont {Pupillo}},\
  }\bibfield  {title} {\bibinfo {title} {Quantum phases of cold polar molecules
  in {2D} optical lattices},\ }\href
  {https://doi.org/10.1103/physrevlett.104.125301} {\bibfield  {journal}
  {\bibinfo  {journal} {Phys. Rev. Lett.}\ }\textbf {\bibinfo {volume} {104}},\
  \bibinfo {pages} {125301} (\bibinfo {year} {2010})}\BibitemShut {NoStop}%
\bibitem [{\citenamefont {Lechner}\ and\ \citenamefont
  {Zoller}(2013)}]{Lechner2013}%
  \BibitemOpen
  \bibfield  {author} {\bibinfo {author} {\bibfnamefont {W.}~\bibnamefont
  {Lechner}}\ and\ \bibinfo {author} {\bibfnamefont {P.}~\bibnamefont
  {Zoller}},\ }\bibfield  {title} {\bibinfo {title} {From classical to quantum
  glasses with ultracold polar molecules},\ }\href
  {https://doi.org/10.1103/physrevlett.111.185306} {\bibfield  {journal}
  {\bibinfo  {journal} {Phys. Rev. Lett.}\ }\textbf {\bibinfo {volume} {111}},\
  \bibinfo {pages} {185306} (\bibinfo {year} {2013})}\BibitemShut {NoStop}%
\bibitem [{\citenamefont {Gorshkov}\ \emph
  {et~al.}(2011{\natexlab{a}})\citenamefont {Gorshkov}, \citenamefont
  {Manmana}, \citenamefont {Chen}, \citenamefont {Ye}, \citenamefont {Demler},
  \citenamefont {Lukin},\ and\ \citenamefont {Rey}}]{Gorshkov2011}%
  \BibitemOpen
  \bibfield  {author} {\bibinfo {author} {\bibfnamefont {A.~V.}\ \bibnamefont
  {Gorshkov}}, \bibinfo {author} {\bibfnamefont {S.~R.}\ \bibnamefont
  {Manmana}}, \bibinfo {author} {\bibfnamefont {G.}~\bibnamefont {Chen}},
  \bibinfo {author} {\bibfnamefont {J.}~\bibnamefont {Ye}}, \bibinfo {author}
  {\bibfnamefont {E.}~\bibnamefont {Demler}}, \bibinfo {author} {\bibfnamefont
  {M.~D.}\ \bibnamefont {Lukin}},\ and\ \bibinfo {author} {\bibfnamefont
  {A.~M.}\ \bibnamefont {Rey}},\ }\bibfield  {title} {\bibinfo {title} {Tunable
  superfluidity and quantum magnetism with ultracold polar molecules},\ }\href
  {https://doi.org/10.1103/physrevlett.107.115301} {\bibfield  {journal}
  {\bibinfo  {journal} {Phys. Rev. Lett.}\ }\textbf {\bibinfo {volume} {107}},\
  \bibinfo {pages} {115301} (\bibinfo {year} {2011}{\natexlab{a}})}\BibitemShut
  {NoStop}%
\bibitem [{\citenamefont {Gorshkov}\ \emph
  {et~al.}(2011{\natexlab{b}})\citenamefont {Gorshkov}, \citenamefont
  {Manmana}, \citenamefont {Chen}, \citenamefont {Demler}, \citenamefont
  {Lukin},\ and\ \citenamefont {Rey}}]{Gorshkov2011b}%
  \BibitemOpen
  \bibfield  {author} {\bibinfo {author} {\bibfnamefont {A.~V.}\ \bibnamefont
  {Gorshkov}}, \bibinfo {author} {\bibfnamefont {S.~R.}\ \bibnamefont
  {Manmana}}, \bibinfo {author} {\bibfnamefont {G.}~\bibnamefont {Chen}},
  \bibinfo {author} {\bibfnamefont {E.}~\bibnamefont {Demler}}, \bibinfo
  {author} {\bibfnamefont {M.~D.}\ \bibnamefont {Lukin}},\ and\ \bibinfo
  {author} {\bibfnamefont {A.~M.}\ \bibnamefont {Rey}},\ }\bibfield  {title}
  {\bibinfo {title} {Quantum magnetism with polar alkali-metal dimers},\ }\href
  {https://doi.org/10.1103/physreva.84.033619} {\bibfield  {journal} {\bibinfo
  {journal} {Phys. Rev. A}\ }\textbf {\bibinfo {volume} {84}},\ \bibinfo
  {pages} {033619} (\bibinfo {year} {2011}{\natexlab{b}})}\BibitemShut
  {NoStop}%
\bibitem [{\citenamefont {Zhou}\ \emph {et~al.}(2011)\citenamefont {Zhou},
  \citenamefont {Ortner},\ and\ \citenamefont {Rabl}}]{Zhou2011}%
  \BibitemOpen
  \bibfield  {author} {\bibinfo {author} {\bibfnamefont {Y.~L.}\ \bibnamefont
  {Zhou}}, \bibinfo {author} {\bibfnamefont {M.}~\bibnamefont {Ortner}},\ and\
  \bibinfo {author} {\bibfnamefont {P.}~\bibnamefont {Rabl}},\ }\bibfield
  {title} {\bibinfo {title} {Long-range and frustrated spin-spin interactions
  in crystals of cold polar molecules},\ }\href
  {https://doi.org/10.1103/physreva.84.052332} {\bibfield  {journal} {\bibinfo
  {journal} {Phys. Rev. A}\ }\textbf {\bibinfo {volume} {84}},\ \bibinfo
  {pages} {052332} (\bibinfo {year} {2011})}\BibitemShut {NoStop}%
\bibitem [{\citenamefont {Hazzard}\ \emph {et~al.}(2013)\citenamefont
  {Hazzard}, \citenamefont {Manmana}, \citenamefont {Foss-Feig},\ and\
  \citenamefont {Rey}}]{Hazzard2013}%
  \BibitemOpen
  \bibfield  {author} {\bibinfo {author} {\bibfnamefont {K.~R.~A.}\
  \bibnamefont {Hazzard}}, \bibinfo {author} {\bibfnamefont {S.~R.}\
  \bibnamefont {Manmana}}, \bibinfo {author} {\bibfnamefont {M.}~\bibnamefont
  {Foss-Feig}},\ and\ \bibinfo {author} {\bibfnamefont {A.~M.}\ \bibnamefont
  {Rey}},\ }\bibfield  {title} {\bibinfo {title} {Far-from-equilibrium quantum
  magnetism with ultracold polar molecules},\ }\href
  {https://doi.org/10.1103/physrevlett.110.075301} {\bibfield  {journal}
  {\bibinfo  {journal} {Phys. Rev. Lett.}\ }\textbf {\bibinfo {volume} {110}},\
  \bibinfo {pages} {075301} (\bibinfo {year} {2013})}\BibitemShut {NoStop}%
\bibitem [{\citenamefont {Greiner}\ \emph {et~al.}(2002)\citenamefont
  {Greiner}, \citenamefont {Mandel}, \citenamefont {Esslinger}, \citenamefont
  {H\"{a}nsch},\ and\ \citenamefont {Bloch}}]{Greiner2002}%
  \BibitemOpen
  \bibfield  {author} {\bibinfo {author} {\bibfnamefont {M.}~\bibnamefont
  {Greiner}}, \bibinfo {author} {\bibfnamefont {O.}~\bibnamefont {Mandel}},
  \bibinfo {author} {\bibfnamefont {T.}~\bibnamefont {Esslinger}}, \bibinfo
  {author} {\bibfnamefont {T.~W.}\ \bibnamefont {H\"{a}nsch}},\ and\ \bibinfo
  {author} {\bibfnamefont {I.}~\bibnamefont {Bloch}},\ }\bibfield  {title}
  {\bibinfo {title} {Quantum phase transition from a superfluid to a mott
  insulator in a gas of ultracold atoms},\ }\href
  {https://doi.org/10.1038/415039a} {\bibfield  {journal} {\bibinfo  {journal}
  {Nature}\ }\textbf {\bibinfo {volume} {415}},\ \bibinfo {pages} {39}
  (\bibinfo {year} {2002})}\BibitemShut {NoStop}%
\bibitem [{\citenamefont {Moses}\ \emph {et~al.}(2015)\citenamefont {Moses},
  \citenamefont {Covey}, \citenamefont {Miecnikowski}, \citenamefont {Yan},
  \citenamefont {Gadway}, \citenamefont {Ye},\ and\ \citenamefont
  {Jin}}]{Moses2015}%
  \BibitemOpen
  \bibfield  {author} {\bibinfo {author} {\bibfnamefont {S.~A.}\ \bibnamefont
  {Moses}}, \bibinfo {author} {\bibfnamefont {J.~P.}\ \bibnamefont {Covey}},
  \bibinfo {author} {\bibfnamefont {M.~T.}\ \bibnamefont {Miecnikowski}},
  \bibinfo {author} {\bibfnamefont {B.}~\bibnamefont {Yan}}, \bibinfo {author}
  {\bibfnamefont {B.}~\bibnamefont {Gadway}}, \bibinfo {author} {\bibfnamefont
  {J.}~\bibnamefont {Ye}},\ and\ \bibinfo {author} {\bibfnamefont {D.~S.}\
  \bibnamefont {Jin}},\ }\bibfield  {title} {\bibinfo {title} {Creation of a
  low-entropy quantum gas of polar molecules in an optical lattice},\ }\href
  {https://doi.org/10.1126/science.aac6400} {\bibfield  {journal} {\bibinfo
  {journal} {Science}\ }\textbf {\bibinfo {volume} {350}},\ \bibinfo {pages}
  {659} (\bibinfo {year} {2015})}\BibitemShut {NoStop}%
\bibitem [{\citenamefont {Reichs\"ollner}\ \emph {et~al.}(2017)\citenamefont
  {Reichs\"ollner}, \citenamefont {Schindewolf}, \citenamefont {Takekoshi},
  \citenamefont {Grimm},\ and\ \citenamefont {N\"agerl}}]{Reichsoellner2017}%
  \BibitemOpen
  \bibfield  {author} {\bibinfo {author} {\bibfnamefont {L.}~\bibnamefont
  {Reichs\"ollner}}, \bibinfo {author} {\bibfnamefont {A.}~\bibnamefont
  {Schindewolf}}, \bibinfo {author} {\bibfnamefont {T.}~\bibnamefont
  {Takekoshi}}, \bibinfo {author} {\bibfnamefont {R.}~\bibnamefont {Grimm}},\
  and\ \bibinfo {author} {\bibfnamefont {H.-C.}\ \bibnamefont {N\"agerl}},\
  }\bibfield  {title} {\bibinfo {title} {Quantum engineering of a low-entropy
  gas of heteronuclear bosonic molecules in an optical lattice},\ }\href
  {https://doi.org/10.1103/PhysRevLett.118.073201} {\bibfield  {journal}
  {\bibinfo  {journal} {Phys. Rev. Lett.}\ }\textbf {\bibinfo {volume} {118}},\
  \bibinfo {pages} {073201} (\bibinfo {year} {2017})}\BibitemShut {NoStop}%
\bibitem [{\citenamefont {Danzl}\ \emph {et~al.}(2010)\citenamefont {Danzl},
  \citenamefont {Mark}, \citenamefont {Haller}, \citenamefont {Gustavsson},
  \citenamefont {Hart}, \citenamefont {Aldegunde}, \citenamefont {Hutson},\
  and\ \citenamefont {N\"{a}gerl}}]{Danzl2010}%
  \BibitemOpen
  \bibfield  {author} {\bibinfo {author} {\bibfnamefont {J.~G.}\ \bibnamefont
  {Danzl}}, \bibinfo {author} {\bibfnamefont {M.~J.}\ \bibnamefont {Mark}},
  \bibinfo {author} {\bibfnamefont {E.}~\bibnamefont {Haller}}, \bibinfo
  {author} {\bibfnamefont {M.}~\bibnamefont {Gustavsson}}, \bibinfo {author}
  {\bibfnamefont {R.}~\bibnamefont {Hart}}, \bibinfo {author} {\bibfnamefont
  {J.}~\bibnamefont {Aldegunde}}, \bibinfo {author} {\bibfnamefont {J.~M.}\
  \bibnamefont {Hutson}},\ and\ \bibinfo {author} {\bibfnamefont {H.-C.}\
  \bibnamefont {N\"{a}gerl}},\ }\bibfield  {title} {\bibinfo {title} {An
  ultracold high-density sample of rovibronic ground-state molecules in an
  optical lattice},\ }\href {https://doi.org/10.1038/nphys1533} {\bibfield
  {journal} {\bibinfo  {journal} {Nat. Phys.}\ }\textbf {\bibinfo {volume}
  {6}},\ \bibinfo {pages} {265} (\bibinfo {year} {2010})}\BibitemShut {NoStop}%
\bibitem [{\citenamefont {Vexiau}\ \emph {et~al.}(2017)\citenamefont {Vexiau},
  \citenamefont {Borsalino}, \citenamefont {Lepers}, \citenamefont {Orbán},
  \citenamefont {Aymar}, \citenamefont {Dulieu},\ and\ \citenamefont
  {Bouloufa-Maafa}}]{Vexiau2017}%
  \BibitemOpen
  \bibfield  {author} {\bibinfo {author} {\bibfnamefont {R.}~\bibnamefont
  {Vexiau}}, \bibinfo {author} {\bibfnamefont {D.}~\bibnamefont {Borsalino}},
  \bibinfo {author} {\bibfnamefont {M.}~\bibnamefont {Lepers}}, \bibinfo
  {author} {\bibfnamefont {A.}~\bibnamefont {Orbán}}, \bibinfo {author}
  {\bibfnamefont {M.}~\bibnamefont {Aymar}}, \bibinfo {author} {\bibfnamefont
  {O.}~\bibnamefont {Dulieu}},\ and\ \bibinfo {author} {\bibfnamefont
  {N.}~\bibnamefont {Bouloufa-Maafa}},\ }\bibfield  {title} {\bibinfo {title}
  {Dynamic dipole polarizabilities of heteronuclear alkali dimers: optical
  response, trapping and control of ultracold molecules},\ }\href
  {https://doi.org/10.1080/0144235X.2017.1351821} {\bibfield  {journal}
  {\bibinfo  {journal} {Int. Rev. Phys. Chem.}\ }\textbf {\bibinfo {volume}
  {36}},\ \bibinfo {pages} {709} (\bibinfo {year} {2017})}\BibitemShut
  {NoStop}%
\bibitem [{\citenamefont {Aldegunde}\ \emph {et~al.}(2008)\citenamefont
  {Aldegunde}, \citenamefont {Rivington}, \citenamefont {\.{Z}uchowski},\ and\
  \citenamefont {Hutson}}]{Aldegunde2008}%
  \BibitemOpen
  \bibfield  {author} {\bibinfo {author} {\bibfnamefont {J.}~\bibnamefont
  {Aldegunde}}, \bibinfo {author} {\bibfnamefont {B.~A.}\ \bibnamefont
  {Rivington}}, \bibinfo {author} {\bibfnamefont {P.~S.}\ \bibnamefont
  {\.{Z}uchowski}},\ and\ \bibinfo {author} {\bibfnamefont {J.~M.}\
  \bibnamefont {Hutson}},\ }\bibfield  {title} {\bibinfo {title} {Hyperfine
  energy levels of alkali-metal dimers: Ground-state polar molecules in
  electric and magnetic fields},\ }\href
  {https://doi.org/10.1103/physreva.78.033434} {\bibfield  {journal} {\bibinfo
  {journal} {Phys. Rev. A}\ }\textbf {\bibinfo {volume} {78}},\ \bibinfo
  {pages} {033434} (\bibinfo {year} {2008})}\BibitemShut {NoStop}%
\bibitem [{\citenamefont {Aldegunde}\ \emph {et~al.}(2009)\citenamefont
  {Aldegunde}, \citenamefont {Ran},\ and\ \citenamefont
  {Hutson}}]{Aldegunde:spectra:2009}%
  \BibitemOpen
  \bibfield  {author} {\bibinfo {author} {\bibfnamefont {J.}~\bibnamefont
  {Aldegunde}}, \bibinfo {author} {\bibfnamefont {H.}~\bibnamefont {Ran}},\
  and\ \bibinfo {author} {\bibfnamefont {J.~M.}\ \bibnamefont {Hutson}},\
  }\bibfield  {title} {\bibinfo {title} {Manipulating ultracold polar molecules
  with microwave radiation: the influence of hyperfine structure},\ }\href@noop
  {} {\bibfield  {journal} {\bibinfo  {journal} {Phys. Rev. A}\ }\textbf
  {\bibinfo {volume} {80}},\ \bibinfo {pages} {043410} (\bibinfo {year}
  {2009})}\BibitemShut {NoStop}%
\bibitem [{\citenamefont {Ran}\ \emph {et~al.}(2010)\citenamefont {Ran},
  \citenamefont {Aldegunde},\ and\ \citenamefont {Hutson}}]{Ran:2010}%
  \BibitemOpen
  \bibfield  {author} {\bibinfo {author} {\bibfnamefont {H.}~\bibnamefont
  {Ran}}, \bibinfo {author} {\bibfnamefont {J.}~\bibnamefont {Aldegunde}},\
  and\ \bibinfo {author} {\bibfnamefont {J.~M.}\ \bibnamefont {Hutson}},\
  }\bibfield  {title} {\bibinfo {title} {Hyperfine structure in the microwave
  spectra of ultracold polar molecules},\ }\href@noop {} {\bibfield  {journal}
  {\bibinfo  {journal} {New J. Phys.}\ }\textbf {\bibinfo {volume} {12}},\
  \bibinfo {pages} {043015} (\bibinfo {year} {2010})}\BibitemShut {NoStop}%
\bibitem [{\citenamefont {Brown}\ and\ \citenamefont
  {Carrington}(2003)}]{BrownandCarrington}%
  \BibitemOpen
  \bibfield  {author} {\bibinfo {author} {\bibfnamefont {J.~M.}\ \bibnamefont
  {Brown}}\ and\ \bibinfo {author} {\bibfnamefont {A.}~\bibnamefont
  {Carrington}},\ }\href
  {https://www.ebook.de/de/product/3510426/john_m_brown_alan_carrington_rotational_spectroscopy_of_diatomic_molecules.html}
  {\emph {\bibinfo {title} {Rotational Spectroscopy of Diatomic Molecules}}}\
  (\bibinfo  {publisher} {Cambridge University Press},\ \bibinfo {year}
  {2003})\BibitemShut {NoStop}%
\bibitem [{\citenamefont {Herzberg}(1950)}]{Herzberg}%
  \BibitemOpen
  \bibfield  {author} {\bibinfo {author} {\bibfnamefont {G.}~\bibnamefont
  {Herzberg}},\ }\href@noop {} {\emph {\bibinfo {title} {Spectra of Diatomic
  Molecules}}},\ \bibinfo {series} {Molecular Spectra and Molecular
  Structure:}, Vol.~\bibinfo {volume} {1}\ (\bibinfo  {publisher} {D. van
  Nostrand Company},\ \bibinfo {address} {Princeton, New Jersey, USA},\
  \bibinfo {year} {1950})\BibitemShut {NoStop}%
\bibitem [{\citenamefont {Aldegunde}\ and\ \citenamefont
  {Hutson}(2017)}]{Aldegunde2017}%
  \BibitemOpen
  \bibfield  {author} {\bibinfo {author} {\bibfnamefont {J.}~\bibnamefont
  {Aldegunde}}\ and\ \bibinfo {author} {\bibfnamefont {J.~M.}\ \bibnamefont
  {Hutson}},\ }\bibfield  {title} {\bibinfo {title} {Hyperfine structure of
  alkali-metal diatomic molecules},\ }\href
  {https://doi.org/10.1103/physreva.96.042506} {\bibfield  {journal} {\bibinfo
  {journal} {Phys. Rev. A}\ }\textbf {\bibinfo {volume} {96}},\ \bibinfo
  {pages} {042506} (\bibinfo {year} {2017})}\BibitemShut {NoStop}%
\bibitem [{\citenamefont {Gregory}\ \emph {et~al.}(2017)\citenamefont
  {Gregory}, \citenamefont {Blackmore}, \citenamefont {Aldegunde},
  \citenamefont {Hutson},\ and\ \citenamefont {Cornish}}]{Gregory2017}%
  \BibitemOpen
  \bibfield  {author} {\bibinfo {author} {\bibfnamefont {P.~D.}\ \bibnamefont
  {Gregory}}, \bibinfo {author} {\bibfnamefont {J.~A.}\ \bibnamefont
  {Blackmore}}, \bibinfo {author} {\bibfnamefont {J.}~\bibnamefont
  {Aldegunde}}, \bibinfo {author} {\bibfnamefont {J.~M.}\ \bibnamefont
  {Hutson}},\ and\ \bibinfo {author} {\bibfnamefont {S.~L.}\ \bibnamefont
  {Cornish}},\ }\bibfield  {title} {\bibinfo {title} {ac stark effect in
  ultracold polar $^{87}${Rb}$^{133}${Cs} molecules},\ }\href
  {https://doi.org/10.1103/PhysRevA.96.021402} {\bibfield  {journal} {\bibinfo
  {journal} {Phys. Rev. A}\ }\textbf {\bibinfo {volume} {96}},\ \bibinfo
  {pages} {021402(R)} (\bibinfo {year} {2017})}\BibitemShut {NoStop}%
\bibitem [{\citenamefont {Gregory}\ \emph {et~al.}(2016)\citenamefont
  {Gregory}, \citenamefont {Aldegunde}, \citenamefont {Hutson},\ and\
  \citenamefont {Cornish}}]{Gregory2016}%
  \BibitemOpen
  \bibfield  {author} {\bibinfo {author} {\bibfnamefont {P.~D.}\ \bibnamefont
  {Gregory}}, \bibinfo {author} {\bibfnamefont {J.}~\bibnamefont {Aldegunde}},
  \bibinfo {author} {\bibfnamefont {J.~M.}\ \bibnamefont {Hutson}},\ and\
  \bibinfo {author} {\bibfnamefont {S.~L.}\ \bibnamefont {Cornish}},\
  }\bibfield  {title} {\bibinfo {title} {Controlling the rotational and
  hyperfine state of ultracold {$^{87}$Rb$^{133}$Cs} molecules},\ }\href
  {https://doi.org/10.1103/physreva.94.041403} {\bibfield  {journal} {\bibinfo
  {journal} {Phys. Rev. A}\ }\textbf {\bibinfo {volume} {94}},\ \bibinfo
  {pages} {041403} (\bibinfo {year} {2016})}\BibitemShut {NoStop}%
\bibitem [{\citenamefont {Will}\ \emph {et~al.}(2016)\citenamefont {Will},
  \citenamefont {Park}, \citenamefont {Yan}, \citenamefont {Loh},\ and\
  \citenamefont {Zwierlein}}]{Will2016}%
  \BibitemOpen
  \bibfield  {author} {\bibinfo {author} {\bibfnamefont {S.~A.}\ \bibnamefont
  {Will}}, \bibinfo {author} {\bibfnamefont {J.~W.}\ \bibnamefont {Park}},
  \bibinfo {author} {\bibfnamefont {Z.~Z.}\ \bibnamefont {Yan}}, \bibinfo
  {author} {\bibfnamefont {H.}~\bibnamefont {Loh}},\ and\ \bibinfo {author}
  {\bibfnamefont {M.~W.}\ \bibnamefont {Zwierlein}},\ }\bibfield  {title}
  {\bibinfo {title} {Coherent microwave control of ultracold
  $^{23}${Na}$^{40}${K} molecules},\ }\href
  {https://doi.org/10.1103/PhysRevLett.116.225306} {\bibfield  {journal}
  {\bibinfo  {journal} {Phys. Rev. Lett.}\ }\textbf {\bibinfo {volume} {116}},\
  \bibinfo {pages} {225306} (\bibinfo {year} {2016})}\BibitemShut {NoStop}%
\bibitem [{\citenamefont {Guo}\ \emph {et~al.}(2018)\citenamefont {Guo},
  \citenamefont {Ye}, \citenamefont {He}, \citenamefont {Qu\'em\'ener},\ and\
  \citenamefont {Wang}}]{Guo2018}%
  \BibitemOpen
  \bibfield  {author} {\bibinfo {author} {\bibfnamefont {M.}~\bibnamefont
  {Guo}}, \bibinfo {author} {\bibfnamefont {X.}~\bibnamefont {Ye}}, \bibinfo
  {author} {\bibfnamefont {J.}~\bibnamefont {He}}, \bibinfo {author}
  {\bibfnamefont {G.}~\bibnamefont {Qu\'em\'ener}},\ and\ \bibinfo {author}
  {\bibfnamefont {D.}~\bibnamefont {Wang}},\ }\bibfield  {title} {\bibinfo
  {title} {High-resolution internal state control of ultracold
  $^{23}${Na}$^{87}${Rb} molecules},\ }\href
  {https://doi.org/10.1103/PhysRevA.97.020501} {\bibfield  {journal} {\bibinfo
  {journal} {Phys. Rev. A}\ }\textbf {\bibinfo {volume} {97}},\ \bibinfo
  {pages} {020501} (\bibinfo {year} {2018})}\BibitemShut {NoStop}%
\bibitem [{\citenamefont {Wei}\ \emph {et~al.}(2011)\citenamefont {Wei},
  \citenamefont {Kais}, \citenamefont {Friedrich},\ and\ \citenamefont
  {Herschbach}}]{Wei2011}%
  \BibitemOpen
  \bibfield  {author} {\bibinfo {author} {\bibfnamefont {Q.}~\bibnamefont
  {Wei}}, \bibinfo {author} {\bibfnamefont {S.}~\bibnamefont {Kais}}, \bibinfo
  {author} {\bibfnamefont {B.}~\bibnamefont {Friedrich}},\ and\ \bibinfo
  {author} {\bibfnamefont {D.}~\bibnamefont {Herschbach}},\ }\bibfield  {title}
  {\bibinfo {title} {Entanglement of polar molecules in pendular states},\
  }\href {https://doi.org/10.1063/1.3567486} {\bibfield  {journal} {\bibinfo
  {journal} {J. Chem. Phys.}\ }\textbf {\bibinfo {volume} {134}},\ \bibinfo
  {pages} {124107} (\bibinfo {year} {2011})}\BibitemShut {NoStop}%
\bibitem [{\citenamefont {Li}\ \emph {et~al.}(2017)\citenamefont {Li},
  \citenamefont {Petrov}, \citenamefont {Makrides}, \citenamefont {Tiesinga},\
  and\ \citenamefont {Kotochigova}}]{Li2017}%
  \BibitemOpen
  \bibfield  {author} {\bibinfo {author} {\bibfnamefont {M.}~\bibnamefont
  {Li}}, \bibinfo {author} {\bibfnamefont {A.}~\bibnamefont {Petrov}}, \bibinfo
  {author} {\bibfnamefont {C.}~\bibnamefont {Makrides}}, \bibinfo {author}
  {\bibfnamefont {E.}~\bibnamefont {Tiesinga}},\ and\ \bibinfo {author}
  {\bibfnamefont {S.}~\bibnamefont {Kotochigova}},\ }\bibfield  {title}
  {\bibinfo {title} {Pendular trapping conditions for ultracold polar molecules
  enforced by external electric fields},\ }\href
  {https://doi.org/10.1103/PhysRevA.95.063422} {\bibfield  {journal} {\bibinfo
  {journal} {Phys. Rev. A}\ }\textbf {\bibinfo {volume} {95}},\ \bibinfo
  {pages} {063422} (\bibinfo {year} {2017})}\BibitemShut {NoStop}%
\bibitem [{\citenamefont {Kotochigova}\ and\ \citenamefont
  {DeMille}(2010)}]{Kotochigova2010}%
  \BibitemOpen
  \bibfield  {author} {\bibinfo {author} {\bibfnamefont {S.}~\bibnamefont
  {Kotochigova}}\ and\ \bibinfo {author} {\bibfnamefont {D.}~\bibnamefont
  {DeMille}},\ }\bibfield  {title} {\bibinfo {title} {Electric-field-dependent
  dynamic polarizability and state-insensitive conditions for optical trapping
  of diatomic polar molecules},\ }\href
  {https://doi.org/10.1103/PhysRevA.82.063421} {\bibfield  {journal} {\bibinfo
  {journal} {Phys. Rev. A}\ }\textbf {\bibinfo {volume} {82}},\ \bibinfo
  {pages} {063421} (\bibinfo {year} {2010})}\BibitemShut {NoStop}%
\bibitem [{\citenamefont {Neyenhuis}\ \emph {et~al.}(2012)\citenamefont
  {Neyenhuis}, \citenamefont {Yan}, \citenamefont {Moses}, \citenamefont
  {Covey}, \citenamefont {Chotia}, \citenamefont {Petrov}, \citenamefont
  {Kotochigova}, \citenamefont {Ye},\ and\ \citenamefont
  {Jin}}]{Neyenhuis2012}%
  \BibitemOpen
  \bibfield  {author} {\bibinfo {author} {\bibfnamefont {B.}~\bibnamefont
  {Neyenhuis}}, \bibinfo {author} {\bibfnamefont {B.}~\bibnamefont {Yan}},
  \bibinfo {author} {\bibfnamefont {S.~A.}\ \bibnamefont {Moses}}, \bibinfo
  {author} {\bibfnamefont {J.~P.}\ \bibnamefont {Covey}}, \bibinfo {author}
  {\bibfnamefont {A.}~\bibnamefont {Chotia}}, \bibinfo {author} {\bibfnamefont
  {A.}~\bibnamefont {Petrov}}, \bibinfo {author} {\bibfnamefont
  {S.}~\bibnamefont {Kotochigova}}, \bibinfo {author} {\bibfnamefont
  {J.}~\bibnamefont {Ye}},\ and\ \bibinfo {author} {\bibfnamefont {D.~S.}\
  \bibnamefont {Jin}},\ }\bibfield  {title} {\bibinfo {title} {Anisotropic
  polarizability of ultracold polar $^{40}${K}$^{87}${Rb} molecules},\ }\href
  {https://doi.org/10.1103/PhysRevLett.109.230403} {\bibfield  {journal}
  {\bibinfo  {journal} {Phys. Rev. Lett.}\ }\textbf {\bibinfo {volume} {109}},\
  \bibinfo {pages} {230403} (\bibinfo {year} {2012})}\BibitemShut {NoStop}%
\bibitem [{\citenamefont {See\ss{}elberg}\ \emph {et~al.}(2018)\citenamefont
  {See\ss{}elberg}, \citenamefont {Luo}, \citenamefont {Li}, \citenamefont
  {Bause}, \citenamefont {Kotochigova}, \citenamefont {Bloch},\ and\
  \citenamefont {Gohle}}]{Seesselberg2018}%
  \BibitemOpen
  \bibfield  {author} {\bibinfo {author} {\bibfnamefont {F.}~\bibnamefont
  {See\ss{}elberg}}, \bibinfo {author} {\bibfnamefont {X.-Y.}\ \bibnamefont
  {Luo}}, \bibinfo {author} {\bibfnamefont {M.}~\bibnamefont {Li}}, \bibinfo
  {author} {\bibfnamefont {R.}~\bibnamefont {Bause}}, \bibinfo {author}
  {\bibfnamefont {S.}~\bibnamefont {Kotochigova}}, \bibinfo {author}
  {\bibfnamefont {I.}~\bibnamefont {Bloch}},\ and\ \bibinfo {author}
  {\bibfnamefont {C.}~\bibnamefont {Gohle}},\ }\bibfield  {title} {\bibinfo
  {title} {Extending rotational coherence of interacting polar molecules in a
  spin-decoupled magic trap},\ }\href
  {https://doi.org/10.1103/PhysRevLett.121.253401} {\bibfield  {journal}
  {\bibinfo  {journal} {Phys. Rev. Lett.}\ }\textbf {\bibinfo {volume} {121}},\
  \bibinfo {pages} {253401} (\bibinfo {year} {2018})}\BibitemShut {NoStop}%
\bibitem [{\citenamefont {Blackmore}\ \emph {et~al.}(2018)\citenamefont
  {Blackmore}, \citenamefont {Caldwell}, \citenamefont {Gregory}, \citenamefont
  {Bridge}, \citenamefont {Sawant}, \citenamefont {Aldegunde}, \citenamefont
  {Mur-Petit}, \citenamefont {Jaksch}, \citenamefont {Hutson}, \citenamefont
  {Sauer}, \citenamefont {Tarbutt},\ and\ \citenamefont
  {Cornish}}]{Blackmore2018}%
  \BibitemOpen
  \bibfield  {author} {\bibinfo {author} {\bibfnamefont {J.~A.}\ \bibnamefont
  {Blackmore}}, \bibinfo {author} {\bibfnamefont {L.}~\bibnamefont {Caldwell}},
  \bibinfo {author} {\bibfnamefont {P.~D.}\ \bibnamefont {Gregory}}, \bibinfo
  {author} {\bibfnamefont {E.~M.}\ \bibnamefont {Bridge}}, \bibinfo {author}
  {\bibfnamefont {R.}~\bibnamefont {Sawant}}, \bibinfo {author} {\bibfnamefont
  {J.}~\bibnamefont {Aldegunde}}, \bibinfo {author} {\bibfnamefont
  {J.}~\bibnamefont {Mur-Petit}}, \bibinfo {author} {\bibfnamefont
  {D.}~\bibnamefont {Jaksch}}, \bibinfo {author} {\bibfnamefont {J.~M.}\
  \bibnamefont {Hutson}}, \bibinfo {author} {\bibfnamefont {B.~E.}\
  \bibnamefont {Sauer}}, \bibinfo {author} {\bibfnamefont {M.~R.}\ \bibnamefont
  {Tarbutt}},\ and\ \bibinfo {author} {\bibfnamefont {S.~L.}\ \bibnamefont
  {Cornish}},\ }\bibfield  {title} {\bibinfo {title} {Ultracold molecules for
  quantum simulation: rotational coherences in {CaF} and {RbCs}},\ }\href
  {https://doi.org/10.1088/2058-9565/aaee35} {\bibfield  {journal} {\bibinfo
  {journal} {Quantum Sci. Technol.}\ }\textbf {\bibinfo {volume} {4}},\
  \bibinfo {pages} {014010} (\bibinfo {year} {2018})}\BibitemShut {NoStop}%
\bibitem [{\citenamefont {{McCarron}}\ \emph {et~al.}(2011)\citenamefont
  {{McCarron}}, \citenamefont {Cho}, \citenamefont {Jenkin}, \citenamefont
  {K\"{o}ppinger},\ and\ \citenamefont {Cornish}}]{McCarron2011}%
  \BibitemOpen
  \bibfield  {author} {\bibinfo {author} {\bibfnamefont {D.~J.}\ \bibnamefont
  {{McCarron}}}, \bibinfo {author} {\bibfnamefont {H.~W.}\ \bibnamefont {Cho}},
  \bibinfo {author} {\bibfnamefont {D.~L.}\ \bibnamefont {Jenkin}}, \bibinfo
  {author} {\bibfnamefont {M.~P.}\ \bibnamefont {K\"{o}ppinger}},\ and\
  \bibinfo {author} {\bibfnamefont {S.~L.}\ \bibnamefont {Cornish}},\
  }\bibfield  {title} {\bibinfo {title} {Dual-species {Bose-Einstein}
  condensate of $^{87}${Rb} and $^{133}${Cs}},\ }\href
  {https://doi.org/10.1103/PhysRevA.84.011603} {\bibfield  {journal} {\bibinfo
  {journal} {Phys. Rev. A}\ }\textbf {\bibinfo {volume} {84}},\ \bibinfo
  {pages} {011603} (\bibinfo {year} {2011})}\BibitemShut {NoStop}%
\bibitem [{\citenamefont {K\"{o}ppinger}\ \emph {et~al.}(2014)\citenamefont
  {K\"{o}ppinger}, \citenamefont {{McCarron}}, \citenamefont {Jenkin},
  \citenamefont {Molony}, \citenamefont {Cho}, \citenamefont {Cornish},
  \citenamefont {{Le Sueur}}, \citenamefont {Blackley},\ and\ \citenamefont
  {Hutson}}]{Koeppinger2014}%
  \BibitemOpen
  \bibfield  {author} {\bibinfo {author} {\bibfnamefont {M.~P.}\ \bibnamefont
  {K\"{o}ppinger}}, \bibinfo {author} {\bibfnamefont {D.~J.}\ \bibnamefont
  {{McCarron}}}, \bibinfo {author} {\bibfnamefont {D.~L.}\ \bibnamefont
  {Jenkin}}, \bibinfo {author} {\bibfnamefont {P.~K.}\ \bibnamefont {Molony}},
  \bibinfo {author} {\bibfnamefont {H.~W.}\ \bibnamefont {Cho}}, \bibinfo
  {author} {\bibfnamefont {S.~L.}\ \bibnamefont {Cornish}}, \bibinfo {author}
  {\bibfnamefont {C.~R.}\ \bibnamefont {{Le Sueur}}}, \bibinfo {author}
  {\bibfnamefont {C.~L.}\ \bibnamefont {Blackley}},\ and\ \bibinfo {author}
  {\bibfnamefont {J.~M.}\ \bibnamefont {Hutson}},\ }\bibfield  {title}
  {\bibinfo {title} {Production of optically trapped $^{87}${RbCs} {Feshbach}
  molecules},\ }\href {https://doi.org/10.1103/PhysRevA.89.033604} {\bibfield
  {journal} {\bibinfo  {journal} {Phys. Rev. A}\ }\textbf {\bibinfo {volume}
  {89}},\ \bibinfo {pages} {033604} (\bibinfo {year} {2014})}\BibitemShut
  {NoStop}%
\bibitem [{\citenamefont {Gregory}\ \emph {et~al.}(2015)\citenamefont
  {Gregory}, \citenamefont {Molony}, \citenamefont {K\"{o}ppinger},
  \citenamefont {Kumar}, \citenamefont {Ji}, \citenamefont {Lu}, \citenamefont
  {Marchant},\ and\ \citenamefont {Cornish}}]{Gregory2015}%
  \BibitemOpen
  \bibfield  {author} {\bibinfo {author} {\bibfnamefont {P.~D.}\ \bibnamefont
  {Gregory}}, \bibinfo {author} {\bibfnamefont {P.~K.}\ \bibnamefont {Molony}},
  \bibinfo {author} {\bibfnamefont {M.~P.}\ \bibnamefont {K\"{o}ppinger}},
  \bibinfo {author} {\bibfnamefont {A.}~\bibnamefont {Kumar}}, \bibinfo
  {author} {\bibfnamefont {Z.}~\bibnamefont {Ji}}, \bibinfo {author}
  {\bibfnamefont {B.}~\bibnamefont {Lu}}, \bibinfo {author} {\bibfnamefont
  {A.~L.}\ \bibnamefont {Marchant}},\ and\ \bibinfo {author} {\bibfnamefont
  {S.~L.}\ \bibnamefont {Cornish}},\ }\bibfield  {title} {\bibinfo {title} {A
  simple, versatile laser system for the creation of ultracold ground state
  molecules},\ }\href {https://doi.org/10.1088/1367-2630/17/5/055006}
  {\bibfield  {journal} {\bibinfo  {journal} {New J. Phys.}\ }\textbf {\bibinfo
  {volume} {17}},\ \bibinfo {pages} {055006} (\bibinfo {year}
  {2015})}\BibitemShut {NoStop}%
\bibitem [{\citenamefont {Molony}\ \emph
  {et~al.}(2016{\natexlab{a}})\citenamefont {Molony}, \citenamefont {Kumar},
  \citenamefont {Gregory}, \citenamefont {Kliese}, \citenamefont {Puppe},
  \citenamefont {Le~Sueur}, \citenamefont {Aldegunde}, \citenamefont {Hutson},\
  and\ \citenamefont {Cornish}}]{Molony2016}%
  \BibitemOpen
  \bibfield  {author} {\bibinfo {author} {\bibfnamefont {P.~K.}\ \bibnamefont
  {Molony}}, \bibinfo {author} {\bibfnamefont {A.}~\bibnamefont {Kumar}},
  \bibinfo {author} {\bibfnamefont {P.~D.}\ \bibnamefont {Gregory}}, \bibinfo
  {author} {\bibfnamefont {R.}~\bibnamefont {Kliese}}, \bibinfo {author}
  {\bibfnamefont {T.}~\bibnamefont {Puppe}}, \bibinfo {author} {\bibfnamefont
  {C.~R.}\ \bibnamefont {Le~Sueur}}, \bibinfo {author} {\bibfnamefont
  {J.}~\bibnamefont {Aldegunde}}, \bibinfo {author} {\bibfnamefont {J.~M.}\
  \bibnamefont {Hutson}},\ and\ \bibinfo {author} {\bibfnamefont {S.~L.}\
  \bibnamefont {Cornish}},\ }\bibfield  {title} {\bibinfo {title} {Measurement
  of the binding energy of ultracold $^{87}${Rb}$^{133}${Cs} molecules using an
  offset-free optical frequency comb},\ }\href
  {https://doi.org/10.1103/PhysRevA.94.022507} {\bibfield  {journal} {\bibinfo
  {journal} {Phys. Rev. A}\ }\textbf {\bibinfo {volume} {94}},\ \bibinfo
  {pages} {022507} (\bibinfo {year} {2016}{\natexlab{a}})}\BibitemShut
  {NoStop}%
\bibitem [{\citenamefont {Molony}\ \emph
  {et~al.}(2016{\natexlab{b}})\citenamefont {Molony}, \citenamefont {Gregory},
  \citenamefont {Kumar}, \citenamefont {Le~Sueur}, \citenamefont {Hutson},\
  and\ \citenamefont {Cornish}}]{Molony2016a}%
  \BibitemOpen
  \bibfield  {author} {\bibinfo {author} {\bibfnamefont {P.~K.}\ \bibnamefont
  {Molony}}, \bibinfo {author} {\bibfnamefont {P.~D.}\ \bibnamefont {Gregory}},
  \bibinfo {author} {\bibfnamefont {A.}~\bibnamefont {Kumar}}, \bibinfo
  {author} {\bibfnamefont {C.~R.}\ \bibnamefont {Le~Sueur}}, \bibinfo {author}
  {\bibfnamefont {J.~M.}\ \bibnamefont {Hutson}},\ and\ \bibinfo {author}
  {\bibfnamefont {S.~L.}\ \bibnamefont {Cornish}},\ }\bibfield  {title}
  {\bibinfo {title} {Production of ultracold $^{87}${Rb}$^{133}${Cs} in the
  absolute ground state: complete characterisation of the {STIRAP} transfer},\
  }\href {https://doi.org/10.1002/cphc.201600501} {\bibfield  {journal}
  {\bibinfo  {journal} {ChemPhysChem.}\ }\textbf {\bibinfo {volume} {17}},\
  \bibinfo {pages} {3811} (\bibinfo {year} {2016}{\natexlab{b}})}\BibitemShut
  {NoStop}%
\bibitem [{\citenamefont {Allouche}\ \emph {et~al.}(2000)\citenamefont
  {Allouche}, \citenamefont {Korek}, \citenamefont {Fakherddin}, \citenamefont
  {Chaalan}, \citenamefont {Dagher}, \citenamefont {Taher},\ and\ \citenamefont
  {Aubert-Fr{\'{e}}con}}]{Allouche2000}%
  \BibitemOpen
  \bibfield  {author} {\bibinfo {author} {\bibfnamefont {A.~R.}\ \bibnamefont
  {Allouche}}, \bibinfo {author} {\bibfnamefont {M.}~\bibnamefont {Korek}},
  \bibinfo {author} {\bibfnamefont {K.}~\bibnamefont {Fakherddin}}, \bibinfo
  {author} {\bibfnamefont {A.}~\bibnamefont {Chaalan}}, \bibinfo {author}
  {\bibfnamefont {M.}~\bibnamefont {Dagher}}, \bibinfo {author} {\bibfnamefont
  {F.}~\bibnamefont {Taher}},\ and\ \bibinfo {author} {\bibfnamefont
  {M.}~\bibnamefont {Aubert-Fr{\'{e}}con}},\ }\bibfield  {title} {\bibinfo
  {title} {Theoretical electronic structure of {RbCs} revisited},\ }\href
  {https://doi.org/10.1088/0953-4075/33/12/312} {\bibfield  {journal} {\bibinfo
   {journal} {J. Phys. B: At., Mol. Opt. Phys.}\ }\textbf {\bibinfo {volume}
  {33}},\ \bibinfo {pages} {2307} (\bibinfo {year} {2000})}\BibitemShut
  {NoStop}%
\bibitem [{\citenamefont {Docenko}\ \emph {et~al.}(2010)\citenamefont
  {Docenko}, \citenamefont {Tamanis}, \citenamefont {Ferber}, \citenamefont
  {Bergeman}, \citenamefont {Kotochigova}, \citenamefont {Stolyarov},
  \citenamefont {de~Faria~Nogueira},\ and\ \citenamefont
  {Fellows}}]{Docenko2010}%
  \BibitemOpen
  \bibfield  {author} {\bibinfo {author} {\bibfnamefont {O.}~\bibnamefont
  {Docenko}}, \bibinfo {author} {\bibfnamefont {M.}~\bibnamefont {Tamanis}},
  \bibinfo {author} {\bibfnamefont {R.}~\bibnamefont {Ferber}}, \bibinfo
  {author} {\bibfnamefont {T.}~\bibnamefont {Bergeman}}, \bibinfo {author}
  {\bibfnamefont {S.}~\bibnamefont {Kotochigova}}, \bibinfo {author}
  {\bibfnamefont {A.~V.}\ \bibnamefont {Stolyarov}}, \bibinfo {author}
  {\bibfnamefont {A.}~\bibnamefont {de~Faria~Nogueira}},\ and\ \bibinfo
  {author} {\bibfnamefont {C.~E.}\ \bibnamefont {Fellows}},\ }\bibfield
  {title} {\bibinfo {title} {Spectroscopic data, spin-orbit functions, and
  revised analysis of strong perturbative interactions for the
  {${A^1}\Sigma^{+}$ and $b^{3}\Pi$} states of {RbCs}},\ }\href
  {https://doi.org/10.1103/PhysRevA.81.042511} {\bibfield  {journal} {\bibinfo
  {journal} {Phys. Rev. A}\ }\textbf {\bibinfo {volume} {81}},\ \bibinfo
  {pages} {042511} (\bibinfo {year} {2010})}\BibitemShut {NoStop}%
\bibitem [{\citenamefont {Gregory}\ \emph {et~al.}(2019)\citenamefont
  {Gregory}, \citenamefont {Frye}, \citenamefont {Blackmore}, \citenamefont
  {Bridge}, \citenamefont {Sawant}, \citenamefont {Hutson},\ and\ \citenamefont
  {Cornish}}]{Gregory2019}%
  \BibitemOpen
  \bibfield  {author} {\bibinfo {author} {\bibfnamefont {P.~D.}\ \bibnamefont
  {Gregory}}, \bibinfo {author} {\bibfnamefont {M.~D.}\ \bibnamefont {Frye}},
  \bibinfo {author} {\bibfnamefont {J.~A.}\ \bibnamefont {Blackmore}}, \bibinfo
  {author} {\bibfnamefont {E.~M.}\ \bibnamefont {Bridge}}, \bibinfo {author}
  {\bibfnamefont {R.}~\bibnamefont {Sawant}}, \bibinfo {author} {\bibfnamefont
  {J.~M.}\ \bibnamefont {Hutson}},\ and\ \bibinfo {author} {\bibfnamefont
  {S.~L.}\ \bibnamefont {Cornish}},\ }\bibfield  {title} {\bibinfo {title}
  {Sticky collisions of ultracold {RbCs} molecules},\ }\href
  {https://doi.org/10.1038/s41467-019-11033-y} {\bibfield  {journal} {\bibinfo
  {journal} {Nat. Commun.}\ }\textbf {\bibinfo {volume} {10}},\ \bibinfo
  {pages} {3104} (\bibinfo {year} {2019})}\BibitemShut {NoStop}%
\bibitem [{\citenamefont {Gregory}\ \emph {et~al.}(2020)\citenamefont
  {Gregory}, \citenamefont {Blackmore}, \citenamefont {Bromley},\ and\
  \citenamefont {Cornish}}]{Gregory2020}%
  \BibitemOpen
  \bibfield  {author} {\bibinfo {author} {\bibfnamefont {P.~D.}\ \bibnamefont
  {Gregory}}, \bibinfo {author} {\bibfnamefont {J.~A.}\ \bibnamefont
  {Blackmore}}, \bibinfo {author} {\bibfnamefont {S.~L.}\ \bibnamefont
  {Bromley}},\ and\ \bibinfo {author} {\bibfnamefont {S.~L.}\ \bibnamefont
  {Cornish}},\ }\bibfield  {title} {\bibinfo {title} {Loss of ultracold
  $^{87}${Rb}$^{133}${Cs} molecules via optical excitation of long-lived
  two-body collision complexes},\ }\href
  {https://doi.org/10.1103/PhysRevLett.124.163402} {\bibfield  {journal}
  {\bibinfo  {journal} {Phys. Rev. Lett.}\ }\textbf {\bibinfo {volume} {124}},\
  \bibinfo {pages} {163402} (\bibinfo {year} {2020})}\BibitemShut {NoStop}%
\bibitem [{\citenamefont {Fahs}\ \emph {et~al.}(2002)\citenamefont {Fahs},
  \citenamefont {Allouche}, \citenamefont {Korek},\ and\ \citenamefont
  {Aubert-Frécon}}]{Fahs2002}%
  \BibitemOpen
  \bibfield  {author} {\bibinfo {author} {\bibfnamefont {H.}~\bibnamefont
  {Fahs}}, \bibinfo {author} {\bibfnamefont {A.~R.}\ \bibnamefont {Allouche}},
  \bibinfo {author} {\bibfnamefont {M.}~\bibnamefont {Korek}},\ and\ \bibinfo
  {author} {\bibfnamefont {M.}~\bibnamefont {Aubert-Frécon}},\ }\bibfield
  {title} {\bibinfo {title} {The theoretical spin-orbit structure of the {RbCs}
  molecule},\ }\href {https://doi.org/10.1088/0953-4075/35/6/307} {\bibfield
  {journal} {\bibinfo  {journal} {J. Phys. B: At., Mol. Opt. Phys.}\ }\textbf
  {\bibinfo {volume} {35}},\ \bibinfo {pages} {1501} (\bibinfo {year}
  {2002})}\BibitemShut {NoStop}%
\bibitem [{\citenamefont {Christianen}\ \emph {et~al.}(2019)\citenamefont
  {Christianen}, \citenamefont {Zwierlein}, \citenamefont {Groenenboom},\ and\
  \citenamefont {Karman}}]{Christianen2019}%
  \BibitemOpen
  \bibfield  {author} {\bibinfo {author} {\bibfnamefont {A.}~\bibnamefont
  {Christianen}}, \bibinfo {author} {\bibfnamefont {M.~W.}\ \bibnamefont
  {Zwierlein}}, \bibinfo {author} {\bibfnamefont {G.~C.}\ \bibnamefont
  {Groenenboom}},\ and\ \bibinfo {author} {\bibfnamefont {T.}~\bibnamefont
  {Karman}},\ }\bibfield  {title} {\bibinfo {title} {Photoinduced two-body loss
  of ultracold molecules},\ }\href
  {https://doi.org/10.1103/PhysRevLett.123.123402} {\bibfield  {journal}
  {\bibinfo  {journal} {Phys. Rev. Lett.}\ }\textbf {\bibinfo {volume} {123}},\
  \bibinfo {pages} {123402} (\bibinfo {year} {2019})}\BibitemShut {NoStop}%
\bibitem [{\citenamefont {Safronova}\ \emph {et~al.}(2006)\citenamefont
  {Safronova}, \citenamefont {Arora},\ and\ \citenamefont
  {Clark}}]{Safronova2006}%
  \BibitemOpen
  \bibfield  {author} {\bibinfo {author} {\bibfnamefont {M.}~\bibnamefont
  {Safronova}}, \bibinfo {author} {\bibfnamefont {B.}~\bibnamefont {Arora}},\
  and\ \bibinfo {author} {\bibfnamefont {C.}~\bibnamefont {Clark}},\ }\bibfield
   {title} {\bibinfo {title} {Frequency-dependent polarizabilities of
  alkali-metal atoms from ultraviolet through infrared spectral regions},\
  }\href {https://doi.org/10.1103/physreva.73.022505} {\bibfield  {journal}
  {\bibinfo  {journal} {Phys. Rev. A}\ }\textbf {\bibinfo {volume} {73}},\
  \bibinfo {pages} {022505} (\bibinfo {year} {2006})}\BibitemShut {NoStop}%
\end{thebibliography}%
\end{document}